\documentclass[12pt,preprint]{aastex}

\def\ltsima{$\;\buildrel < \over \sim \;$}
\def\simlt{\lower.5ex \hbox{\ltsima}}
\def\lesssim{\simlt}
\def\gtsima{$\;\buildrel > \over \sim \;$}
\def\simgt{\lower.5ex \hbox{\gtsima}}
\def\gtrsim{\simgt}

\begin{document}

\title{Atomic Hydrogen Gas in Dark-Matter Minihalos 
and the Compact High Velocity Clouds}

\vspace{1.5cm}

\author{Amiel Sternberg}
\vspace{0.5cm}
\affil{School of Physics and Astronomy and the Wise Observatory,
        The Beverly and Raymond Sackler Faculty of Exact Sciences,
        Tel Aviv University, Tel Aviv 69978, Israel;
        amiel@wise.tau.ac.il}
\vspace{0.5cm}
\author{Christopher F. McKee}
\vspace{0.5cm}
\affil{Physics Department, and Astronomy Department, 
University of California at Berkeley, Berkeley, CA 94720}
\vspace{0.5cm}
\and
\vspace{0.5cm}
\author{Mark G. Wolfire}
\vspace{0.5cm}
\affil{Astronomy Department, University of Maryland, College Park, MD 20742}

\received{}
\accepted{}

\begin{abstract}

We calculate the coupled hydrostatic and ionization structures of
spherically symmetric, pressure-supported gas
clouds that are confined by gravitationally dominant
dark-matter (DM) mini-halos and by an
external bounding pressure provided by a hot medium.
We focus on clouds that are
photoionized and heated by the present-day background
metagalactic field and determine the conditions for the
formation of warm (WNM), and multi-phased (CNM/WNM)
neutral atomic hydrogen (HI) cores
in the DM-dominated clouds.

We consider $\Lambda$CDM 
dark-matter halos with cuspy (NFW) and constant density (Burkert)
cores. 
We compute models for a wide range of halo masses, total cloud gas masses,
and external bounding pressures. 

We present models for the pressure-supported HI structures
observed in the Local Group dwarf irregular galaxies Leo A and Sag DIG.
We find that the hydrogen gas becomes neutral for
projected HI column densities exceeding $10^{19}$ cm$^{-2}$.
We identify the HI cloud boundaries observed in Leo A and Sag DIG with
the ionization fronts, and we derive an upper limit of 
$P_{\rm HIM}/k \lesssim 100$ cm$^{-3}$ K for the ambient 
pressure of the intergalactic medium in the Local Group. 
The observed HI gas scale heights in Leo A and Sag DIG imply
characteristic DM densities of $1.2$ amu cm$^{-3}$
(or $0.03 M_\odot$ pc$^{-3}$),
consistent with the DM densities previously 
inferred via HI rotation curve
studies of dwarf and low-surface brightness galaxies.
Leo A and Sag DIG obey the scaling correlations
that are expected for typical (median) 
DM halos in a $\Lambda$CDM cosmology, provided the
halos contain constant density cores,
as suggested by Burkert.

We construct explicit ``mini-halo'' models for the multi-phased
(and low-metallicity)
compact high-velocity HI clouds (CHVCs).
If the CHVC halos are drawn from the same family of
halos that successfully reproduce the dwarf galaxy observations,
then the CHVCs must be ``circumgalactic'' objects, 
with characteristic distances of 150 kpc. 
For such systems
we find that multi-phased behavior occurs for
peak WNM HI column densities between $2\times 10^{19}$ and
$1\times 10^{20}$ cm$^{-2}$, consistent with observations.
In contrast, if the CHVCs are ``extragalactic'' objects
with distances $\gtrsim 750$ kpc, then their associated halos
must be very ``underconcentrated'', with characteristic
DM densities $\lesssim 0.08$ cm$^{-3}$, much
smaller than expected for their mass, and significantly
smaller than observed in the dwarf galaxies. Furthermore,
multi-phased cores then require higher shielding columns.
We favor the circumgalactic hypothesis. 
If the large population of CHVCs represent ``missing
low-mass DM satellites'' of the Galaxy, then
these HI clouds must be pressure-confined to keep the gas neutral.
For an implied CHVC mini-halo scale velocity of $v_s=12$ km s$^{-1}$, the
confining pressure must exceed $\sim 50$ cm$^{-3}$ K.
A hot ($\sim 2\times 10^6$ K) Galactic corona
could provide the required pressure at 150 kpc.

Our static mini-halo models are able to account for many
properties of the CHVCs, including their observed
peak HI columns, core sizes, and multi-phased behavior. However,
important difficulties remain, including the presence in some objects of
extended low column density HI wings, and
H$\alpha$ emission line fluxes in several CHVCs that are
significantly larger than expected.

\end{abstract}


\section{Introduction}

High velocity clouds (HVCs) are atomic hydrogen (HI) clouds with
radial velocities inconsistent with gas
in differential circular rotation in the Galactic
disk. Since their discovery via 21 cm observations
by Muller, Oort \& Raimond (1963) the HVCs have been 
observed, surveyed, and catalogued in ever increasing detail
(Bajaja et al. 1987; Hulsbosch \& Wakker 1988; 
Wakker \& van Woerden 1991; Hartmann \& Burton 1997; Braun \& Burton 2000;
Br\"uns et al. 2000; 
Burton, Braun \& Chengalur 2001; Putman et al. 2002; Lockman et al. 2002; 
de Heij, Braun \& Burton 2002a).
However, the nature and origin of the HVCs
have remained the subject of considerable and unresolved debate
(Oort 1966; Wakker \& van Woerden 1997; Blitz et al. 1999;
Wakker, van Woerden \& Gibson 1999; Gibson et al.~2002; Sembach 2002).

The HVCs consist of several hundred compact, kinematically distinct,
clouds with typical angular diameters of $\sim 1-2^\circ$
(Braun \& Burton 2000; Putman et al. 2002), as well as
several large (diffuse) and kinematically
continuous complexes (e.g., the ``A, B, C, H, \& M'' clouds)
that extend over $\sim 100$ square degrees.
The HVCs appear to be devoid of any stellar counterparts.
The radial velocities of the HVCs identified in the
Wakker \& van Woerden (1991) catalogue range from
$-$464 to +297 km s$^{-1}$ relative to the local standard of rest.
de Heij, Braun \& Burton (2002b) 
identified 67 distinct compact high velocity clouds
(CHVCs) in the single-dish 
``Leiden/Dwingeloo'' HI survey (Hartmann \& Burton 1997).
Putman et al. (2002) identified an additional 179 
CHVCs in the (more sensitive)
southern-hemisphere ``HI Parkes All Sky Survey'' (HIPASS).
Most of the CHVCs are infalling. 
The ``Leiden/Dwingeloo'' survey,  
carried out with angular and velocity resolutions
of $0\fdg5$ and 1 km s$^{-1}$ respectively,
shows that the typical (one-dimensional) velocity
dispersions of the CHVCs are $\sim 14$ km s$^{-1}$ with
little scatter about this value (Blitz et al. 1999).
The velocity dispersions imply gas temperatures $\sim 10^4$ K,
consistent with a warm medium (WM) possibly consisting of a mixture
of a warm neutral (WNM) and ionized (WIM) gas.
Typically, N$_{\rm HI}\sim 5\times 10^{18}$ cm$^{-2}$
averaged over the extents of the CHVCs. However,
the CHVCs are barely resolved with the (25-meter) Dwingeloo 
and (64-meter) Parkes telescopes. 
More recently Burton et al. (2001) were able to resolve the spatial
structures of ten CHVCs in higher resolution ($3.5 \times 3.7$ arcmin)
observations with the (305-meter) Arecibo telescope.
They found that the typical 
observed (projected) $1/e$ exponential scale-lengths of
the HI distributions equal 690 arcsec, with central HI columns 
ranging from $2\times 10^{19}$ to $2\times 10^{20}$ cm$^{-2}$.
In some objects, 
gas was detected out to $\sim 1^\circ$ from the cloud centers
down to column densities of $\sim 2\times 10^{17}$ cm$^{-2}$.
The line widths are slightly narrower in the Arecibo
data set compared to the Leiden/Dwingeloo observations,
with corresponding velocity dispersions of $\sim 11$ km s$^{-1}$.
The velocity gradients across the clouds are
relatively small, $\sim 10$ km s$^{-1}$ degree$^{-1}$,
and the WNM cloud dynamics are controlled  
primarily by thermal motions. 

In another important development
Braun \& Burton (2000) and de Heij, Braun \& Burton (2002c)
carried out high resolution Westerbork
interferometric observations (with $\sim 1$ arcmin synthesized beams)
and found that at least some
CHVCs contain high column density (up to $\sim 10^{21}$ cm$^{-2}$)
cores, with typical radii $\sim 10$ arcmin, and line profiles as narrow as
2 km s$^{-1}$.
The narrow line widths imply
HI gas in a cold neutral medium  
(CNM; $T\lesssim 100$ K). Previous to this work
CNM cores were known to exist only in the extended
A, H, and M complexes (Wakker \& Schwarz 1991).
The recent single-dish and interferometric 
observations reveal that ``core/envelope''
CNM/WNM structures may be a common feature of the CHVCs.

The distance is the critical unknown for the HVCs. The
large complexes are plausibly nearby objects, and stellar
absorption-line observations in the large M and A clouds
(Danly, Albert \& Kuntz 1993; van Woerden et al. 1999) constrain their
distances to within 5 and 10 kpc respectively. However, 
no direct distance determinations are yet available for the large
population of CHVCs. 

Many theories for the origin of the HVCs have been proposed over
the years (Wakker \& van Woerden 1997). 
Oort (e.g.~1966; 1970) recognized
that if the HVCs are gravitationally bound objects with masses
dominated mainly by the observed HI gas they
must be distant, with distances 
$d\gtrsim \Delta v^2f/\pi Gm_{\rm H}N_{\rm HI}\theta$,
where $\Delta v$ is the observed velocity dispersion,
$N_{\rm HI}$ is the mean HI column within angular radius $\theta$, 
and $f$ is the fraction of the total dynamical mass present as
HI gas. If the clouds are bound, then the required
value of $f$ is proportional to the assumed distance $d$,
since the HI mass varies as $d^2$, whereas the dynamical mass 
$M_{\rm dyn}\equiv \Delta v^2d\theta/G \propto d$.
For $f=1$ the observations imply
implausibly large ``binding distances'' $d\gtrsim 10$ Mpc,
and HI masses $M_{\rm HI}\equiv \pi d^2\theta^2m_{\rm H} N_{\rm HI}\ \gtrsim 10^9 M_\odot$.
Oort's preferred interpretation was that the HVCs are
transient ``circumgalactic'' ($\lesssim 100$ kpc) 
primordial gas clouds that have recently condensed 
out of the intergalactic medium,
and are only now being accreted onto the disk
in the end-stages of the Galaxy formation process.
Shapiro \& Field (1976) introduced the idea of
the Galactic fountain (see also Bregman 1996)
in which hot gas ejected from the disk cools and condenses
into neutral clouds that fall back onto the Galaxy in
a continuing cycle. Giovanelli (1981)
suggested that the HVCs are scattered fragments of the Magellanic 
Stream.\footnote{
The ``Magellanic Stream" (Mathewson, Cleary \& Murray 1974)
is an example of extended HVC gas that is relatively
well understood. It consists of a narrow trail of HI gas that emanates
from the Magellanic Clouds, and stretches over 100$^\circ$
across the sky.  The stream is almost certainly gas that has
been tidally (Gardiner \& Noguchi 1996)
or ram-pressure stripped (Moore \& Davis 1994)
from the Magellanic Clouds as these satellite galaxies orbit the Milky Way.}  
A Local Group\footnote{
We recall that the LG is a dynamically bound system of galaxies consisting
of the Milky Way and its two largest neighbors M31 and M33, and 
40 or more smaller and fainter dwarf galaxies (Mateo 1998). 
The LG is typical of the many small groups and clusters of
galaxies situated along the outer boundary of the Virgo supercluster
(van den Bergh 1999). The LG radius 
defined as the zero-velocity surface relative to the Hubble flow
is $\sim 1.5$ Mpc, and the characteristic 
velocity dispersion is $\sim 60$ km s$^{-1}$ (Sandage 1986). 
The mass of the LG is dominated by the Milky Way and M31, and equals about 
$3.3\times 10^{12} M_\odot$.}
 (LG) origin for the HVCs was also proposed and
discussed by various authors in the years immediately following
their discovery (Oort 1966; Verschuur 1969; Einasto et al. 1974; Eichler 1976).
In this picture the HVCs are genuine 
extragalactic objects associated with the LG
system, and are at typical distances of $\sim 1$ Mpc, 
rather than being a structural feature of the Galaxy 
and its immediate surroundings.

The distances to the clouds may be constrained via
observations of optical H$\alpha$ recombination 
line emission that provides a measure of the
ionizing radiation field incident on the clouds
(Weiner \& Williams 1996; Bland-Hawthorn et al. 1998;
Tufte et al. 1998; Weiner, Vogel \& Williams 2001). 
At sufficiently large distances from the Galaxy,
the dominant radiation field becomes the weak metagalactic background
and the clouds should be weak H$\alpha$ emitters.
Nearby clouds might be ionized by more intense ``leakage 
radiation'' from the Galactic disk, and could be more intense 
H$\alpha$ sources.  
Sensitive optical Fabry-Perot H$\alpha$ 
observations have been reported and compiled
by Weiner et al. (2001) for several HVC complexes
and one isolated CHVC. Tufte et al. (2002) have reported 
H$\alpha$ observations in five CHVCs.
For the complexes Weiner et al. find that the
measured H$\alpha$ surface brightnesses range from
41 to 1680 milli-Rayleighs
\footnote{
1 Rayleigh = $10^6/4\pi$ photons cm$^{-2}$ s$^{-1}$ sr$^{-1}$.
For case-B recombination at a temperature of 10$^4$ K,
a surface brightness of 1 Rayleigh is produced by a slab
with emission measure $EM = 2.8$ cm$^{-6}$ pc,
which for two-sided illumination in ionization
equilibrium, is produced by a Lyman continuum photon
intensity $J^* = 3.6\times 10^5$ photons cm$^{-2}$ s$^{-1}$ sr$^{-1}$.}
(mR), implying photoionizing Lyman continuum (Lyc) intensities
of 1.5$\times 10^4$ to 6.0$\times 10^5$
photons cm$^{-2}$ s$^{-1}$ sr$^{-1}$. 
 Weiner et al. detected an H$\alpha$ intensity
of 48 mR in CHVC 230+61+165 (an object included in the 
Burton et al. [2001] Arecibo data set).
Tufte et al. (2002) report
H$\alpha$ intensities
in the range of 20 to 140 mR in four CHVCs, and an upper limit
of 20 mR in an additional CHVC. The detected fluxes are larger 
than the (2$\sigma$) upper limit
of 20 mR set by Vogel et al.~(1995) for the H$\alpha$ intensity of the
isolated extragalactic Giovanelli \& Haynes (1989) cloud that Vogel et al.
argue is probing the metagalactic field (see also Madsen et al. 2001;
Weymann et al. 2001).
Tufte et al. argue that the relatively large CHVC H$\alpha$ fluxes
imply that the CHVCs are located within the Galactic halo.
However, the distance estimates are uncertain due to 
uncertainties in the actual
strength of the metagalactic field, the Galactic radiation escape fraction,
and the role of collisional ionization.

The metallicity and dust content are additional potentially
important clues to the origin of the HVCs.
Thermal ($\sim 100$ $\mu$m)
emission signatures of dust grains have been searched for
but not detected (Wakker \& Boulanger 1986). However,
a few metallicity measurements are available,
mainly via ultraviolet absorption line studies toward
background quasars and Seyfert galaxies.
Significantly subsolar (though not
purely primordial) metal abundances are indicated,
primarily via observations of gas phase
sulfur
\footnote{
Sulfur is generally considered to be
robust probe of the intrinsic metallicity since S
remains undepleted in diffuse gas (Jenkins et al. 1987).}.
Low metallicities would appear to be inconsistent with a Galactic
fountain, and favor models in which the gas originates outside
the Galaxy. 
Lu et al. (1998) observed HVC 287+22+240 using the
Goddard High Resolution Spectrograph (GHRS) on board
{\it Hubble Space Telescope} (HST) and
detected several S{\small II} and
Fe{\small II} absorption lines. Lu et al.  
derived S/H=$0.25\pm 0.07$, Fe/H=$0.033\pm 0.006$,
and S/Fe=$7.6\pm 2.2$ (relative to solar abundances).  
They interpreted the supersolar S/Fe ratio as
being due to significant iron depletion and therefore evidence for
the presence of dust grains. In view of similar abundances observed
in the Magellanic Clouds, as well as the proximity of 287+22+240
to these objects, Lu et al. concluded that this HVC is
a ``leading arm'' component of the Magellanic stream.
Wakker et al. (1999) carried out GHRS
spectroscopy of HVC gas
in complex C and inferred S/H=$0.089\pm 0.024$.
They favor the Oort hypothesis,
and interpret complex C as infalling primordial gas that has
been ``contaminated'' by metals as it interacts with the Galactic halo.
Braun \& Burton (2000)
derived a similar low metallicity for one of the CHVCs in their sample 
(125+41-207).  They combined their HI observations together with Mg{\small II}
absorption measurements reported by Bowen \& Blades (1993), 
and inferred a Mg/H abundance in the range of 0.04 to 0.07 relative
to solar.

Recently, Murphy et al. (2000) and Sembach et al. (2000) 
carried out {\it Far Ultraviolet Spectroscopic Explorer} (FUSE)
absorption-line studies of several HVCs.
In complex C Murphy et al. found
Fe/H $\sim 0.5$ (relative to solar) that may indicate
undepleted iron and hence 
a very low dust content.
Sembach et al.~(2002) set a metallicity limit of $O/H < 0.46$
in CHVC 224-83-197 via the detection of O{\small I} absorption. 
An important result of the FUSE observations was the detection of
O{\small VI} absorption tracing highly ionized gas in several HVCs. 
These observations complement earlier  
GHRS observations of C{\small IV}, Si{\small IV} and N{\small V}
absorption in high velocity gas that Sembach et al. (1999)
suggested is produced in the ionized envelopes of neutral
hydrogen HVCs (as the ionized clouds do not appear to be 21 cm sources).
Sembach et al. (1999) argued that 
the observed C$^{+3}$, Si$^{+3}$ and N$^{+4}$ abundances
are consistent with low pressure ($P/k \sim 2$ cm$^{-3}$ K)
gas photoionized by an assumed Lyc flux at a level
plausibly consistent with the metagalactic background. 
They concluded that the C{\small IV}, Si{\small IV} and N{\small V}
absorbers are located at large distances and are 
Local Group objects. The O{\small VI}  
abundances are too large to
be compatible with photoionization, and may 
be produced collisionally at the interfaces between the
HVCs and a tenuous Galactic corona or Local Group medium
through which the they are moving (Sembach et al. 2000).
We note that collisional ionization (in gas with pressures
considerably larger than 2 cm$^{-3}$ K)
may also contribute to the production
of the C$^{+3}$, Si$^{+3}$ and N$^{+4}$ ions.  

Evidence of an ambient medium in which the CHVCs may be embedded
could also provide clues to their origin. Br\"uns, Kerp \& Pagels (2001)
mapped one of the CHVCs (125+41-207) in the Burton et al. sample 
with the 100m Effelsberg telescope. They found that the
WNM component is asymmetric with a ``cometary'' appearance.
The head-tail morphology suggests that gas is being stripped off
of the main body of the cloud as it moves through an
enveloping medium. 

The Local Group hypothesis for the HVCs, particularly the CHVCs,
 has received renewed interest
recently for additional observational
and theoretical reasons.  Bajaja et al. (1987) noted that
asymmetries in the ``position-velocity'' 
diagram of the HVC ensemble could be
understood if the HVCs are moving within the LG as a whole.  More recently
Blitz et al. (1999) and Braun \& Burton (1999) demonstrated that the
dispersion in CHVC velocities is minimized if the cloud velocities are
measured relative to the LG barycenter (see however, Gibson et al. 2000).
Furthermore, Blitz et al. presented a simulation of the LG dynamics that 
shows that ``test particles'' interacting with the Milky Way, M31,
and the distant Virgo cluster, 
are (during a Hubble time) drawn into an elongated
filamentary structure that
reproduces the observed distribution and velocities of HVCs
in the directions of the LG barycenter and anti-barycenter
(see however Moore \& Putman 2001; de Heij et al. 2002a).
These kinematic
considerations are strong evidence for a LG origin,
although the cloud distances are not very well constrained.
Blitz et al. concluded that the typical distances
of the CHVCs are $d \gtrsim 750$ kpc. For $d=750$ kpc, a 
characteristic radius of $\sim 2.5$ kpc
and a central gas density of $\sim 6\times 10^{-3}$ cm$^{-3}$
are implied by the Burton et al. (2001) observations. The 
HI mass in each cloud is $\sim 1\times 10^7 M_\odot$,
and the total HI mass in the system of CHVCs is $\gtrsim 10^9 M_\odot$.
Such large masses may be in conflict with the apparent lack
of similar objects around other galaxies and groups 
(Zwaan \& Briggs 2000; Rosenberg \& Schneider 2002).  
If the CHVCs are long-lived, gravitationally 
bound objects as conjectured by Blitz et al., then 
the measured velocity dispersions imply total dynamical masses
$M_{\rm dyn}\approx  5\times 10^8 M_\odot$,
comparable to dwarf galaxy masses.   In this picture the HI mass is
a fraction $f=0.02$ of the total dynamical mass,
and the 3.4 kpc radius corresponds to the
``scale height'' $\sim GM_{\rm dyn}/\Delta v^2$ of the gravitationally
confined gas.  Blitz et al. suggested that the 
dynamical mass is dominated by dark-matter and that the CHVCs are tracing
individual dark-matter ``halos''.  

The notion that the CHVCs represent dark-matter halos
is motivated by the cosmological theory of
hierarchical structure formation.
In this theory, 
bound systems (``halos'') of non-baryonic,
dark matter evolve in a hierarchical collapse, 
initiated by small gravitationally
unstable primordial density fluctuations
(Gunn \& Gott 1972; Press \& Schechter 1974;
Blumenthal et al. 1984; Navarro, Frenk \& White 1997; Bullock et al. 2001). 
Low-mass halos virialize first, 
and then merge into progressively more massive systems.
Galaxies form as gas accumulates, cools, 
and collapses inside the virialized DM halos (White \& Rees 1978).
Detailed numerical simulations of the evolving collapse
of dissipationless cold (non-relativistic) dark matter (CDM)
have been carried out by several investigators 
(e.g.~Navarro et al. 1997; hereafter NFW). 
The simulations predict the existence of many
more low-mass halos than observed at the faint-end
(i.e. dwarf galaxy scale) of the galaxy luminosity function.
Various suggestions have been made to account for this discrepancy, including:
blow out of gas by winds and supernovae produced in small initial starbursts
(Dekel \& Silk 1986; Tegmark, Silk \& Evrard 1993); photoionization heating 
of the baryonic component by the metagalactic
radiation field, that inhibits the collapse of gas into the low-mass halos 
(Efstathiou 1992; Thoul \& Weinberg 1996; Kepner, Babul \& Spergel 1997;
Kitayama \& Ikeuchi 2000; Bullock Kravtsov \& Weinberg 2000)
and expulsion of gas trapped within the low-mass halos during the epoch
of ``reionization'' (Barkana \& Loeb 1999).  In all of these scenarios
the low-mass halos (or ``mini-halos'' as coined by Rees 1986)
continue to exist, but because star-formation has
been inhibited in them, they are not observable as galaxies.

In two recent studies, Klypin et al. (1999) and Moore et al. (1999) 
presented high-resolution simulations of the evolution
of dark matter halos at the mass
and size scales of the Milky Way and M31 galaxies that
dominate the LG.  They found
that many low-mass sub-halos survive and 
continue to accrete into each of the dominating galaxy 
halos to the present day.  
In particular, the predicted number of low-mass satellite 
halos appears to exceed
the number of observed satellite dwarf galaxies within $\sim 300$ kpc
of the Milky Way.  Klypin et al. suggested that the CHVCs are in fact the 
``missing DM satellites'' that appear in their simulations.
In this picture the CHVCs are ``circumgalactic'' mini-halos
associated with the Milky Way, as opposed to ``extragalactic''
objects associated with the LG as proposed by Blitz et al.
The presence of DM substructure may be detectable
via the gravitational lensing of background quasars (Mao \& Schneider 1998;
Metcalf \& Madau 2001; Dalal \& Kochanek 2002). 

The CNM/WNM structures observed in the
CHVCs are in many ways similar to the multi-phased HI distributions observed
in gas-rich dwarf irregular (dIrrs) galaxies
\footnote{
Blitz \& Robishaw (2000) have recently shown that some dwarf
spheroidals (dSphs) may also be gas-rich, with the gas
being present at large offsets from the optical stellar light.}. 
As summarized by Mateo et al. (1998), the observed HI masses in 
the dIrrs range from 0.1 to 0.5 of the total galaxy mass
(including the dark matter). 
The HI gas is generally much more extended than 
the ``Holmberg radii" (0.1$-$1 kpc) of the 
stellar components as traced in the optical. 
Young \& Lo (1996; 1997) carried out detailed 
{\it Very Large Array} (VLA)
interferometric studies of the dIrrs Leo A and Sag DIG and
found that the observed HI line profiles
can be separated into broad and narrow components.  
The broad (WNM) components appear as ``envelopes'' 
throughout the mapped regions of the galaxies, 
whereas the narrow (CNM) components are associated 
with small clumps or cores located near star-forming regions.
The WNM envelopes appear to be pressure supported
as opposed to rotationally supported.  In these respects
the HI structures in the
dwarfs are similar to those observed in the CHVCs. 
However, there are important differences (in addition to the
fact that star forming regions are absent in the CHVCs).
First, the peak HI column densities of the WNM gas in the dwarfs
(up to a few 10$^{21}$ cm$^{-2}$) are significantly larger than
in the CHVCs (up to $\sim 10^{20}$ cm$^{-2}$). 
Second, the angular scale sizes of the HI distributions
are smaller in the dwarfs, $\sim 70$ arcsec in Sag DIG,
and $\sim 160$ in Leo A, compared to the CHVCs where the
characteristic size is $\sim 690$ arcsec. These facts will
be important in our analysis of the CHVCs and dwarf galaxies
as possibly related objects.  Another key difference is that
the distances to the dwarfs are well determined (via stellar
photometry) whereas the distances to the CHVCs are unknown.
 
In this paper we quantitatively examine the hypothesis
that the CHVCs are stable HI clouds confined by the gravitational
potentials of dark-matter halos.  We
wish to determine whether viable ``mini-halo'' 
models for the CHVCs can be
constructed as either ``circumgalactic'' 
($d\sim 100$ kpc) or ``extragalactic'' ($d\sim 1$ Mpc) LG objects.
Such clouds will be exposed to the background
metagalactic radiation field that
ionizes and heats the gas. The dark-matter confined clouds may
also be subjected to bounding pressures, provided for example, by
a hot circumgalactic Galactic corona, or by an intergalactic
medium filling the Local Group.

We address several questions.
Given the observed HI column densities and distributions what are the
halo virial masses and characteristic dark matter densities 
required to confine the gas clouds?
Are the required halo properties consistent with those
expected from cosmological structure formation simulations?
Under the assumption of photoionization by the metagalactic field, what are
the minimum gas masses and column densities
required for the formation
of neutral hydrogen cores within the dark-matter halos?
Are the observed HI distributions consistent with photoionization
by the metagalactic field?  Under what conditions do the cloud cores
become multi-phased? Can halo models be constructed for the
pressure supported HI clouds in the Local Group dwarf galaxies?
If the CHVCs are the starless ``cousins'' of the dwarf galaxies,
and the CHVC and dwarf galaxy halos are similar,
what are the implied distances to the CHVCs?

We construct general purpose models in which we
compute the gas density distributions
and ionization structures of pressure supported hydrostatic
hydrogen gas clouds that are
trapped in the potential wells of DM halos, 
are simultaneously heated and photoionized by an external radiation
field, and are subjected to specified bounding pressures.
Computations along these lines have been presented
in the literature (with varying degrees of sophistication)
mainly in the context of the mini-halo model
for the intergalactic Ly$\alpha$ clouds
(e.g.~Rees 1986; Murakami \& Ikeuchi 1990; 
Miralda-Escude \& Rees 1993; Kepner et al. 1997).
Of most relevance here are the
computations presented by Kepner et al. (see also Corbelli \& Salpeter 1993)
who studied the phase transitions (ionized/neutral/molecular)
of hydrogen gas clouds in DM halos.  
Their main focus was to show that low-mass halos  
form shielded neutral gas cores (a precondition for star-formation)
only at late ($z\lesssim 1$) cosmic times when the
metagalactic photoionizing radiation field has become sufficiently diluted
(Efstathiou 1992).  However, Kepner et al. made several
simplifying assumptions that make it difficult to apply
their models to either the CHVCs or the LG dwarf galaxies.
These included assuming a fixed relation between the halo mass
and the characteristic DM density (see \S 2),
adopting an arbitrary functional relationship
between the total gas mass and halo mass, and the
neglect of external pressure.
In our analysis we relax all of these assumptions.

In addition, we determine the conditions
required for the formation of thermally unstable multi-phased cores.
To do this we employ and extend 
the methods presented by Wolfire et al. (1995a)
in their study of the multi-phased Galactic disk,
and calculate the thermal equilibrium properties
of low metallicity HI gas heated by the
metagalactic field. We then incorporate the results of
these computations into our dark-matter halo models.   
We note that Wolfire et al. (1995b) (see also Ferrara \& Field 1994)
carried out an explicit analysis of the
HVC complexes observed by Wakker \& Schwarz (1991). In particular,
the cloud distances were constrained
by arguing that the CNM detected by Wakker \& Schwarz
is possible only when the ambient
gas pressure, exceeds a well defined minimum $P_{\rm min}$.
In this analysis the gas pressure
is provided by a Galactic X-ray corona. 
However, Wolfire et al. did not consider dark-matter dominated clouds. In such systems
the central cloud pressures can become much larger than the ambient pressures.
The transition from the WNM to the CNM phase 
then becomes possible in the cloud cores even when the ambient 
pressure is much smaller than $P_{\rm min}$.

We first discuss in \S 2 the basic parameters and properties of the
dark-matter halos we consider in our study, including also a discussion
of the relevant cosmological scaling relations.
In \S 3 we then describe our model clouds
and discuss the methods we use to compute the
cloud hydrostatics, radiative transfer, and HI phase
structure.  In \S 4 we discuss our representation
of the metagalactic background field.
In \S 5 we derive several useful analytic formulae
for the gas distributions in DM halos. In \S 6 we present
the results of our numerical computations. In \S 7
we model the HI gas structures in the dwarf galaxies Leo A and Sag DIG.
In \S 8 we consider models of the CHVCs as either circumgalactic
or extragalactic mini-halos.
A summary is presented in \S 9.

\section{Dark Matter Halos}

The analytic theory (Gunn \& Gott 1972)
of non-linear (``spherical top-hat'')
gravitational collapse shows that
individual DM halos are most simply characterized by two independent 
parameters (e.g.~the characteristic density and halo radius).
Numerical simulations of the dissipationless collpase of
cold dark matter (NFW; Bullock et al. 2001)
show that the DM distributions within the halos can 
be well represented by a spherically symmetric universal density profile
(the ``NFW profile'') over a wide range of mass scales. 
The simulations also show that for any particular
cosmological model the two halo parameters are highly
correlated. The correlation exists because a given mass-scale
is associated with a particular virialization red-shift, with
narrow dispersion.

An important feature of the NFW profiles is 
that they are cuspy with diverging
central DM densities. Central cusps in CDM halos
are a predicted theoretical result of
the dissipationless collapse.
However, HI rotation curves in dark-matter dominated dwarfs
and in several low surface brightness galaxies
imply the presence of constant density cores
(Moore 1994;  Burkert 1995;
Blais-Ouellette, Amram \& Carignan 2001; Firmani et al. 2001),
although this conclusion may be hampered by beam smearing
effects (van den Bosch \& Swaters 2001; van den Bosch et al. 2000). 
In view of the HI rotation curves,
Burkert (1995) introduced a semi-empirical profile (the ``Burkert profile'')
in which the cuspy NFW core is replaced with a uniform density core.
In our analysis we consider both NFW and Burkert profiles.

Realistic dark matter halos may be complex triaxial structures,
containing embedded clumps (sub-halos) and elongated filaments. 
In our analysis we represent the halos
assuming they are spherically symmetric and smooth.  
We begin by defining the basic halo
parameters for spherically symmetric systems, and we then discuss
the cosmological correlation relation between the halo scale parameters.

\subsection{Halo Parameters}
 
For spherically symmetric halos, 
the dark-matter density distribution may be written as
\begin{equation}
\label{e:rho_d}
\rho_d = \rho_{ds} f_\rho(x)
\end{equation}
where $\rho_{ds}$ is the characteristic ``scale density'' of the halo,
$f_\rho$ is the density profile, and $x\equiv r/r_s$ is
the dimensionless radial coordinate, where $r$ is the distance
from the halo center and
$r_s$ is the ``scale radius'' of the DM distribution.
For realistic density profiles, the DM density is large
for $x<1$ and becomes small when $x\gg 1$.

The enclosed DM mass at any radius is then
\begin{equation}
M_d = M_{ds} f_M(x) \ \ \ ,
\end{equation}
where 
\begin{equation}
\label{e:fMx}
f_M(x) = 3\int_0^x{f_\rho(x')x'^2dx'} \ \ \ ,
\label{e:fM}
\end{equation}
and
\begin{equation}
M_{ds}\equiv {4\pi \over 3}\rho_{ds} r_s^3 \ \ \ ,
\end{equation}
is the characteristic or ``scale mass'' of the halo.

The gravitational potential of the halo may be written as
\begin{equation}
\label{e:fvarphi}
\varphi = v_s^2 f_\varphi(x)
\end{equation}
where
\begin{equation}
f_\varphi(x) = \int_0^x{{f_M(x')\over x'^2}dx'} \ \ \ ,
\label{e:phi}
\end{equation}
and where we define the ``scale (circular) velocity''
\begin{equation}
v_{s}^2\equiv {GM_{ds}\over r_s}={4\pi G \over 3}\rho_{ds}r_s^2  \ \ \ ,
\label{e:vsdef}
\end{equation}
where $G$ is the gravitational constant.

The circular velocity at any distance from the halo center is
\begin{equation}
\label{e:vcircx}
v=v_s f_v(x) \ \ \ 
\end{equation}
where
\begin{equation}
f_v(x)=\biggl[{f_M(x)\over x}\biggr]^{1/2} \ \ \ .
\end{equation}

It is evident from equations (1)-(3) that for an assumed 
density profile $f_\rho$ the physical scales 
of a halo are fixed by two independent structural parameters.
These may be chosen as two out of the four mutually
dependent parameters $\rho_{ds}$, $r_s$, $v_s$ or
$M_{ds}$.
E.g., a halo may be defined
by the scale density and radius, $\rho_{ds}$ and $r_s$
(using eq.~[\ref{e:rho_d}]), or by the scale velocity and radius, $v_s$ 
and $r_s$ (using eq.~[\ref{e:fvarphi}]). The functions
$f_\rho$, $f_M$, and $f_\varphi$ for  
NFW and Burkert halos are listed in Table 5
in Appendix A.

\subsection{Cosmological Relations}

In cosmological studies the two independent
halo scale parameters are often given in terms of the
virial radius, $r_{\rm vir}$ (or the virial mass $M_{\rm vir}$), and the
``concentration parameter,'' 
$x_{\rm vir}\equiv r_{\rm vir}/r_s$ (often denoted by ``$C$").
The virial radius is defined as the radius within which the
mean dark-matter density of the halo exceeds the background
matter density of the universe, $\rho_u$, 
by a factor $\Delta \sim 200$, corresponding to
the overdensities expected in 
collapsed and virialized objects (Gunn \& Gott 1972).
The virial mass, 
\begin{equation}
\label{e:mvir}
M_{\rm vir}\equiv {4\pi \over 3}\Delta\rho_u r_{\rm vir}^3
\end{equation}
is the DM mass enclosed within the virial radius.
Thus, $M_{\rm vir}=M_{ds}f_M(x_{\rm vir})$.

With these definitions it follows that
\begin{equation}
\label{e:rhos}
\rho_{ds} = \Delta\rho_u{x_{\rm vir}^3 \over f_M(x_{\rm vir})} \ \ \ \ ,
\end{equation}
\begin{equation}
\label{e:rs}
r_s = \biggl({3 \over 4\pi \Delta \rho_u}\biggr)^{1/3} 
      M_{\rm vir}^{1/3}{1\over x_{\rm vir}} \ \ \ \ ,
\end{equation}
and
\begin{equation}
\label{e:vs}
v_s = \biggl({4\pi \over 3}G^3 \Delta\rho_u\biggr)^{1/6}M_{\rm vir}^{1/3}
      \biggl[{x_{\rm vir} \over f_M(x_{\rm vir})}\biggr]^{1/2} \ \ \ \ .
\end{equation}
In these expressions,
the matter density $\rho_u$ and the overdensity factor $\Delta$
are fixed by the underlying cosmological model. The halo
scale parameters $\rho_s$, $r_s$, and $v_s$, are then
determined by a choice of $M_{\rm vir}$ and $x_{\rm vir}$.

An important feature of the
CDM collapse simulations is the finding that
at any cosmic epoch
the two independent halo scale parameters are correlated,
over a wide range of halo masses (Navarro, Frenk \& White 1996, 1997;
Bullock et al. 2001). 
The correlation can be expressed in
several ways, e.g., as
a relation, $x_{\rm vir}(M_{\rm vir})$, between the concentration parameter
and the virial mass, or as a relation, $r_s(\rho_{ds})$, between the
scale radius and scale density.  The correlation
exists because a given mass scale (as defined for example by
$r_s$ and $\rho_{ds}$) is associated
with a narrow range of virialization redshifts, while
the halo scale density $\rho_{ds}$ is effectively determined
by the value of the cosmic matter density at the epoch of virialization.
The numerical simulations show that for any assumed cosmological
model the correlation is
well approximated by a simple
power-law relation $x_{\rm vir}\sim M_{\rm vir}^{\alpha}$, where
for low masses 
the index $\alpha$ is a weak function of the mass.
For a given virial mass a dispersion of
concentrations is found, with increasing departures from
the typical concentration being less likely.

We adopt the results of Bullock et al. (2001)
for simulations carried out assuming the ``concordance''
$\Lambda$CDM cosmological model (Bahcall et al. 1999). 
In this model the matter and vacuum density parameters
$\Omega_m = 0.3$ and $\Omega_\Lambda = 0.7$,
the Hubble constant $H_0 = 70$ km s$^{-1}$ Mpc$^{-1}$,
the root-mean-square (rms) amplitude of
mass fluctuations in spheres of radius $8 h^{-1}$ Mpc 
$\sigma_8 = 1$, and the present day matter
density of the universe is
$\rho_u = 2.76\times 10^{-30}$ g cm$^{-3}=
1.66\times 10^{-6}$ amu cm$^{-3}$.
\footnote{
A density of 1 amu cm$^{-3}$ equals
$1.66\times 10^{-24}$ g cm$^{-3}=$ 
0.93 GeV cm$^{-3}=
2.45\times 10^{-2}M_\odot$ pc$^{-3}$.}
For this model 
the overdensity factor $\Delta$ for virialized halos equals 340
(Bryan \& Norman 1997). The fraction of matter contained in
baryons is $\Omega_b/\Omega_m=0.088$ (assuming $\Omega_b h^2=0.013$).

We are interested in the predicted
halo properties at the current epoch ($z=0$). 
Bullock et al. carried out numerical simulations for
halos with present-day
virial masses ranging from 10$^{14}$ $M_\odot$ down to 10$^{11}$ 
$M_\odot$.  They also presented a ``toy model''
for the halo evolution that accurately reproduces the
relation between concentration parameter and virial mass
found in the collapse simulations.  
The numerical results show that the dispersion in concentrations
(for a given halos mass)
is well approximated by a log-normal distribution 
in which $\sigma$ standard deviations correspond to 
departures $\Delta({\rm log} x_{\rm vir}) = 0.14\sigma$
from the median concentration
\footnote{Due to a typographical error the $1\sigma$ deviations
are incorrectly given as $\Delta({\rm log} x_{\rm vir}) = 0.18$ 
in Bullock et al. (2001) (Bullock priv. comm.).}.
For halos between $10^{11}$ and $10^{14}$ $M_\odot$,
Bullock et al. find that the expression
\begin{equation}
\label{e:correl}
x_{\rm vir} = 10.3 \times 10^{0.14\sigma} \biggl({M_{\rm vir} \over 
           10^{13} \  M_{\odot}}\biggr)^{-0.13} \ \ \ 
\end{equation}
provides an accurate fit to the results of their numerical
simulation and associated toy model.

In our study we are interested in low-mass halos
with virial masses in the range $\sim 10^8$ to $10^{10}$ $M_\odot$,
at and somewhat below the ``dwarf galaxy'' mass scale. This mass range is
well below the resolution of the numerical simulations presented
by Bullock et al.  We therefore
use their $\Lambda$CDM toy model (Bullock priv.~comm.) 
to compute the concentrations for low-mass halos. In Figure 1 we plot
the computed concentrations for median ($\sigma=0$) halos
with virial masses equal to $10^8$, $10^9$, $10^{10}$ and $10^{11}M_\odot$. 
We find that the fit
\begin{equation} 
\label{e:fit}
x_{\rm vir} = 27\times 10^{0.14\sigma} 
             \biggl({M_{\rm vir} \over 10^9 \ M_\odot}\biggr)^{-0.08} \ \ \ 
\end{equation}
is accurate to within 3\% of the toy-model computations for median halos.
In this expression we have also assumed that, as for the high-mass halos,
one-sigma deviations correspond to $\Delta({\rm log} x_{\rm vir}) = 0.14$
departures from median halos.
In Figure 2 we plot the $x_{\rm vir}(M_{\rm vir})$ relation,
as given by equation~(\ref{e:fit}) 
for families of $+3\sigma$ overconcentrated, median, and
$-3\sigma$ underconcentrated halos.
The weak mass dependence, and the 
slight flattening of the power-index, from
$\alpha=-0.13$ in equation~(\ref{e:correl}) to $\alpha=-0.08$
in equation~(\ref{e:fit}), is consistent with the behavior of the
(post-recombination)
cold dark matter fluctuation spectrum, which remains flat at low masses
(Peebles 1982; Blumenthal et al. 1984; Eke, Navarro \& Steinmetz 2001).

For a given value of $\sigma$, equation~(\ref{e:fit}) together with 
equations (\ref{e:rhos}),
(\ref{e:rs}), and (\ref{e:vs}) may be used to compute all of the
halo properties as functions of a single halo parameter.
As is indicated by these equations, the resulting scale
parameters depend on the enclosed mass function $f_M(x)$. We find that
virtually identical results are obtained for NFW halos and Burkert halos.
In Figure 2 we plot the halo concentration, $x_{\rm vir}$,
the virial mass, $M_{\rm vir}$, the 
scale density, $n_{ds}\equiv \rho_{ds}/m_{\rm H}$ (amu cm$^{-3}$),
and the halo scale radius, $r_s$ (kpc), as functions of the halo
and scale velocity, $v_s$ (km s$^{-1}$)
\footnote{In cosmological studies, halos are often parameterized
by the circular velocity at the virial radius, 
$v_{\rm vir}\equiv v_s [f_M(x_{\rm vir})/x_{\rm vir}]^{1/2}$,
or the maximum circular velocity $v_{\rm max}=0.8v_s$
(for NFW and Burkert potentials).  We prefer
to use $v_s$ as it directly fixes the potential-well depth
(see eq.~[\ref{e:fvarphi}]).}.
We display these quantities for $+3\sigma$, median, and $-3\sigma$ halos.
For median halos, the scale radius $r_s$ ranges from 0.29 to 6.4 kpc,
for $v_s$ between 10 and 100 km s$^{-1}$.
The scale density increases with decreasing $v_s$,
reflecting the earlier collapse epoch of lower mass halos.
However, as expected, the density dependence
is weak. For median halos, $n_{ds}$ decreases from 
2.7 to 0.55 amu cm$^{-3}$ for $v_s$ between 10 and 100 km s$^{-1}$.
For a narrow range of scale velocities, large changes in the
characteristic DM densities are possible only by varying $\sigma$,
i.e., by moving to over-concentrated or underconcentrated halos.

In his study of rotationally supported dwarf galaxies, 
Burkert (1995) found that the
radii of the inferred (constant density) DM halo cores
are correlated with the magnitudes of the
observed circular velocities, as expected theoretically.
In Figure 2 we plot the observed
\footnote{Burkert (1995) actually plotted the correlation
between $r_s$ and the observed circular velocity at $r_s$.
We have transformed the observed velocities to scale velocities
using equation~(\ref{e:vcircx}) and Burkert's profile. 
This gives $v(r_s)=0.618v_s$.}
correlation between $r_s$ and $v_s$ for the four low-mass dwarf spiral
galaxies (DDO 154, DDO 170 NGC 3109, and DDO 105) in Burkert's sample
(see his Fig.~2).  It is evident that these objects
lie precisely on the expected relation for median ($\sigma=0$)
halos. The dwarf galaxy rotation curve data thus
provide empirical support for the theoretically
derived correlation relation given by equation~(\ref{e:fit}).  

Convenient analytic scaling relations for the 
various halo parameters as
functions of $v_s$, may be developed by setting
$f_M(x)\simeq 1.40 x^{1/2}$ in equations~(\ref{e:rhos})
and (\ref{e:vs}).  This approximation for 
$f_M$ is accurate to within 12\%,
for both Burkert and NFW halos, for $4.5 < x < 50$. 
It then follows that
\begin{equation}
\label{e:xvfit}
x_{\rm vir} = 32.1\times 10^{0.149\sigma}
           v_{s6}^{-0.225} \ \ \ ,
\end{equation}
\begin{equation}
\label{e:mvfit}
M_{\rm vir} = 5.55\times 10^7 \times 10^{-0.112\sigma}
            v_{s6}^{3.191}
        \ \ \ {\rm M}_\odot \ \ \ ,
\end{equation}
\begin{equation}
\label{e:ndsfit}
n_{ds} = 2.71\times 10^{0.373\sigma}
              v_{s6}^{-0.638} 
     \ \ \ {\rm amu \ cm}^{-3} \ \ \ ,
\end{equation}
and
\begin{equation}
\label{e:rsfit}
r_s = 0.29\times 10^{-0.186\sigma}
              v_{s6}^{1.319} 
        \ \ \ {\rm kpc} \ \ \ ,
\end{equation}
where $v_{s6}\equiv v_s/10$ km s$^{-1}$.
These expressions are plotted as the dashed lines in Figure 2.
They provide excellent analytic representations (to within
1\% accuracy) of the halo scalings for median and
underconcentrated halos, and are less accurate 
($\sim 30\%$) for over-concentrated halos.

In our computations we adopt the
$x_{\rm vir}(M_{\rm vir})$ correlation 
relation as given by expression~(\ref{e:fit}),
and the resulting halo parameters displayed in Figures 1 and 2,
as theoretically motivated guides to the properties of ``median''
low-mass halos.  In our study,
we also explicitly consider halos that are substantially
``over-concentrated'' or ``under-concentrated'' with respect to the 
median halos.

\section{Clouds}

We are interested in determining the structural 
properties of hydrostatic gas clouds that are
confined by gravitationally dominant dark-matter halos, and
are in ionization and thermal equilibrium with an external source of 
Lyman continuum (Lyc) and far-ultraviolet (FUV) radiation.
We focus on ionization and heating by
the metagalactic field.
We envision clouds consisting of a
``warm medium'' (WM) with up to three
components;
(a) an outer layer of warm photoionized gas (WIM),
that serves as a shielding layer for 
(b) a neutral core 
of warm HI gas (WNM), and possibly (c) a still smaller region within which 
the HI can exist as a
multi-phased (WNM/CNM) mixture of warm and cold gas.
When ionizing radiation is present, the WIM component 
will always be present. However,
the WNM and WNM/CNM components will not always exist, and one
of our goals is to determine the specific conditions required for
their formation.

We consider the possibility that the halo-cloud systems are
embedded within a low-density hot ionized medium (HIM),
provided for example by a Galactic X-ray corona in the circumgalactic
environment, or an intergalactic medium in the
Local Group environment. The HIM 
exerts a bounding pressure,
$P_{\rm HIM}$, on the outer surface of the WM cloud 
and serves to further confine the WM gas, particularly
in the limit of weak DM halo potentials.
We assume that the HIM is sufficiently hot such that it 
is completely unbound to any of the halos we consider.
We also assume that the HIM 
is optically thin to the background radiation field.

We wish to address several questions. For a given
dark-matter halo and a total mass of warm gas, what are
the radial extents and associated column densities of the HI gas
distributions?
For small gas masses we expect the WM clouds to be
photoionized throughout. In such WIM dominated clouds, any
HI is present as a 
small neutral fraction within the ionized gas. 
As the gas mass is increased the WIM component can be expected
to become optically thick to the ionizing radiation, leading to
the formation of a neutral
hydrogen WNM core. 
For an assumed halo and background field, what is the
minimum WM mass required for the formation of a neutral WNM core?
When WNM cores exist, how large are they? How do the core sizes
and associated HI column densities depend on the halo scale parameters? 
If the hydrostatic
pressure gradients are sufficiently large the central gas pressures
may become large enough to allow a multi-phased WNM/CNM medium
(or a complete conversion to CNM) in the neutral cores. 
For clouds in which the ionization and heating is
dominated by the metagalactic field, when is multi-phased
behavior possible in the neutral cores?

To address these questions, we first construct models in which we 
solve the coupled equations of hydrostatic equilibrium and radiative transfer
for {\it warm} spherically symmetric clouds that are 
(a) confined by gravitationally dominant halos, (b) subjected to
a bounding pressure, and (c)
exposed to an external source of Lyc and FUV radiation
that ionizes the gas.
We focus on irradiation by the metagalactic field.
  Because the temperatures of the warm
components are expected to lie
in a narrow range ($\sim 10^4$ K) we specify the gas temperatures
for the WIM and WNM components in advance in these computations.
Given the resulting radial profiles for the hydrostatic gas pressures,
densities, and fractional ionizations, we compute various quantities
of observational interest including the HI gas masses
and column densities, gas scale heights, locations of the
ionization fronts (if present), cloud emission
measures, H$\alpha$ surface brightnesses, and other quantities of interest.
Our treatment of the cloud hydrostatics and radiative transfer
is described in \S 3.1.

We then determine the phase states of the HI gas
in the neutral cores (when such cores exist).
The thermal phase state of the neutral gas at each cloud
radius is determined by comparing the local value
of the hydrostatic pressure $P(x)$ to the local
values of two critical pressures $P_{\rm min}(x)$ and
$P_{\rm max}(x)$. At low pressures
$P < P_{\rm min}$ the gas is WNM, at intermediate pressures
$P_{\rm min} < P < P_{\rm max}$ the gas is multi-phased,
and at high pressures $P > P_{\rm max}$ the gas must convert to CNM.
\footnote{
The fact that interstellar HI gas can exist as WNM
and CNM has been the subject of many theoretical investigations
(e.g., Field, Goldsmith \& Habing 1969; 
Zeldovich \& Pilkelner 1969; McKee \& Ostriker 1977;
de Jong 1977; Draine 1978; Ferriere, Zweibel \& Shull 1988;
see also the reviews by Field 1975, and Kulkarni \& Heiles 1987, 1988).
Recent comprehensive theoretical studies have been
presented by Wolfire et al. (1995a; 2002). 
The two distinct phases are possible because the
gas heating rates are generally proportional to the gas
density, $n_{\rm H}$, whereas the gas cooling rates are proportional
to $n_{\rm H}^2$. Thus, at low densities cooling becomes
inefficient, and the temperature is driven to maximal (WNM) values of
$\sim 10^4$ K set by Ly$\alpha$ cooling.  At high densities,
cooling by fine structure line emission,
primarily 158$\mu$m [CII] emission, becomes much more
effective and the gas can cool to CNM temperatures $\lesssim 100$ K.
At intermediate densities a thermal instability occurs as
the dominant cooling mechanism switches from Ly$\alpha$ to
[CII] emission. The instability and multiphased behavior occurs within
a range of pressures -- $P_{\rm min}$ to $P_{\rm max}$ -- that depend on
the detailed heating and cooling rates.
For pressures less than $P_{\rm min}$ only the warm phase is possible,
and for pressures greater than $P_{\rm max}$ 
only the cold phase is possible.}.

As we discuss in \S 3.2 for a given incident radiation field
the critical pressures $P_{\rm min}$ and
$P_{\rm max}$  for shielded HI gas can be
expressed as functions of the primary ionization rate $\zeta_{\rm p}$,
independent of the cloud geometry (e.g. spherical vs.~plane parallel).
In \S 3.2 we present the results of thermal equilibrium
computations which yield the required
$P_{\rm min}(\zeta_{\rm p})$ and $P_{\rm max}(\zeta_{\rm p})$.

Given $P_{\rm min}(\zeta_{\rm p})$ and $P_{\rm max}(\zeta_{\rm p})$  
we determine the phase states of the HI gas in our halo-cloud models.
In these models the local hydrostatic pressure depends 
on the depth of the halo potential well and the value
of the bounding pressure, and
increases from the surface to the core. Conversely,
$P_{\rm min}$ and $P_{\rm max}$
decrease from the surface to the core
as the ionization rate $\zeta_{\rm p}$ 
decreases with increasing cloud depth and 
opacity. Thus, any neutral gas may be expected to be pure WNM at large
radii where $P < P_{\rm min}$, but may become multi-phased 
or fully converted to CNM at small radii where  
$P_{\rm min} < P < P_{\rm max}$ or $P > P_{\rm max}$.  
For sufficiently deep
halo potentials the conversion to CNM may occur
even if the ambient bounding pressure is negligibly small.

We determine the conditions
for which $P_{\rm min}(0) < P_0 < P_{\rm max}(0)$ where
$P_0$ is the central pressure. In such clouds the multi-phased
cores will extend out to a radius $x$ 
where $P(x) = P_{\rm min}(x)$.  We assume that in the multiphased zones the
CNM occupies negligible volume
but remains uniformly mixed with the WNM.
When $P_0 > P_{\rm max}(0)$ the HI in the cores 
will be driven entirely into the
CNM phase.  When this happens, thermal support is lost,
and our hydrostatic solution in which warm
gas is assumed to be present from the cloud center to the 
surface, becomes inconsistent with gas heating by the external field.
Either additional (internal) heat sources must be
postulated, or the cloud must collapse. Thus, in our computations
we find the maximum 
masses, $M_{\rm WMmax}$, of warm gas that can be maintained
by the external field within a given halo and an assumed bounding pressure. 
The total mass of CNM is unconstrained in our models, except that we
assume it cannot dominate the gravitational potential.

\subsection{Hydrostatics and Radiative Transfer}

In hydrostatic equilibrium
\begin{equation}
dP=-\rho d\varphi \ \ \ ,
\label{e:hbasic}
\end{equation}
where $P$ is the gas pressure, $\rho$ is the gas density, and
$\varphi$ is the gravitational potential.  
The pressure $P=\rho c_g^2$, where $c_g^2\equiv kT/\langle m\rangle$
is the isothermal sound speed of the gas, $T$ is the gas temperature, 
and $\langle m\rangle$ is the mean mass per gas particle.
We assume that the potential is dominated by the dark-matter halo
so that $\varphi=v_s^2f_\varphi(x)$, where $v_s$ is the halo
scale velocity and $f_\varphi$ is the dimensionless potential
for either NFW or Burkert halos.
It follows that the gas pressure $P(x)$ is given by
\begin{equation}
\label{e:sbasic}
P(x) = P_0{\rm exp}\biggl[-\int_0^x {v_s^2\over c_g^2}
{f_M(x^\prime)\over x^{\prime 2}}dx^\prime\biggr]
\end{equation}
where $f_M(x)$ is the (dimensionless)
DM mass distribution function, and $P_0$ is the central gas pressure. 
It is evident that the hydrostatic structure is controlled by 
the ratio, $v_s/c_g$, of the halo scale velocity to the gas sound speed.
The hydrostatic and ionization structures of the cloud are coupled 
since the sound speed $c_g$ depends on $T$ and $\langle m\rangle$, 
which vary with the ionization state.

In ionization equilibrium
\begin{equation}
\label{e:ioneq}
\zeta n({\rm H^0}) = \alpha_Bn({\rm H^+})n(e),
\end{equation}
where $\zeta$ is the total local photoionization rate, and
$n({\rm H^0})$, $n({\rm H^+})$ and $n(e)$ are the atomic hydrogen, proton
and electron densities respectively.
We assume a case-B recombination coefficient $\alpha_B$.
The local primary photoionization rate is 
\begin{equation}
\label{e:zeta}
\zeta_{\rm p}(x) = 4\pi \int_{{\nu_0}}^\infty \sigma_\nu J^*(\nu ,x) d\nu
\end{equation}
where $\nu_0$ is the Lyman limit frequency,
$\sigma_\nu$ is the photoabsorption
cross section, and $J^*(\nu ,x)$ is the local mean photon intensity
\begin{equation}
\label{e:anglej}
J^*(\nu ,x) \equiv {1\over 2}J^*_0(\nu )\int_{-1}^1 
{\rm exp}\biggl[-\sigma_\nu N_{\rm HI}(x,\mu)\biggr]d\mu,
\end{equation}
where $J^*_0(\nu )$ is the unattenuated isotropic radiation field
(in units of photons cm$^{-2}$ s$^{-1}$ Hz$^{-1}$ sr$^{-1}$),
$N_{\rm HI}(x,\mu)$ is the HI absorbing
column from the cloud surface to $x$ along
a ray that is inclined by an angle $\theta$ 
relative to the outward radial direction, and where
$\mu\equiv {\rm cos}\theta$.

For the range of hydrogen gas columns (up to $\sim 10^{22}$ cm$^{-2}$)
that we consider in our computations, the gas opacity is dominated
by a combination of hydrogen and helium photoabsorptions.
We assume a primoridal helium abundance $n_{\rm He}/n_{\rm H}=1/12$
(Olive \& Steigman 1995; Ballantyne, Ferland \& Martin 2000)
where $n_{\rm H}$ and $n_{\rm He}$ are the total hydrogen and helium particle
densities respectively.   
Photoabsorptions by heavier ``metals'' may be neglected for the
low metallicities ($Z \lesssim 0.1$) we assume
(Morrison \& McCammon 1983; Balucinska-Church \& McCammon 1992).
In our treatment of the radiative transfer we make
the simplifying approximation that the hydrogen and
helium ionization fractions are equal at all locations.
We also assume that every 
helium recombination photon is absorbed in a hydrogen ionization. 
With these assumptions 
$\sigma_\nu= \sigma_{\nu,{\rm H}} + (1/12)\sigma_{\nu,{\rm He}}$.
We adopt the Balucinska-Church \& McCammon expressions
for $\sigma_{\nu,{\rm H}}$ and $\sigma_{\nu,{\rm He}}$.
In computing the total ionzation rate we also include the contributions
of the secondary collisional ionizations 
produced by the primary photoelectrons (Shull \& van Steenberg 1985).

We construct models assuming that the background metagalactic
field is the dominant source of radiation impinging on the clouds.
Our representation of this field is displayed in 
Figure 3 in a $\nu J_\nu$ plot.
We discuss our fit to the metagalactic field in
Appendix B. For our assumed field  
$4\pi J^*=1.28\times 10^4$ photons s$^{-1}$ cm$^{-2}$,
where the Lyc integral
\begin{equation}
\label{e:lint}
4\pi J^*\equiv 
\int d\Omega \int_{{\nu_0}}^\infty d\nu {I_\nu \over h\nu}  \ \ \ .
\end{equation} 
Most (93\%) of the photons are emitted between 1 and 4 ryd.
The unattenuated primary hydrogen ionization rate is 
$\zeta_{{\rm p},0}=4\times 10^{-14}$ s$^{-1}$.

For our assumed helium abundance,
\begin{equation}
\label{e:pbasic}
P={13 \over 12}(1+x_p)n_{\rm H} kT
\end{equation}
where $x_p\equiv n(H^+)/n_{\rm H}$ is the proton fraction.
The electron fraction $x_e\equiv n(e)/n_{\rm H}={13\over 12}x_p$.
The mean mass per gas particle is
\begin{equation}
\langle m\rangle = {16\over 13}{1\over 1 + x_p} m_{\rm H}
\end{equation}
where $m_{\rm H}$ is the atomic hydrogen mass.

In our computations
we assume $T=10^4$ K for both the WNM and the WIM.
To verify this assumption for the WIM we have computed
the gas temperature of low-metallicity photoionized
gas using CLOUDY 
(version C90.04, Ferland et al. 1998)\footnote{See
www.pa.uky.edu/$\sim$gary/cloudy/} assuming an
ionizing field as given by our representation of the
metagalactic field discussed in Appendix B. We find
that the gas temperature of the ionized gas
ranges from $8.0\times 10^3$
to $1.2\times 10^4$ K for gas densities in the range
$10^{-3}$ to 0.1 cm$^{-3}$. The HI thermal phase computations
we discuss in \S 3.2 show that the WNM gas temperature is
close to $10^4$ K. For $10^4$ K gas,
the associated gas sound speeds are
$c_{g,{\rm WNM}}=8.19$ km s$^{-1}$ for the WNM, 
and $c_{g,{\rm WIM}}=11.6$ km s$^{-1}$
for the WIM.  

The local gas pressure, fractional ionization, and gas temperature,
determine the local density of hydrogen nuclei $n_{\rm H}$ via
equation~(\ref{e:pbasic}). We define
the hydrogen gas (ionized plus neutral) 
density distribution $f_{\rm gas}$ such that
\begin{equation} 
\label{e:dbasic}
n_{\rm H}=n_{\rm H,0}f_{\rm gas}(x)
\end{equation}
where $n_{\rm H,0}$ is the central density.  We define the outer edge
of the WM cloud as the radius $x_{\rm W/H}$ at which the cloud pressure
is equal to the pressure $P_{\rm HIM}$ of the 
enveloping hot ionized medium. Thus, $x_{\rm W/H}$ is located at
the interface between the WIM and the HIM.
The total (hydrogen) cloud mass is then
\begin{equation}
\label{e:mwm}
M_{\rm WM} = 4\pi r_s^3 m_{\rm H}  
n_{\rm H,0} \int_0^{{x_{\rm W/H}}} f_{\rm gas}(x)x^2 dx \ \ \ \ .
\end{equation}
For a given halo potential and background field, a cloud model
is selected by choosing a value for the 
total mass $M_{\rm WM}$ given an assumed value
for $P_{\rm HIM}$, or equivalently, setting the boundary condition
that the pressure at a selected value of $x_{\rm W/H}$ equals $P_{\rm HIM}$.

We have constructed a code to solve the coupled hydrostatic and
ionization equilibrium equations.
In our numerical solution we specify $P_{\rm HIM}$ and $x_{\rm W/H}$, and
then adopt an iterative procedure.
First, we assume an ionization and opacity structure for the gas 
(initially by setting $x_e=1$ everywhere, and 
assuming that the cloud is optically thin).
We then integrate equation~(\ref{e:sbasic}) subject to the boundary
condition $P(x_{\rm W/H})=P_{\rm HIM}$. This yields a
hydrostatic pressure distribution $P(x)$, and a value for the central
pressure. $P(x)$ and the assumed ionization structure
determine the local values for the hydrogen gas densities. 
We then compute the local photoionization rates 
given the absorbing columns and opacities to the cloud surface.
Given the ionization rates and the hydrogen gas densities
we solve the equation of ionization equilibrium
and find new values for the ionization fractions $x_e$
and cloud opacities. The local gas sound speeds are then updated, and 
the hydrostatic structure is recomputed etc.
We iterate until the local ionization fractions have converged
to within 1\%.  

Given the solutions to the hydrostatic and ionization structures,
several quantities of interest may be computed, including
the total gas column densities, the HI masses and columns,
the cloud emission measures and H$\alpha$ surface brightnesses,
the gas scale heights, and the location of the ionization fronts.

The total column density of hydrogen nuclei along a line-of-sight with
impact parameter $x$ from the cloud center is 
\begin{equation}
N_{\rm WM}(x) \equiv  2n_{\rm H,0}r_s\int_x^{{x_{\rm W/H}}} f_{\rm gas}(x')
{x'\over[x'^2-x^2]^{1/2}}dx' \ \ \ ,
\label{e:colden}
\end{equation} 
and the HI column density along a line-of-sight a distance
$x$ from the cloud center is 
\begin{equation}
N_{{\rm HI}}(x) 
\equiv  2n_{\rm H,0}r_s\int_x^{{x_{\rm W/H}}} (1-x_p) f_{\rm gas}(x')
{x'\over[x'^2-x^2]^{1/2}}dx' \ \ \ \ ,
\end{equation}
where $x_p$ is the local proton fraction.

The emission measure along a 
line-of-sight with impact parameter $x$ is
\begin{equation}
\label{e:em}
EM(x) = 2n^2_{\rm H,0}r_s \int_x^{{x_{\rm W/H}}} x_ex_p f^2_{\rm gas}(x')
{x'\over[x'^2-x^2]^{1/2}}dx' 
\end{equation}
where $x_e$ and $x_p$ are the local electron and proton fractions.
The associated H$\alpha$ surface brightness in milli-Rayleighs
may be written as
\begin{equation}
I^*_{\rm H\alpha}(x)=10^9 \alpha_{\rm eff} EM
       =364 EM_{\rm pc}~~~{\rm mR},
\label{e:Halpha-gen}
\end{equation}
where $\alpha_{\rm eff}=8.7\times 10^{-14}$ cm$^3$ s$^{-1}$
is the H$\alpha$ effective recombination coefficient (Osterbrock 1989)
and $EM_{\rm pc}$ is the emission measure in cm$^{-6}$ pc. 

While formal solutions may be found for all values of 
$P_{\rm HIM}$ and $x_{\rm HIM}$, several additional
conditions must be satisfied for physically realizeable
solutions. First, while the halos
formally extend to infinite radius,  
we assume that effective physical boundaries exist beyond which both
the baryons and dark-matter particles are subjected to
external forces that readily detach them from the halo systems.
As a plausible limit
we assume that the radii of the halos and trapped clouds cannot
exceed the virial radius. Thus, we require that
$x_{\rm W/H} < x_{\rm vir}$ (where $x_{\rm vir}$
is the halo concentration parameter).  If the halos are
tidally truncated within the virial radius, then the
limit becomes $x_{\rm W/H} < r_t/r_s$, where $r_t$ is the tidal
radius.

Second, self-consistent solutions require that the
DM mass dominates the gas mass at all radii,
otherwise the assumption that the halo gravity
dominates is violated. Self-consistent
solutions must necessarily satisfy the condition
$M_{\rm WM} < M_{\rm DM}(x_{\rm W/H})$, where
$M_{\rm DM}(x_{\rm W/H})$ is the dark-matter mass enclosed within
the entire WM cloud.

Third, the WM gas must be well bound to the halos.
Otherwise, mass loss (especially from the outer
WIM layers) will occur.  We define bound
WM clouds as those that satisfy the condition 
\begin{equation}
\label{e:bound}
{{\cal W}\over{\cal T}}\equiv {v_s^2[f_\varphi(x_{\rm W/H})-f_\varphi(\infty)] 
\over {3\over 2}c_{\rm WIM}^2} > 1
\end{equation}
where ${\cal W}$ is the gravitational potential energy per particle
at $x_{\rm W/H}$, and
${\cal T}=(3/2)c_{\rm WIM}^2$ is the mean thermal energy per particle
in the outer WIM layer
\footnote{
A similar condition was adopted by Barkana \& Loeb (1998)
in their study of the photoevaporation of gas in low-mass halos
at the reionization epoch.}.

We conclude this section by making an important distinction
between {\it pressure-confined} and {\it gravitationally-confined} clouds.
Because the depths of the halo potential wells
we are considering are finite 
(as $x\rightarrow \infty$, $f_\varphi \rightarrow 3$ for NFW halos,
and $f_\varphi \rightarrow 3\pi/4$ for Burkert halos,
see Appendix A) a maximum possible pressure (density) contrast 
between large and small radii exists for a given value
of $v_s/c_g$. 
When $v_s/c_g$ is small the maximum pressure (density)
contrast is small, and the gas
extends to large halo radii $x\gg 1$. In this limit a cloud of
finite mass must be pressure-confined, and the halo gravity
provides just a small pressure enhancement towards the cloud center.
In pressure-confined clouds, properties such as the total
cloud mass, or the internal gas pressure, are determined
by conditions at the cloud surface via
by the bounding pressure $P_{\rm HIM}$. 
Pressure-confined clouds will tend to become unbound
as the confining pressure becomes small. 

When $v_s/c_g$ is large the pressure contrast becomes large, and
the gas is effectively restricted to small halo radii $x \ll 1$.
In this limit, which we refer to as the ``small-$x$'' limit,
the cloud is gravitationally confined, and for any plausible
value of the central pressure (or density), the
total gas mass remains finite for arbitrarily small
bounding pressures. Clouds
may be assumed to be in the small-$x$ limit when 
$v_s/c_g \gtrsim 1.5$ (see Appendix A).
In gravitationally confined clouds, properties
such as the total mass or internal gas pressure are 
determined by the central gas pressure, that is, by conditions
at the cloud centers. Gravitationally confined
clouds become increasingly bound as $v_s/c_g$ becomes large
and the potential wells deepen.
When the observable gas is in the small-$x$
limit, the implied dynamical mass
$M_{\rm dyn}=c_g^2 r_{\rm gas}/G$ (where $r_{\rm gas}$ is the
observed gas scale height) is a small fraction of the halo
scale mass, and an even smaller fraction of the
virial mass. 
In gravitationally confined clouds,
much more dark-matter must exist beyond the observed gas
structures, unless the halos are truncated at radii much less
than their virial radii.
 
In Appendix A we 
present simple analytic expresssions for the gas
scale heights, gas masses, and column densities, 
for clouds in the small-$x$ limit, and we derive an analytic
condition for the transition from fully ionized clouds
to clouds containg neutral cores. These analytic
results (for both NFW and Burkert halos) are
summarized in Table 6, and may be used as guides
to understanding the numerical model results we present in
\S 4, \S 5, and \S 6.

\subsection{$P_{\rm min}$ and $P_{\rm max}$} 

We employ the methods described
in Wolfire et al. (1995a, 2002) to compute
$P$ vs.~$n_H$ ``thermal phase diagrams'' for
low-metallicity HI gas heated by metagalactic radiation
for a wide range of assumed HI shielding columns.
We use the results of these computations to determine  
the critical pressures 
$P_{\rm min}$ and $P_{\rm max}$
as functions of the shielding column and associated
primary ionization rate $\zeta_{\rm p}$.

In our phase diagram computations  
the equations of thermal balance and of
ionization and chemical equilibrium are solved
self-consistently.  
In these computations we adopt a plane parallel geometry
in treating the radiative transfer through the absorbing columns,
and we assume an incident flux equal to $2\pi J^*_\nu$
where $J_\nu$ is the 
mean intensity assumed in our spherically symmetric halo cloud models
\footnote{In spherical clouds the ionization rate near the
surface of optically thick clouds
is proportional to $2\pi J^*_\nu$, whereas near the cloud center
the ionization rate is proportional to $4\pi J^*_\nu$
(see eq.~[\ref{e:anglej}]). In our 
plane-parallel computations we adopt an incident flux equal to 
a value of $2\pi J^*_\nu$, with the recognition that the resulting
$P_{\rm min}$ and $P_{\rm max}$ as functions of the
{\it ionization rate} are insensitive to the assumed incident flux.}.
We calculate the hydrogen ionization rates
using the same approximations for the Lyc absorption and opacity
that we use in computing the ionization structure of the spherical
halo clouds.

The three major heating processes included in our HI phase computations
are UV and X-ray ionization, FUV photoelectric emission, 
and molecular hydrogen (H$_2$) photodissociation.
Their relative importance depends on the Lyc vs.~FUV
intensities, on the dust-to-gas mass ratio $D/G$,
and the metallicity $Z$.  The metallicity also controls the
gas cooling efficiencies.
In view of the generally low (but non-negligible) 
metal abundances observed in the HVCs and CHVCs
we assume $Z=0.1$, where $Z=1$ corresponds to the Galactic
(gas-phase) abundances of
$n_{\rm O}/n_{\rm H}=3.2\times 10^{-4}$ for oxygen
(Meyer, Jura \& Cardelli 1998) 
and $n_{\rm C}/n_{\rm H}=1.4\times 10^{-4}$
for carbon (Cardelli et al. 1996).  The dust content and grain
properties of the CHVCs are unknown, and we assume that
the dust-to-gas ratio is proportional to the metallicity,
and set $D/G=0.1$.  Following Wolfire et al. (2002)
we adopt a slightly modified form of the Bakes \& Tielens (1994)
photoelectric heating efficiencies.
The FUV field, that leads to photoelectric emission,
produces the C$^+$ cooling ions and photodissociates H$_2$, is attenuated
by dust absorption. 
For $D/G=0.1$ a photoelectric heating
FUV optical depth of 1
occurs at an HI column density of $1.1\times 10^{22}$ cm$^{-2}$
(Roberge, Dalgarno \& Flannery 1981).  
We assume that that H$_2$ forms on grain surfaces with 
a rate coefficient $R=3\times 10^{-17}Z$ cm$^3$ s$^{-1}$ proportional
to the metallicity, and also via the standard gas phase
sequence H+$e^- \rightarrow$ H$^- + \nu$, H$^- + $H$^+ \rightarrow
$H$_2 + \nu$.
The H$^-$ abundance is limited by photodetachment, mainly by
far-red photons, with a cross section that peaks near 8500 \AA\ 
(Wishart 1979).
Detailed expressions for the various heating and cooling rates, and
the basic atomic and molecular data,
are given in Wolfire et al. (1995a) with some updates
in Kaufman et al. (1999) and Wolfire et al. (2002).

As a specific example we show in Figure 4 our numerical
results for a shielding column
$N_{\rm HI}=5\times 10^{19}$ cm$^{-2}$.  This particular
shielding column is important since as we will show
this is the typical peak HI column density for WM clouds
in which the gas becomes multi-phased in the cloud cores.
For our assumed representation of the metagalactic field 
(see Appendix B),
and a shielding column of $5\times 10^{19}$ cm$^{-2}$, 
the primary hydrogen ionization rate is equal to $1.1\times 10^{-18}$ s$^{-1}$.

The resulting $P$ vs.~$n_{\rm H}$ phase diagram is
shown in Figure 4a for densities ranging from
10$^{-3}$ to 100 cm$^{-3}$. The thermally unstable region, defined as 
the zone where $d({\rm log}P)/d({\rm log}n_{\rm H}) < 0$, is visible
between the WNM and CNM branches, and a
two-phased medium is possible for pressures between 
$P_{\rm min}=45.0$ cm$^{-3}$ K, and $P_{\rm max}=310$ cm$^{-3}$ K.
The dominant heating and cooling rates per hydrogen nucleus,
$\Gamma$ and $n_{\rm H}\Lambda$, are displayed in Figure 4b.
UV/X-ray ionization dominates the heating of the WNM.  
The UV/X-ray heating rate decreases from
$2\times 10^{-28}$ to $2\times 10^{-29}$ ergs s$^{-1}$ H$^{-1}$ in the
density range 10$^{-3}$ to 100 cm$^{-3}$. The decline is due
to a decrease in the fractional ionization. As $x_e$ decreases an 
increasing percentage of the UV/X-ray photoelectron energy 
is lost in atomic excitations and
secondary ionizations rather than being deposited as heat.  In contrast,
the photoelectric heating rate (and efficiency) remains approximately
constant, and is equal to $\sim 2\times 10^{-29}$ ergs s$^{-1}$ H$^{-1}$,
across the entire density range.  Consequently, the CNM is heated by
a combination of UV/X-ray and photoelectric heating.  At the highest
densities the molecular formation rate becomes sufficiently
large such that H$_2$ photodissociation heating dominates.
The WNM is cooled by Ly$\alpha$ emission, and the CNM is cooled
by [CII] 158$\mu$m fine structure emission.
At high densities carbon remains neutral and
[CI] 609$\mu$m fine-structure line emission dominates the cooling. 
The computed fractional ionizations and gas temperatures
are shown in panels
Figures 4c and 4d. In the WNM, $T\sim 10^4$ K, and in the
CNM, $T=10-100$ K.

In Figure 5a we display the $P$ vs. $n_{\rm H}$ phase diagrams for
shielding columns ranging from $1.0\times 10^{18}$ cm$^{-2}$
to $1.0\times 10^{21}$ cm$^{-2}$.  As the shielding column
increases, the ionization and heating rates
decrease, and the critical pressures $P_{\rm max}$ and
$P_{\rm min}$ both decrease. In Figure 5b we plot $P_{\rm min}$
and $P_{\rm max}$ as functions of the primary ionization rate, and the 
associated values of the shielding column. 
For our assumed field the ionization rate 
ranges from $8.9\times 10^{-16}$ s$^{-1}$ to $9.1\times 10^{-21}$ s$^{-1}$
for shielding columns between $1.0\times 10^{18}$ cm$^{-2}$
and $3.0\times 10^{21}$ cm$^{-2}$. For this range,
$P_{\rm min}/k$ varies from $6.4\times 10^3$ to 5.7 cm$^{-3}$ K, and
$P_{\rm max}/k$ varies from $1.2\times 10^4$ to 11.2 cm$^{-3}$ K.

For a fixed heating rate,
$P_{\rm min}$ and $P_{\rm max}$ scale approximately as $1/Z$
(where $Z$ is the metallicity). This is because the fine-structure
and metastable metal line cooling rates are 
proportional to $Z$, and the metal line cooling
dominates at both $P_{\rm min}$ and $P_{\rm max}$ (see Fig.~4b).

We use the computations of $P_{\rm min}(\zeta_{\rm p})$ and
$P_{\rm max}(\zeta_{\rm p})$
displayed in Figure 5b to determine the
HI phase states of the gas within the halo clouds.
As we have discussed, 
we compare the hydrostatic halo cloud
pressure $P$ to the values of $P_{\rm min}$ and $P_{\rm max}$ 
both of which vary with the local primary ionization rate 
within the halo clouds. When
$P < P_{\rm min}$ the gas is WNM, when
$P_{\rm min} < P < P_{\rm max}$ the gas is multi-phased,
and when $P > P_{\rm max}$ the gas must convert to CNM.

\section{Numerical Computations: $M_{\rm gas}$ vs. $M_{\rm vir}$ Diagrams}

We have carried out computations of the hydrostatic gas
distributions, ionization structures, and HI thermal phase properties,
of dark-matter dominated clouds for a wide range of 
assumed total gas masses and halo masses. 
We consider both NFW and Burkert potentials,
and a range of HIM bounding pressures.
The computational results may be used to select models
that fit the observed HI distributions and thermal phase
properties observed in the Local Group dwarf galaxies
and in the CHVCs. 

In Figures 6 through 11 we display a variety of computed 
cloud properties as functions of the total gas
mass, $M_{\rm WM}$, and halo virial mass $M_{\rm vir}$. We emphasize that
the total gas mass $M_{\rm WM}$ refers to ``warm gas'' only,
and includes the WIM and WNM, but excludes CNM.
As discussed in \S 3, in computing the hydrostatic and ionization
structures we adopt a metallicity of 0.1 $Z_\odot$, and
we assume that $T=10^4$ K in both the WNM and the WIM.
We assume irradiation by the metagalactic field. We focus on 
the behavior of median ($\sigma=0$) Burkert halos and $-4\sigma$ NFW halos,
since as we discuss in \S 7, 
for such halos the computed gas scale heights and WNM cloud
sizes correspond to those observed in Local Group dwarf galaxies.
For each family of halos we display results for HIM bounding pressures,
equal to 1, 10, and 100 cm$^{-3}$ K.

The quantities shown in the 
$M_{\rm gas}$ vs. $M_{\rm vir}$ plots
are; (a) the total cloud radius $r_{\rm W/H}$ (kpc);
(b) the total gas to DM mass ratio $M_{\rm WM}/M_{\rm DM}$
within $r_{\rm W/H}$;
(c) the radius
$r_{\rm WNM}$ (kpc) of the WNM core (if present);
(d) the total HI gas mass $M_{\rm HI}$ (M$_\odot$) contained in
the WNM and WIM components;
(e) the peak HI column density $N_{\rm HI}$ (cm$^{-2}$) integrated
across the entire cloud diameter,  
and (f) the central gas pressure $P_0/k$ (cm$^{-3}$ K).

We recall (see \S 2) that for a given DM density profile
(NFW or Burkert) and an assumed correlation relation,
$x_{\rm vir}(M_{\rm vir})$, between the halo concentration 
$x_{\rm vir}\equiv r_{\rm vir}/r_s$
and the virial mass, a choice for $M_{\rm vir}$ 
uniquely determines the halo structure and scale parameters.
We consider halos with $M_{\rm vir}$ ranging
from $10^8$ to $10^{10}$ $M_\odot$. For median
Burkert halos this mass range corresponds to scale velocities
from 12 to 50 km s$^{-1}$. For the $-4\sigma$ NFW halos this
mass range corresponds to $v_s$ ranging from 8 to 40 km s$^{-1}$.

In Figures 6 through 11  
the WM gas masses for which results are shown range from
$5\times 10^6$ to $5\times 10^7M_\odot$. 
This range was selected such that
peak HI columns of $10^{19}$ to $10^{21}$ cm$^{-2}$,
representative of the columns 
observed in the CHVCs and dwarfs, appear within
the displayed $M_{\rm WM}$ vs. $M_{\rm vir}$ plots.

For the purposes of the discussion that follows we refer to halos
approaching $\sim 10^8M_\odot$ as ``low-mass halos'', and halos
approaching $\sim 10^{10}M_\odot$ as ``high-mass halos''.

The total cloud radius $r_{\rm W/H}$ is located at the outer edge of 
the WIM envelope where the cloud pressure equals the bounding pressure 
$P_{\rm HIM}$. For fixed $P_{\rm HIM}$, and a given halo mass 
$M_{\rm vir}$, the WM cloud radius increases with increasing
cloud mass $M_{\rm WM}$. In low-mass halos the clouds
are pressure-confined, and $M_{\rm WM}$ 
is proportional to the cloud volume, so that
$r_{\rm W/H}\sim M_{\rm WM}^{1/3}$. The cloud radius is less
sensitive to $M_{\rm WM}$ in high-mass halos as the clouds
become gravitationally confined to approximately fixed volumes.
For a given halo mass and cloud mass, the cloud radius $r_{\rm W/H}$
decreases as $P_{\rm HIM}$ is increased and
the clouds are pressure confined to smaller volumes.

The vertically running curve in panels b, c, and f of Figures 6 through 11
identify the locus of marginally bound clouds as defined by
equation~(\ref{e:bound}).  Clouds to the left of this curve 
are unbound and gas particles with thermal
energies gas can escape the halo potential wells.
Clouds to the right of this curve satisfy equation~(\ref{e:bound}) and are 
fully bound. The position of the ``bounding'' curve shifts towards
lower virial masses as $P_{\rm HIM}$ increases and the WM gas
is pressure-confined to the inner parts of the potential wells.
For example,
$10^4$ K clouds in median Burkert halos, with
$P_{\rm HIM}=10$ cm$^{-3}$ K, are bound if 
$M_{\rm vir} \gtrsim  
1.5\times 10^8M_\odot$ (or $v_s \gtrsim 15$ km s$^{-1}$). 

We define $r_{\rm WNM}$ as the radius
at which the fractional ionization $x_e=0.5$. This radius effectively
marks the location of the ionization fronts (when present) and
represents the transition point from the WIM to the WNM.
In critical clouds, $r_{\rm WNM}=0$.
Critical clouds are indicated by the dashed curves in 
the panels 6c through 11c.  
For example, for $P_{\rm HIM}=10$ cm$^{-3}$ K, and a median
$3\times 10^8M_\odot$ Burkert halo,
a critical cloud forms for $M_{\rm WM}=2\times 10^6M_\odot$.
Clouds to the left and below the dashed
curves are fully ionized, while clouds to the right and above 
are predominantly neutral. 
For fixed $M_{\rm WM}$, the clouds are ionized
when $M_{\rm vir}$ is small since the clouds are then 
extended low-density objects.
As $M_{\rm vir}$ is increased the clouds
are gravitationally confined to smaller volumes, and become denser.
Recombination becomes more effective, and a neutral core forms.
For high-mass halos, neutral cores exist when $M_{\rm WM}$
is large. However, as the WM gas mass is reduced
the cloud mass finally becomes too small to maintain an optically
thick WIM envelope and the cloud becomes fully ionized.
As $P_{\rm HIM}$ is increased the clouds are compressed, and
cloud neutralization occurs at smaller virial masses. 

We note that gas close to $r_{\rm W/H}$ will be undetectable since by
definition the gas columns fall to zero at the cloud boundaries.
The {\it observed} cloud sizes are better represented by the gas scale 
height $r_{\rm gas}$, or the neutral boundary $r_{\rm WNM}$.
Depending on the bounding pressure, $r_{\rm gas}$ and
$r_{\rm WNM}$ can be significantly smaller than $r_{\rm W/H}$. 
In gravitationally confined clouds $r_{\rm WNM}$
is generally a few times larger than the gas scale height, 
which may be estimated using
the analytic small-$x$ expressions for $r_{\rm gas}$ (see Appendix A).
E.g., for median Burkert halos, and assuming
$c_{g,{\rm WNM}}=8.2$ km s$^{-1}$, the analytic gas scale-height
ranges from 0.3 to 0.5 kpc, for $M_{\rm vir}$ between $10^8$ and $10^{10}$
M$_\odot$.  For $-4\sigma$ NFW halos, $r_{\rm gas}$ 
ranges from 0.8 to 0.3 kpc.
Our numerical results are consistent with these analytic estimates.
In median Burkert halos, and $-4\sigma$ NFW halos,
$r_{\rm WNM}$ ranges from 0.5 to 2 kpc.

The dependence of the peak HI column density
and total HI cloud mass on 
$M_{\rm vir}$ and $M_{\rm WM}$ can
now be readily understood. In pressure-confined clouds, the
gas densities are approximately constant, and the
cloud volumes are proportional to the WM gas mass.
Thus, $N_{\rm HI,peak} \sim M_{\rm WM}^{1/3}$.
The pressure-confined clouds tend to be ionized when $P_{\rm HIM}$
is low, and the HI then exists as a small neutral fraction
within the WIM dominated gas. The peak HI column is therefore small
($< 3\times 10^{18}$ cm$^{-2}$)
and decreases rapidly 
as $M_{\rm vir}$ decreases, since the gas then
becomes more highly ionized. In ionized clouds, the total HI mass
is a small fraction of the total gas mass. For example, for 
$P_{\rm HIM}=1$ cm$^{-3}$ K, and a 
$M_{\rm vir}=4\times 10^8M_\odot$ median Burkert halo with 
$M_{\rm WM}=8\times 10^5M_\odot$, the neutral
HI mass $M_{\rm HI}=10^3M_\odot$, or
$\sim 0.1\%$ of the total WM mass.
Neutral clouds in high-mass halos are in the small-$x$ limit
with approximately constant volumes. 
In such clouds the neutral gas dominates
the total gas mass, and $N_{\rm HI,peak}$ is
proportional to $M_{\rm WM}$.

A sharp increase in $N_{\rm HI,peak}$
is associated with the transition from ionized to neutral clouds
as the halo mass increases. For 
critical clouds, the peak HI column density equals
$3\times 10^{18}$ cm$^{-2}$ for both Burkert and NFW halos,
independent of $M_{\rm vir}$ or $M_{\rm WM}$.
The fact that the peak HI column density in critical clouds equals
a nearly constant value of $3\times 10^{18}$ cm$^{-2}$, independent
of $M_{\rm vir}$, is due to the weak dependence 
of the critical central densities, $n_{\rm H,0crit}$, on the halo
mass or scale velocity. This is also indicated 
analytically (see Table 6 in Appendix A).
For example, for median Burkert halos, a ``universal''
central gas density $n_{\rm H,0crit}\approx 5.5\times 10^{-3}$ cm$^{-3}$
is associated with critical clouds,
assuming a WIM sound speed $c_{g,{\rm WIM}}=11.6$ km s$^{-1}$,
and $J_3^*=1$ (see Table 6).
This then implies a fixed ionization rate 
$\zeta_{\rm crit}=n_{\rm H,0crit}\alpha_B=1.4\times 10^{-15}$ s$^{-1}$
at the center of critical clouds [where by definition $n(e)=n(H)$].  
Since for our assumed field the unattenuated ionization rate 
$\zeta_0=4\times 10^{-14}$ s$^{-1}$, 
a well specified opacity is required
to reduce the ionization rate to $\zeta_{\rm crit}$.
The neutral fraction then approaches unity for clouds with peak columns
exceeding 10$^{19}$ cm$^{-2}$.

The condition that $M_{\rm WM}/M_{\rm DM} < 1$, where $M_{\rm DM}$ is the
dark-matter mass within $r_{\rm W/M}$, must be satisfied for
self consistent dark-matter dominated solutions.
This condition is met for the entire parameter
space displayed in Figures 6 through 11. The ratio $M_{\rm WM}/M_{\rm DM}$
is smallest for low mass clouds in high mass halos. In such
objects the gas is restricted to the small-$x$ regions where
the DM densities are large.  A small gas mass then implies
$n_{\rm H,0}/n_{ds} \ll 1$.
For fixed $M_{\rm vir}$, the mass ratio increases
with $M_{\rm WM}$.
For a fixed gas mass, $M_{\rm WM}/M_{\rm DM}$ increases
as $M_{\rm vir}$ becomes small, since the
gas clouds then extend to large radii ($x\gg 1$) where the DM density
becomes small.  Thus,  $M_{\rm WM}/M_{\rm DM}$ is largest in
high-mass clouds in low-mass halos. The  
bound clouds are also dark-matter dominated
for the range of gas masses we consider. Thus, all regions
to the right of the bounding curves (indicating the marginally
bound clouds) in Figures 6 through 11, represent self-consistent,
dark-matter dominated, hydrostatic solutions for the WM cloud structure. 

In pressure-confined clouds, the central pressure
remains small and is insensitive to $M_{\rm WM}$
because $P_0/P_{\rm HIM}$ is not large
in such objects. For example, for low-mass halos,
with $v_s \lesssim 15$ km s$^{-1}$, 
$P_0/k\lesssim 10$ cm$^{-3}$ K, when $P_{\rm HIM}=1$ cm$^{-3}$ K.
In gravitationally confined clouds, 
where the cloud volumes
are approximately constant,
$P_0$ is proportional to $M_{\rm WM}$ and can become very large.
E.g., $P_0/k\gtrsim 10^4$ cm$^{-3}$ K in high-mass
halos ($v_s \gtrsim 30$ km s$^{-1}$)
containing $M_{\rm WM} \gtrsim 5\times 10^6M_\odot$.
In critical clouds,
$P_0/k\approx 100$ cm$^{-3}$ K, consistent with
our analytic estimate of 
$n_{\rm H,0crit}\approx 5\times 10^{-3}$ cm$^{-3}$, 
and equation~(\ref{e:pbasic})
for the gas pressure.

We also show
(in Figs. 6f-11f displaying the $P_0/k$ contours)
the regions of parameter space where the gas becomes
multi-phased at the cloud centers.
These regions are indicated by the narrow strip defined by
the two dashed curves. On the left curve,
$P_0=P_{\rm min}(0)$, and multi-phased gas is just possible at $r=0$.
On the right curve, $P_0=P_{\rm max}(0)$, and such 
clouds contain the largest possible
multi-phased cores. Since both $P_{\rm min}$ and $P_{\rm max}$ were
computed assuming heating by the external field, our results 
reveal the values of $M_{\rm vir}$ and $M_{\rm WM}$ for which
multi-phased behavior is possible in clouds heated by the external field only. 
It is evident that multi-phased cores occur for values of $M_{\rm vir}$
and $M_{\rm WM}$ that are only slightly larger than those required for the
transition from ionized to neutral clouds. Multi-Phased behavior occurs
for clouds with peak HI column densities between $2\times 10^{19}$ and
$1\times 10^{20}$ cm$^{-2}$.  

To the right of, or above, the multi-phased strip,
we have $P_0 > P_{\rm max}(0)$.
The warm gas in the central regions of such clouds must fully convert to CNM,
unless additional internal heating mechanisms are invoked
to alter the local values of $P_{\rm min}$ and $P_{\rm max}$.
Thus, for a given halo mass,
the $P_0=P_{\rm max}(0)$ contour delineates the maximal
mass of warm gas that can exist as thermally supported WNM
heated by the external field.
Larger WM masses can exist only if 
internal heat sources drive the gas to the WNM phase.
For example, for $P_{\rm HIM}/k=10$ cm$^{-3}$ K, and 
a median Burkert halo with 
$M_{\rm vir}=2\times 10^9M_\odot$
the maximal WM mass is $2\times 10^6M_\odot$ (see Fig.~7f).
If the gas mass is increased above this value,
the heating rates at the cloud center
are decreased due to the increased cloud
opacity, and the gas in the core must convert to CNM. This would then
drive the total mass of warm gas back to the maximum value of    
$2\times 10^6M_\odot$ (leading also to a reduction in the cloud opacity).
More generally, if the gas becomes multi-phased, with CNM clouds that
are not necessarily at the center, the total gas mass (including the CNM)
can increase 
beyond the maximum WNM mass set by the condition $P_0=P_{\rm max}(0)$.
The maximal WM masses increase without
limit as $M_{\rm vir}$ becomes small, but these
clouds are unbound.  For example, for {\it bound} clouds in
median Burkert halos, with $P_{\rm HIM}/k=10$ cm$^{-3}$ K,
the largest possible WM mass is $1\times 10^7M_\odot$ (see Fig.~7f).

The region of the $M_{\rm vir}$ vs. $M_{\rm WM}$ plane
within the ``multi-phased strip'', and to the right of the
``bounding line'', represents the region of parameter space
within which acceptable models for the CHVCs may be found.
In this region the WM clouds are bound to gravitationally dominant
dark-matter halos, and contain neutral multi-phased cores.
As we discuss further below, a
key result is our finding that the location of the
multi-phased strip coincides with the range of peak
HI column densities observed in the CHVCs.

\section{Modelling the Dwarf Galaxies Leo A and Sag DIG}

Our main goal is to find halo-cloud models that provide acceptable
fits to the HI properties of the CHVCs.
However, to validate our approach, we first 
model the pressure-supported 
HI structures observed in the dwarf 
irregular (dIrr) galaxies Leo A and Sag DIG,
both of which are well studied members of the Local Group.
Modelling the dwarfs is an important
auxiliary step because their distances are 
well determined, so that the true physical sizes (as opposed to just
the angular sizes) of the WNM clouds are known.  
A specific family of halos (e.g.~median, $-1\sigma$, $-2\sigma$ etc.)
can therefore be selected using the observed gas scale heights.
We list the basic optical properties (Mateo 1998) and
21 cm HI properties (Young \& Lo 1996; 1997) of Leo A and Sag DIG in Table 1.  

For Leo A we adopt
the recently revised distance of $690\pm 60$ kpc determined
by Tolstoy et al. (1998).
Young \& Lo (1996) assumed a much larger distance of 2.2 Mpc
to Leo A based on earlier and less certain observations by
Hoessel et al. (1994).   For Sag DIG we adopt Mateo's (1998)
recommended distance of $1061\pm 160$ kpc (also assumed by
by Young \& Lo 1997). 
The total V-band luminosities and half-light core radii
are $3.0\times 10^7$ L$_\odot$ and $0.185$ kpc for Leo A,
and $6.9\times 10^7$ L$_\odot$ and $0.125$ kpc for Sag DIG (Mateo 1998).
Both galaxies appear to have undergone several episodes of star-formation
(Tolstoy et al. 1998; Cook 1988). In both Leo A and Sag DIG
the measured oxygen abundances are O/H$\approx 2\times 10^{-5}$, implying
low metallicities $Z\approx 0.07Z_\odot$
(Skillman, Kennicutt \& Hodge 1989;  Skillman, Terlevich \& Melnick 1989).

Young \& Lo (1996) carried out high resolution ($15^{\prime\prime}$) 
interferometric VLA 21 cm observations of Leo A.  They found that
the optical galaxy is surrounded by a much 
larger $2.8 \times 1.7$ kpc envelope of HI gas.
The total VLA 21 cm flux is $65 \pm 3$ Jy km s$^{-1}$, 
just about equal to the single-dish (43-meter Green Bank)
flux of $69 \pm 3$ Jy km s$^{-1}$, implying that
the interferometer detected all of the HI emission.
The total HI mass is $7.9\times 10^6M_\odot$ (for $d=690$ kpc).
The peak HI column density is $2.7\times 10^{21}$ cm$^{-2}$, and
emission was detected down to a minimum HI column of 
$N_{{\rm HI,min}}\sim 2\times 10^{19}$ cm$^{-2}$, at
a (mean) radius of $r(N_{{\rm HI,min}})=1.2$ kpc from the galaxy center.
A key finding of the Young \& Lo study is that
the global line profile can be separated into broad
($9.6\pm 0.2$ km s$^{-1}$) and narrow ($5.2\pm 0.3$ km s$^{-1}$) components. 
The broad (WNM) component contains 80\% of the flux and 
is produced throughout the HI image,
whereas the narrow (CNM) component is localized near 
isolated star-forming regions within the optical galaxy.
In view of these observations we set
$N_{{\rm HI,peak}}=2.2\times 10^{21}$ cm$^{-2}$
for the WNM distribution in Leo A.

In their VLA study of Sag DIG, Young \& Lo (1997) discovered
a similar $3.3 \times 3.0$ kpc HI envelope
surrounding the much smaller optical image. The VLA flux
is $30.3\pm 1.5$ Jy km s$^{-1}$, close to the single dish
flux of $32.6\pm 1.2$ Jy km s$^{-1}$. The total implied
HI mass (for $d=1.1$ Mpc) is $9.3\times 10^6M_\odot$.
The HI structure in Sag DIG is somewhat more complex than
in Leo A, and appears as an asymmetric ring with
several clumps of high column density gas
near the optical galaxy.  The peak column density is
$1.2\times 10^{21}$ cm$^{-2}$, and emission was detected
down to $N_{{\rm HI,min}}=5\times 10^{18}$ cm$^{-2}$,
at $r_{{\rm HI,min}}=1.6$ kpc from 
the galaxy center.  The Sag DIG HI line profile can also 
be decomposed into extended broad WNM (9-10 km s$^{-1}$) 
and localized narrow CNM (4-6 km s$^{-1}$) components.
Young \& Lo find that 83\% of the total flux is contained in the WNM
component. Thus, $N_{{\rm HI,peak}}=1.0\times 10^{21}$ cm$^{-2}$ 
for the WNM in Sag DIG

Given the interferometric HI maps
(Fig.~1 in Young \& Lo 1996, and Fig.~3 in Young \& Lo 1997)
it is possible to determine the 
(projected) 1/e gas scale heights for Leo A and Sag DIG.
In Leo A, averaging over the long and short axes,   
the {\it projected} angular scale height is
$\theta_{\rm gas}\approx 160$ arcsec, corresponding to 
a projected gas scale height of $0.5$ kpc
for a distance of 690 kpc.
Determining a gas scale height for Sag DIG is more
complicated due to the inhomogeneities in the HI emission.  
However, based on the Young \& Lo (1997) map
we estimate that $\theta_{\rm gas}\approx 70$ arcsec, 
implying a projected scale height $\sim 0.4$ kpc.
It is significant that we find comparable scale heights for
Leo A and Sag DIG.

Can a DM halo model be found, that reproduces (a) the total 
extents of the HI clouds, (b) the observed $1/e$ gas scale
heights, and (c) the peak WNM HI column densities, assuming
photoionization by the metagalactic field?
What is the maximum possible Local Group
HIM pressure that is consistent with the dwarf galaxy observations?

For the dwarf galaxies we adopt a representative 
projected gas scale height of 0.5 kpc, and a WNM velocity dispersion
of 9 km s$^{-1}$. The observed 
velocity dispersion is consistent with $\sim 10^4$ K gas. 
Crucially, the observed scale height
effectively determines the relative concentration
of the DM halo, since as is indicated by our analytic expressions
for $r_{\rm gas}$ (see Table 6, and Figs.~16 and 17, in Appendix A)  
the intrinsic scale height 
$r_{\rm gas}$ is weakly dependent
on the halo scale velocity. Thus, the observed scale height may
be used to select an appropriate value of $\sigma$ for
the dwarf galaxy halos.  In particular, it is clear that
if the dwarfs are embedded in NFW halos, these
halos must be significantly
underconcentrated, with $\sigma \approx -4$ relative to the
(theoretical) $\Lambda$CDM $x_{\rm vir}(M_{\rm vir}$) relation
for median halos as specified by equation~(\ref{e:fit}).
In contrast, {\it median} Burkert halos will produce scale
heights very close to those observed in the dwarfs.
It is also evident from our $M_{\rm vir}$ vs.~$M_{\rm WM}$
computations (Figs.~6-11) for median Burkert halos and 
$-4\sigma$ NFW halos, that $r_{\rm WNM}\approx 1$ to 2 kpc for 
HI gas column densities $\sim 10^{21}$ cm$^{-2}$, 
consistent with the observed cloud sizes and peak HI column densities.
These results guide the choice of appropriate dwarf DM halo models.

However, a complication arises in that the {\it optical} galaxies do not
appear to be dark-matter dominated. In Leo A and Sag DIG,
$(M/L_V)_\odot \approx 3.5$ and 1.4
within the optical radii (Mateo 1998).
These mass-to-light ratios are within the range expected for evolved stellar
systems (e.g.~Bruzual \& Charlot 1993; Sternberg 1998),
and are, for example, much smaller than observed
in the dark-matter dominated dwarf-spheroidals Ursa Minor
and Draco where $(M/L_V)_\odot \sim 80$ (Aaronson 1983; Gallagher 1994). 

Thus, in modelling the hydrostatic HI gas distributions
in Leo A and Sag DIG we
consider the possibility that stars contribute to the gravitational potential.
We make the simplifying approximation that the total gravitational
potential is the sum 
\begin{equation}
\varphi = \varphi_{\rm DM} + \varphi_{*}
\end{equation}
of (independent) DM and stellar potentials. For the stellar potential 
$\varphi_*$ we adopt a Plummer model (see Spitzer 1987).
In this model the enclosed stellar mass
\begin{equation}
\label{e:plummass}
M_{*}(y)=M_{s*}{y^3 \over (1+y^2)^{3/2}}\equiv M_{s*}f_{M*}(y) \ \ \ ,
\end{equation}
where $y\equiv r/r_{s*}$, and where $r_{s*}$ is
the scale radius of the stellar density distribution,
and $M_{s*}$ is the total stellar mass.
We assume that $r_{s*}$ is equal to the
observed (projected) half-light core radius
\footnote{
For many stellar density distributions, including the Plummer model,
the projected half-mass radius $r_{hp}=0.77r_h$, where $r_h$ is the
half-mass radius. For the Plummer distribution $r_h=1.3r_{s*}$. 
Thus, $r_{s*}\simeq r_{hp}$ (Spitzer 1987).}, 
and we fix the total stellar mass
assuming $(M/L_V)_\odot=1$. As representative
values for Leo A and Sag DIG we set $r_{s*}=0.185$ kpc 
and $M_{s*}=3\times 10^6M_\odot$ (Mateo 1998).

We augment the equation of hydrostatic equilibrium 
to include the stellar mass. 
Expression~(\ref{e:sbasic}) for the gas pressure $P(x)$ becomes
\begin{equation}
\label{e:hplum}
P(x) = P_0\,{\rm exp}\biggl\{-\int_0^x  {v_s^2\over c_g^2}
\biggl[{f_M(x^\prime) \over x^{\prime 2}} + {M_{s*}\over M_{ds}}{f_{M*}(r_sx/r_{s*})
\over x^{\prime 2}}\biggr]dx^\prime\biggr\}
\end{equation}
where $x=r/r_s$ is the radius in units of the DM halo scale radius $r_s$,
and $M_{ds}$ is the halo scale mass.

In Figure 12 we show a specific gravitationally confined
model for the dwarfs, in which we combine
a stellar system (with $r_{s*}=0.185$ kpc, and $M_{s*}=3\times 10^6M_\odot$)
with a median Burkert halo with a scale velocity $v_s=30.0$ km s$^{-1}$
corresponding to $M_{\rm vir}=2.0\times 10^{9}M_\odot$. 
We assume a low bounding pressure
and set $P_{\rm HIM}=1$ cm$^{-3}$ K. A good fit is
obtained for a total gas mass $M_{\rm WM}=2.2\times 10^7M_\odot$.
Additional model parameters are listed in Table 2.

In this model the combined stellar and DM mass dominates the enclosed
mass at all radii. The stellar component (i.e. the optical galaxy)
dominates out to the stellar core radius of 0.2 kpc. At larger radii
the DM halo dominates the mass. The WM cloud consists
of a WNM core with $r_{\rm WNM}=1.4$ kpc, surrounded by a WIM envelope
extending to $r_{\rm W/H}=4.4$ kpc. The WM cloud satisfies 
equation~({\ref{e:bound}) and
is bound. The total HI mass equals $1.3\times 10^7M_\odot$, or
61\% of the total hydogen mass. Virtually all of the HI mass is
contained within $r_{\rm WNM}$.
The computed peak HI column density is $1.4\times 10^{21}$ cm$^{-2}$,
consistent with the observed peak WNM column densities.
The computed
$1/e$ scale height for the projected HI column density 
distribution is $r_{\rm {N_{\rm HI}}}=0.6$ kpc, in good agreement with the observed
scale heights of 0.5 kpc. 

The HI column falls sharply near $r_{\rm WNM}=1.4$ kpc.
The HI column decreases from $2\times 10^{19}$ to $1\times 10^{18}$
cm$^{-2}$, between 1.3 and 1.5 kpc.  
The sharp decline in the HI column densities that occurs between
$10^{19}$ and $10^{18}$ cm$^{-2}$   
is a general feature of the computed gas
distributions, for both Burkert and NFW potentials, and is due
to the photoionization cut-off of the HI gas density distribution.
The cut-off column of $\sim 10^{19}$ cm$^{-2}$
is similar to those predicted for 
flattened, rotationally supported, disk systems 
(e.g.~Corbelli \& Salpeter 1993; Maloney 1993).  
The radius at which this cut-off occurs is insensitive to
the intensity of the ionizing radiation field
because the ionization front
is located at a radius where the total gas density 
is decreasing exponentially.

For Leo A we are essentially able to reproduce $r_{{\rm HImin}}=1.2$ kpc for
$N_{{\rm HImin}}=2\times 10^{19}$ cm$^{-2}$ 
as found by Young \& Lo in Leo A. In our interpretation
this observational cut-off radius is very close to $r_{\rm WNM}$,
and corresponds to the location of the ionization front.
However, if in Sag DIG the neutral boundary also occurs near 1.24 kpc,
we are unable to account for the column of $5\times 10^{18}$ cm$^{-2}$
reported by Young \& Lo at the significantly larger radius of 1.6 kpc.
In our model we predict a column of only $6\times 10^{17}$ cm$^{-2}$
at this radius.
As we discuss in \S 6, we encounter a similar difficulty in fitting
the extended low column density distributions in the CHVCs. 

In Figure 12d we plot the H$\alpha$ recombination line surface brightness
$I_{\rm H\alpha}$ as a function of cloud radius. The H$\alpha$ emission
is produced in the outer WIM envelope, and
the computed surface brightness
includes the equal contributions
from both sides of cloud. At the cloud center
$I_{\rm H\alpha}=4.3$ mR. The surface brightness rises 
to 8 mR near $r_{\rm WNM}$ due to limb-brightening, and
then declines rapidly as the projected columns of ionized gas become small.

The HI mass in our model clearly exceeds 
the maximum WNM mass, of about $5\times 10^5M_\odot$, 
(see Fig.~6f) that can be maintained
by the metagalactic field.  This fact is indicated by
the $P_{\rm min}$ and $P_{\rm max}$ curves (shown in Fig.~12e) that were 
computed assuming heating and ionization by the external field.
It is evident that $P > P_{\rm max}$ within 0.9 kpc. The central pressure
$P_0/k= 6.0\times 10^{3}$ cm$^{-3}$ K
(corresponding to a gas density $n_0=0.5$ cm$^{-3}$) is much greater
than $P_{\rm max}(0)/k=71.5$ cm$^{-3}$ K.
Thus, our model for the WNM in the dwarf galaxies
implicitly assumes that
additional heat sources (presumably associated with the
optical components) are available to produce the observed mass of warm HI
in these objects. We note that
the computed $P_{\rm min}$ and $P_{\rm max}$ are coupled to the
ionzation rate which declines with increasing cloud opacity. 
Figure 12f shows that
the total hydrogen ionization rate decreases from the unattenuated optically
thin value of $4.0\times 10^{-14}$ s$^{-1}$ at the cloud edge, to
$1.0\times 10^{-18}$ s$^{-1}$ in the optically thick core.
At $r=0$ the total ionization rate is $10$ times larger than the primary
rate, due to the dominating effect of secondary ionizations
(which are negligible in the ionized envelopes).

The model displayed in Figure 18 is representative. As is indicated
by the parameter-space study presented in \S 4, acceptable models
for the dwarf galaxy halos can be found for other values of 
$M_{\rm vir}$, $M_{\rm WM}$ and $P_{\rm HIM}$,
and for (underconcentrated) 
NFW halos rather than Burkert halos.  It is evident
from Figures 6e and 9e that for a low bounding pressure 
($P_{\rm HIM}/k=1$ cm$^{-3}$ K) acceptable solutions require
$v_s \gtrsim 25$ km s$^{-1}$. In such halos, as in our particular model,
the WNM and WIM components are
gravitationally confined.  If larger bounding pressures
are assumed (e.g.~$P_{\rm HIM}/k=100$ cm$^{-3}$)
lower values of $v_s$ are allowed. 
We note that for our model,
the effect of increasing $P_{\rm HIM}$ is 
simply to truncate the outer part of the WM cloud at the
radius where $P_{\rm gas}$ equals the increased value of
$P_{\rm HIM}$. The inner
part remains largely unaffected by the truncation.
For example, if the bounding pressure is increased
from 1 to 100 cm$^{-3}$ s$^{-1}$, then
the WM cloud boundary moves inwards to $r_{\rm W/H}\approx 1.3$ kpc.
Most of the tenuous WIM mass is removed, but the inner neutral core
is preserved.  However, if $P_{\rm HIM}$ is increased above $\sim 100$
cm$^{-3}$ K, 
the neutral core is also truncated, and its size shrinks below the
observed values.  We conclude that the dwarf
galaxy HI structures are consistent with an upper-limit of 
$P_{\rm HIM}/k\approx 100$ cm$^{-3}$ K for 
the intergalactic medium in the Local Group.

Our model fit demonstrates that the observed extended neutral HI clouds
in Leo A and Sag DIG may be readily interpreted as 
dark-matter dominated, pressure-supported gas clouds (although
we may have difficulty in accounting for extended HI at a level
$\lesssim 5\times 10^{18}$ cm$^{-2}$). In our model,  
the central gas density $n_{H,0}=0.46$ cm$^{-3}$ is much larger
than the central gas density
of $n_{\rm H,0crit}=5\times 10^{-3}$ cm$^{-3}$ in critical clouds, 
and is sufficiently large to keep the gas cloud 
neutral against the metagalactic radiation field out to $\gtrsim 1$ kpc. 
The characterstic (or central) DM density in our model
is $n_{ds}=1.24$ cm$^{-3}$ (or $3.1\times 10^{-2}M_\odot$ pc$^{-3}$),
in excellent agreement with the
typical central DM densities of $2.0\times 10^{-2}M_\odot$ pc$^{-3}$
found in HI rotation curve studies of rotationally supported
dwarfs and low surface brightness galaxies
(de Blok \& McGaugh 1997; Firmani et al. 2000; de Blok, McGaugh \& Rubin 2001).
Finally, we note that while both NFW and Burkert halos may be found that 
fit the dwarf galaxy observations, the Burkert solutions require
median halos with ``typical'' dark-matter concentration parameters
(in our model $x_{\rm vir}=25.6$). However, NFW solutions are
problematic in that they must 
be extremely underconcentrated, by $\sim -4\sigma$
relative to the $\Lambda$CDM $x_{\rm vir}(M_{\rm vir})$ relation,  
to match the observed scale heights in the dwarf galaxies.

\section{Modelling the CHVCs}

We have demonstrated that mini-halo models can be constructed for
the HI distributions in the
dwarf galaxies Leo A and Sag DIG.  Can similar models
be found for the CHVCs? Here, we focus on the HI properties of the
CHVCs as most recently probed by the high resolution
single dish observations and interferometric observations
presentd by Braun \& Burton (2000) and Burton et al. (2001)
(see also Br\"uns et al. 2001; Putman et al. 2002).

In their Arecibo observations, Burton et al. (2001) were able to resolve
the WNM HI gas distributions in a sample of ten CHVCs.
They constructed $1^\circ \times 1^\circ$
sized maps of each object at a spatial resolution of $3.1 \times 3.7$
arcmin, and determined the locations of the peak HI flux concentrations.
They then carried out deep integration ``cross-cut'' observations to 
study the detailed gas distributions around the central peaks.
Burton et al. found that the peak WNM
HI column densities range from $2\times 10^{19}$ to 
$2\times 10^{20}$ cm$^{-2}$, and that the projected $1/e$ scale-lengths
of the inner parts of the clouds range from 412 to 908 arcsec, 
with a typical value of 690 arcsec.
It is significant that the peak HI columns in the CHVCs are much
smaller than those observed in
Leo A and Sag DIG.  In their cross-cut observations, Burton et al.
detected gas down to column densities of $\sim 2\times 10^{17}$ cm$^{-2}$
(corresponding to the $1\sigma$ sensitivity limit). In some cases
low column density ``wings'' extend
out to $\sim 1^\circ$ from the cloud centers.
In some clouds the HI gas distributions are quite asymmetric,
with emission at a level of $10^{18}$ cm$^{-2}$ persisting out
to several gas scale heights on one side of the emission peak,
with a more rapid decline on the opposite side.
The asymmetries could be due to the presence of head-tail
structures as observed by Br\"uns et al (2001) in CHVC 125+41-207,
or to an imprecise positioning of the
emission centroids.
Burton et al. argued that intrinsically exponential 
gas density distributions (but observed in projection)
provide a reasonable description of
the data, although significant departures from this
simple density law are clearly present. Burton et al. also found that
the velocity gradients across the clouds are
relatively small  ($\sim 10$ km s$^{-1}$ degree$^{-1}$)
implying that rotation is not dynamically important
in supporting the WNM clouds.

CNM cores were detected by
Braun \& Burton (2000) and de Heij et al. (2002c) in 11 CHVCs
(including two that were subsequently observed at Arecibo)
in high resolution Westerbork interferometric observations. 
These observations revealed the presence of
high column density (up to $\sim 10^{21}$ cm$^{-2}$)
cores displaying narrow ($2-15$ km s$^{-1}$ FWHM) HI line profiles.
The narrow linewidth emission clearly traces a
CNM component, and typically
contributes $\sim 30\%$ of the total CHVC HI luminosity.
Rotation may be important in the cores,
where velocity gradients are $\sim 1$ km s$^{-1}$ arcmin$^{-1}$. 
The radii of the CNM cores range from 5 to 17 arcmin, 
with an average value of 10 arcmin. It is significant that the
CNM core sizes are comparable to the WNM gas scale heights.

Given these observational results we adopt the following
properties as ``typical'' for an indivdual CHVC, and which
should be reproduced in any successful model for the CHVCs;
(1) a peak WNM HI column of $5\times 10^{19}$ cm$^{-2}$,
(2) a projected (1/e) WNM gas scale height of 690 arcsec
for an assumed distance $d$,
(3) HI columns of $\sim 1\times 10^{18}$ cm$^{-2}$ extending
to $\sim 3$ scale heights,
and (4) a multi-phased core region with projected radius
equal to or slightly smaller than the gas scale height.
Properties (1)-(3) are similar to those that fix
the dwarf halos, with the critical difference that
the distances to the CHVCs are not known a priori. Property (4)
provides an additional constraint, since we wish to construct
models in which the multi-phased cores arise in gas
heated exclusively by the metagalactic field. We will find that
minihalo models can be constructed that readily reproduce
properties (1), (2) and (4), while we have difficulty in
accounting for property (3).

The total number of CHVCs is also an important constraint.
We recall that de Heij et al. (2002b) identified 67 CHVCs
within the Leiden-Dwingeloo survey, and that    
Putman et al. (2002) found an additional 179 CHVCs
in their Parkes HI survey, giving a total of 246 objects.
If the CHVCs represent ``missing satellites'' associated
with either the Local Group (Blitz et al. 1999),
or with the Milky Way (Klypin et al. 1999), then  
the total number of objects should be related to the typical 
mass of the CHVC mini-halos. The numerical 
simulations of halo evolution predict that 
the number of embedded dark-matter
sub-halos contained within any
given parent halo increases rapidly with decreasing sub-halo mass
(Klypin et al. 1999; Moore et al. 1999).
As a guide to this predicted behavior we adopt the 
results of Klypin et al. for their $\Lambda$CDM simulation
of the Local Group. In their computations two massive
halos are produced, representing the Milky Way and M31.
Each massive object contains a large number of
embedded sub-halos.  Klypin et al. find that the number
density of sub-halos is proportional to $v_s^{-2.75}$,
where $v_s$ is the sub-halo scale velocity
\footnote{
We note that Klypin et al. present their results in terms
of the maximum circular velocities, $v_{\rm max}$, of
the subhalos. We recall that for NFW and Burkert profiles
$v_{\rm max}=0.8v_s$.}. They also find that for a given $v_s$,
the volume density of objects decreases as $\sim 1/d^2$,
where $d$ is the distance from the center of the parent halo.
We reexpress the results of Klypin et al. 
(their eq. [3] and [4]) as 
\begin{equation}
\label{e:klypin}
{\cal N}(>v_s,<d) = 1.06\times 10^3 
\biggl({M_{{\rm vir,p}} \over 10^{12} \ {\rm M_\odot}}\biggr)
\biggl({v_s \over 10 \ {\rm km} \ {\rm s}^{-1}}\biggr)^{-2.75}
\biggl({d \over 1 \ {\rm Mpc}}\biggr) \ \ \ ,
\end{equation}
where ${\cal N}(>v_s,<d)$ is the total number of sub-halos with
scale velocities greater than $v_s$, contained within a distance $d$
from the center of a parent halo with virial mass $M_{{\rm vir,p}}$.

With the above observational and theoretical facts in mind, we now
consider two basic ``scenarios'' for the CHVCs.  We first consider
the CHVCs as ``circumgalactic'' objects associated with the Milky Way
halo. In this scenario the characteristic distance to the CHVCs is
$\sim 150$ kpc.
We then consider the possibility that the CHVCs are 
more distant ($d\gtrsim 750$ kpc) ``extragalactic'' 
Local Group objects as proposed by Blitz et al. (1999).
As we will show, minihalo models for the CHVCs can be constructed
for both scenarios, but our analysis strongly favors the circumgalactic model.

\subsection{CHVCs as Circumgalactic Minihalos}

Circumgalactic distances for the CHVCs
arise naturally if the CHVC halos are drawn from the same
family of median Burkert halos (or $-4\sigma$ NFW halos)
that successfully reproduce the dwarf galaxy observations.
If the CHVC halos are not significantly over- or under-concentrated
with respect to the dwarf halos, the gas scale heights in the
CHVCs and in the dwarfs must be comparable, since 
the scale heights are insensitive to the halo mass
(see the expressions for $r_{\rm gas}$ in Table 6, and
Figures 16 and 17, in Appendix A).
If the scale-heights are in fact
comparable, then the CHVCs must be significantly
closer to the Milky Way than Leo A or Sag DIG. This is because
the 1/$e$ angular sizes
of the dwarfs (160 and 70 arcsec) as observed by Young \& Lo
are significantly smaller than the angular scale heights (690 arcsec)
of the CHVCs as observed by Burton et al.  Given the known distances
to the dwarfs, the larger CHVC angular scale heights place
the CHVCs at distances of 100 to 200 kpc. 
At $d=150$ kpc, 690 arcsec corresponds to a linear scale of 0.5 kpc.

The $M_{\rm WM}$ vs.~$M_{\rm vir}$ computations displayed in
Figures 6 through 11 show that in clouds heated by the
metagalactic field, multi-phased cores exist for
peak HI column between $2\times 10^{19}$ to $1\times 10^{20}$ cm$^{-2}$.
The fact that multi-phased cores are predicted for the range
of peak WNM column densities observed in the CHVCs
strongly suggests that the thermal states of the CHVCs are controlled
by the external field. 
For a distance of 150 kpc the external field may already be dominated by
the metagalactic background
with a possible (though uncertain)
contribution from Galactic stellar Lyc photons (see Appendix B).
For those models with multi-phased cores
(i.e., those within the ``multi-phased strip'' in Figs.~6f through 11f)
the boundaries of the WNM clouds are predicted to occur at 
$r_{\rm WNM}\approx 0.6$ kpc.  It is evident 
that many such halo ``solutions'' for the CHVC behavior
can be found for a wide range of halo masses and bounding pressures.
For example, for $P_{\rm HIM}/k=10$ cm$^{-3}$ K, bound cloud models 
can be constructed for $v_s > 17$ km s$^{-1}$ 
(see Fig.~7) in Burkert halos.  Clearly, 
however, gravitationally
confined solutions in very massive halos
(e.g., setting $M_{\rm vir}=10^{10}M_\odot$, 
with $M_{\rm WM}=10^6M_\odot$) are not relevant, since $\sim 200$
such objects would exceed the total mass of the Milky Way.

In fact, the large number of CHVCs suggests that they must
be pressure-confined in low-mass halos. For example, if half
of the CHVCs lie within the characteristic
distance $d$, then it follows from equation~(\ref{e:klypin}) that
the CHVC halos must be low-mass objects. In particular,
for scale velocities $v_s\sim 12$ km s$^{-1}$,
${\cal N}(>v_s)=100$ within $d=150$ kpc, where we
have assumed $M_{\rm vir,p}=10^{12}M_\odot$ for the Milky Way (see below).
Such low scale velocities require large confining HIM pressures to keep
the gas both neutral and bound to the halos.

As a specific example of such a pressure-confined solution,
we consider a median Burkert halo with $v_s=12$ km s$^{-1}$.
For this choice $M_{\rm vir}=1\times 10^{8}M_\odot$.
We set $P_{\rm HIM}/k=50$ cm$^{-3}$ K, which is the minimum possible bounding 
pressure that allows $10^4$ K WM gas to remain bound
to such halos. We fix $M_{\rm WM}=1.1\times 10^6M_\odot$
in order to produce a peak HI column within the range of observed values.
For this gas mass, the baryon to DM mass
ratio for the mini-halos is $M_{\rm WM}/M_{\rm vir}=1.1\times 10^{-2}$,
and is significantly smaller than the cosmic baryon fraction
$\Omega_b/\Omega_m=0.09$. 
The resulting model structure is displayed in Figure~13,
and the various model
parameters are summarized in Table 3.

The WM cloud consists
of a WNM core with $r_{\rm WNM}=0.67$ kpc, surrounded by a WIM envelope
extending to $r_{\rm W/H}=1.3$ kpc. The gas at $r_{\rm W/H}$
is marginally bound with ${\cal W}/{\cal T}$=1.0 at the cloud edge (a condition
fixed by our choice for $P_{\rm HIM}$).
The central gas density $n_{\rm H,0}=2.1\times 10^{-2}$ cm$^{-3}$, and 
is sufficiently large for the formation of a neutral core. 
The computed HI mass is $2.8\times 10^5M_\odot$, or about 20\% of the
total WM mass.

The computed peak HI column is $5.1\times 10^{19}$ cm$^{-2}$.
The HI column falls to 1/$e$ of the central value at 
$r_{\rm {N_{\rm HI}}}=0.50$ kpc.
For an angular $1/e$ size of 690 arcsec the implied distance
is $d=150$ kpc.
The pressure contrast $P_0/P_{\rm HIM}=4.7$, so that $P_0/k=235$
cm$^{-3}$ K. The central pressure satisfies the condition
$P_{\rm min}(0) < P_0 < P_{\rm max}(0)$, where $P_{\rm min}(0)/k=95$
and $P_{\rm max}(0)/k = 730$ cm$^{-3}$ K. A multi-phased medium is possible
out to 0.33 kpc, at which point $P=P_{\rm min}$.  A multi-phased core
of this size is also consistent with the observations (again
assuming $d\approx 150$ kpc). For radii greater than 0.33 kpc,
the local gas pressure $P$ falls below the local values of
$P_{\rm min}$ and the neutral gas must be entirely in the WNM phase.

We note that for a distance of 150 kpc,
clouds in median $v_s=12$ km s$^{-1}$ NFW halos 
(as opposed to Burkert halos)
would be much smaller than observed. In such systems
the projected scale height is only 0.2 kpc.
In the circumgalactic model, $v_s=12$ km s$^{-1}$ NFW halos
would have to be significantly underconcentrated,
with $\sigma \approx -2$, to give clouds 
with projected scale heights of 0.5 kpc.

An important prediction of the model displayed 
in Figure 13 is the presence of a sharp
drop in the HI column at $r_{\rm WNM}$. 
As in the case of our dwarf galaxy models,
the ionization front is located near the radius where the projected
column falls to $10^{19}$ cm$^{-2}$. In our CHVC model
$N_{\rm HI}=5\times 10^{18}$ cm$^{-2}$ at $r_{\rm WNM}=0.7$ kpc, 
and decreases below $1\times 10^{18}$ cm$^{-2}$ at 0.8 kpc.
We note that in the CHVC model $r_{\rm WNM}/r_{\rm gas}=1.2$,
whereas in our dwarf galaxy model $r_{\rm WNM}/r_{\rm gas}=2.5$.
This difference is due to the significantly larger HI gas mass,
and associated peak HI column, in the dwarf compared to the CHVC.

At this point we encounter a difficulty in accounting for
the extended HI gas ``wings'' found by Burton et al. (2001),
in at least some of the CHVCs that they observed. 
To illustrate this difficulty, we plot in Figure 13b 
(as the dashed curve) the projected HI column density distribution 
assuming that the intrinsic
HI gas density profile varies as a pure exponential
out to several scale heights. We recall that Burton et al. 
suggested that such exponentials are representative
of the observed gas distributions. We normalize the
exponential distribution such that the peak HI column density and
the projected $1/e$ scale height are equal to those in our CHVC model.
In our model, the HI column density decreases to a
value of $10^{18}$ cm$^{-2}$ at 1.2 (projected) scale heights, whereas
for the pure exponential distribution 
the HI column falls to $10^{18}$ cm$^{-2}$
at 4 scale heights. Thus, if an exponential density law
represents the observations down to columns of 10$^{18}$ cm$^{-2}$,
then the photoionization feature expected
near a column of 10$^{19}$ cm$^{-2}$ would appear to be absent in the data
\footnote{
We remark that purely exponential gas density distributions are expected
in NFW halos in the small-$x$ limit (see Appendix A).
However, this requires massive halos. Furthermore,
eliminating the photoionization cutoff to a column
below $10^{18}$ would require reducing our estimate
for the metagalactic Lyc flux (for photons between 1 and $\sim 2$
ryd) by a factor of at least $\sim 10^3$.}. 
We caution however, that the exponential forms proposed
by Burton et al. are only approximate representations.
In fact, in some objects (e.g.~in the CHVCs 186-31-206 and
230+61+165; see their Fig.~11) features that resemble
our predicted photoionization cutoffs may be present.

As is indicated by Figure 13d the predicted H$\alpha$ surface
brightness ranges from 4.3 mR at the cloud center, to a
maximum limb-brightened value of 6.6 mR at a projected radius
of 0.7 kpc from the cloud center. The predicted H$\alpha$ surface
brightnesses of the CHVCs and dwarf galaxies are equal
because both cloud systems are optically thick.  
The predicted surface brightness is proportional to
our assumed metagalactic field intensity, and is consistent with
the expectation that in optically thick clouds
the emission measure along a cloud
diameter equals $2\pi J^*/\alpha_B$ (plus an additional
contribution due to the secondary ionizations). 
Direct measurements of the H$\alpha$ fluxes from the population
of CHVCs are largely unavailable. However, 
Weiner et al. (2001) and Tufte et al. (2002) 
report H$\alpha$ intensities of 20 to 140 mR in five CHVCs.
These values are considerably larger
than our predicted surface brightnesses.  We consider it unlikely
that the metagalactic Lyc field is more than $\sim 3$
times greater than our estimated strength,
and additional sources of ionization appear to be required.
Galactic radiation could contribute if these particular objects
are considerably closer to the Galaxy than 150 kpc.
Collisional ionization at the interfaces between the
clouds and a dense HIM is another possibility.  
Further observations are required to
determine whether the fluxes found by 
Weiner et al. and Tufte et al. are
typical of the entire population of CHVCs.

For $v_s=12$ km s$^{-1}$ halos, neutral bound clouds require
large bounding pressures, with $P_{\rm HIM}/k 
\gtrsim 50$ cm$^{-3}$ K
at distances of $\sim 150$ kpc from the Galaxy. We now argue that
such pressures
could plausibly be provided by a hot
($T_{\rm HIM} \sim 2\times 10^6$ K) Galactic corona.
To demonstrate this we construct a simple hydrostatic model for 
a corona within the Galactic potential. We model the 
Galactic potential as the combination
\begin{equation}
\label{e:galpot}
\varphi_{\rm Gal} = \varphi_{\rm disk} + \varphi_{\rm DM}
\end{equation}
of a baryonic disk and a dark-matter halo. 
The coronal HIM pressure at a distance $d$ is then given by 
\begin{equation}
\label{e:galp}
P_{\rm HIM} = P_{\rm mid} e^
{[\varphi_{\rm Gal}(0)-\varphi_{\rm Gal}(d)]/c_{\rm HIM}^2}
\end{equation}
where $P_{\rm mid}$ is the mid-plane thermal pressure,
and $c_{\rm HIM}=116T_6^{1/2}$ km s$^{-1}$ is the HIM
sound speed (where $T_6\equiv T/10^6$ K).
For the disk we assume a
cylindrical $(R,z)$  Miyamoto-Nagai potential
\begin{equation}
\label{e:miyamoto}
\varphi_{\rm disk} = -{C_1 v_{\rm circ}^2 \over
[R^2 + (a_1 + \sqrt{z^2 + b_1^2})^2]^{1/2}}
\end{equation}
where we set $C_1=8.887$ kpc, $v_{\rm circ}=225$ km s$^{-1}$,
$a_1=6.5$ kpc, and $b_1=0.26$ kpc (Wolfire et al. 1995b).  
In using this expression we fix $R$ to a constant value of 8.5 kpc.
At large distances from the disk, 
$z/R \gg 1$, and $z\approx d$ where $d$ is 
the distance from the Galactic center.  
For the DM halo we adopt a median Burkert halo
with $M_{\rm vir}=10^{12} M_\odot$. The associated
halo parameters are $x_{\rm vir}=13.9$, $r_s=18.6$ kpc,
$n_{ds}=0.26$ amu cm$^{-3}$, and $v_s=200$ 
km s$^{-2}$. The virial radius is $r_{\rm vir}=258$ kpc,
and the circular
velocity at the virial radius $v_{\rm vir}=129$ 
km s$^{-1}$.
Our model for the Galactic DM halo is very similar to the
model presented by Moore et al. (2001).
It satisfies the requirements (e.g.~Navarro \& Steinmetz 2000)
that (a) the disk mass, $M_{\rm disk}\lesssim 10^{11}$ M$_\odot$,
is consistent with the mass of baryons
that could have cooled out of the Galactic DM halo, i.e., the requirement that 
$M_{\rm disk} \lesssim (\Omega_b/\Omega_m)M_{\rm vir}$
(where $\Omega_b/\Omega_m = 0.09$), and (b)  
the dark matter mass within 10 kpc ($1.8\times 10^{10} M_\odot$)
is less than the upper limit of $\sim 3\times 10^{10} M_\odot$
inferred from Galaxy mass models (Dehnen \& Binney 1998).
Following Wolfire et al. (1995b), we fix the mid-plane pressure such that
the coronal HIM produces a total X-ray emission
measure of $5\times 10^{-3}$ cm$^{-6}$ pc
(Garmire et al. 1992; Kuntz \& Snowden 2000).   
For a coronal HIM temperature
of $2\times 10^6$ K (corresponding to $c_{g,\rm HIM}
=164$ km s$^{-1}$)
we find that $P_{\rm mid}/k = 2.5\times 10^3$ cm$^{-3}$ K,
consistent with the observed mid-plane thermal pressures. The
total mass of hot coronal gas equals $2.5\times 10^{10} M_\odot$,
and is a significant fraction of the baryonic mass of the Galaxy.  
In Figure 14 we display the resulting coronal
pressure as a function of the distance $d$.
It is evident that $P_{\rm HIM}$ remains large beyond 100 kpc, 
and ranges from 99 to 43 cm$^{-3}$ K between 100 and 260 kpc.
These pressures are sufficiently large to confine and neutralize
the gas contained within a population of low-mass minihalos located
in the circumgalactic environment.

We emphasize that the coronal HIM pressure maintains
the WM gas within the low-mass halos in neutral form, and therefore
visible as 21 cm sources. If, for example, our 
$v_s=12.5$ km s$^{-1}$ mini-halo containing
$1.1\times 10^6 M_\odot$ of gas is moved out
to larger distances where the enveloping HIM pressure is 
presumably significantly reduced, 
the cloud expands and becomes ionized.
Thus, if the CHVCs are infalling objects, a
picture emerges in which a population of low-mass halos could
exist beyond the immediate circumgalactic environment, but would
be ionized by the metagalactic field and therefore remain
invisible as HI sources. It is possible that such objects
could be detected via absorption line studies of ionized
gas tracers (e.g.~Sembach et al. 1999). As such objects
approach the higher pressure environment of the Galaxy
the clouds are compressed and neutralized, and become visible
HI sources.

A distance of $d=150$ kpc is likely well within the
virial radius of Galactic DM halo (for our 
Galaxy model, $r_{\rm vir}=260$ kpc). Our model CHVC mini-halo
must therefore be tidally truncated at a radius that is smaller than the 
(original) mini-halo virial radius. 
What is the tidal truncation radius at 150 kpc? 
Assuming the CHVC is on a circular orbit at a distance $d$
around the Milky Way, the truncation radius $r_t$ 
is set by the tidal criterion 
$3{\bar n}_{\rm MW}(d) = {\bar n}_{\rm CHVC}(r_t)$
where ${\bar n}_{\rm MW}(d)$ is the mean density of the
Galaxy within $d$, and ${\bar n}_{\rm CHVC}(r_t)$
is the mean DM density of the CHVC halo within $r_t$.
For our Galaxy model,
$3{\bar n}_{\rm MW}=2.4\times 10^{-3}$ cm$^{-3}$
at 150 kpc. For our $v_s=12$ km s$^{-1}$ minihalo,
the implied tidal radius $r_t=6.8$ kpc,
well within the mini-halo virial radius of 12.0 kpc.
However, $r_t$ is still significantly larger than the
WM cloud radius $r_{\rm W/H}=1.3$ kpc set by the confining
HIM pressure. The actual truncation radius will depend on
the orbital parameters, and could be significantly smaller
than our estimate if the periastron distances are sufficiently
small.  
If the CHVCs come too close to the Galaxy, they
will be tidally disrupted. As we now discuss, they may also be
destroyed by ram-pressure stripping.  

While low-mass halos require a large HIM pressure to keep the
gas bound and neutral, if the HIM becomes too dense
the CHVC gas will be ram-pressure stripped out of the mini-halos.
Stripping occurs when
the ram-pressure exceeds the gravitational force per unit area
on the gas (Gunn \& Gott 1972; Blitz \& Robishaw 2000;
Mori \& Burkert 2000; Quilis \& Moore 2001). 
For neutral clouds moving
with orbital velocity $v$ through a
hot ionized medium, stripping will occur when
\begin{equation}
\label{e:stripping}
\biggl({n_{\rm HIM} \over 10^{-4} \ {\rm cm^{-3}}}\biggr) \gtrsim 4.4
\biggl({\Sigma_{\rm DM} 
         \over 10^7 \ { M_\odot} \ {\rm kpc}^{-2}}\biggr)
\biggl({N_{\rm HI} \over 10^{20} \ {\rm cm}^{-2}}\biggr)
\biggl({v \over 100 \ {\rm km} \ {\rm s}^{-1}}\biggr)^{-2}
\end{equation}
where $n_{\rm HIM}$ is the HIM gas density, $\Sigma_{\rm DM}$ is the
dark-matter surface density, and $N_{\rm HI}$ is the HI column density.
For our model CHVC the DM surface density within the projected $1/e$
scale height is $\Sigma_{\rm DM}=1.0\times 10^{7} M_\odot$ kpc$^{-2}$.
Setting $N_{\rm HI}$ equal to the peak column density of 
$5.1\times 10^{19}$ cm$^{-2}$, and assuming a typical CHVC velocity
of 100 km s$^{-1}$, it follows that stripping will occur
for $n_{\rm HIM} \gtrsim 2\times 10^{-4}$ cm$^{-3}$. This is 
much larger than a HIM density of 
$2.5\times 10^{-5}$ cm$^{-3}$ for a pressure 
$P_{\rm HIM}/k=50$ cm$^{-3}$ K, and 
$T_{\rm HIM}=2\times 10^6$ K.  We conclude that our pressure-confined 
$v_s=12$ km s$^{-1}$ mini-halo model clouds
(and the cores of the observed clouds) would be stable against 
rapid stripping by our postulated coronal HIM at 150 kpc.
We note, however, that at HIM densities of $2.5\times 10^{-5}$ cm$^{-3}$,
the observed (but not predicted)
low-column density ($\lesssim 5\times 10^{18}$ cm$^{-2}$) wings, 
should be stripped off the clouds. Their presence thus poses
a challenge to our view of the CHVCs as pressure-confined objects.
The cloud wings could conceivably represent
stripped gas, but then
the question remains as to why this low column density gas
is not photoionized by the background field (and by the
additional sources of ionization indicated by the
Weiner et al. observations).

We mention one additional problem. If the bounding pressure
is reduced sufficiently, not only do the clouds become ionized,
they also become unbound. For example, it is clear from
Figure 6c that for $P_{\rm HIM}/k=1$ cm$^{-3}$ K, our CHVC model
lies in the unbound portion of the $M_{\rm WM}$ vs $M_{\rm vir}$
diagram, indicating that mass-loss from the outer layers
of the cloud would occur for such a low HIM pressure.
Thus, the question arises as to how the gas was retained
by the mini-halos prior to their approach into the circumgalactic
environment.
This problem can be alleviated if the graviational potential
well is deepened by increasing the
mini-halo mass. E.g., increasing $v_s$ to 17 km s$^{-1}$
is sufficient to keep our WM mass of $1.1\times 10^6 M_\odot$
bound (although still ionized) for $P_{\rm HIM}/k=1$ cm$^{-3}$ K
(see Fig.~6c).
However, the expected number of objects decreases rapidly
as $v_s$ is increased
as indicated by equation~(\ref{e:klypin}).

\subsection{CHVCs as Extragalactic Minihalos}

We now consider the possibility that the CHVCs are more
distant ``extragalactic'' objects, and represent
sub-structure on the scale of the entire Local Group (LG).
This is the basic picture proposed by Blitz et al.,
who placed the CHVCs at characteristic distances of $\gtrsim 750$ kpc.
As we will show, 
models for the CHVCs can be constructed for such distances.
However, there are two major difficulties with this picture.
First, the mini-halos must be
extremely underconcentrated.
Second, the reduced central gas pressures
suppress the formation of multi-phased cores.

For $d=750$ kpc, the observed CHVC HI gas scale height of 690 arcsec
corresponds to a linear scale of $2.5$ kpc. 
Such large values for the
gas scale height require extremely underconcentrated halos,
for both Burkert and NFW potentials (see Table 6 and Figs.~16 and 17).
Such objects must deviate very substantially from the 
correlation $x_{\rm vir}(M_{\rm vir})$ expected for 
median $\Lambda$CDM halos,
with $\sigma \lesssim -4$ for Burkert halos, and $\sigma \lesssim -6$
for NFW halos. 
This is a major problem with the LG hypothesis.  
At distances $\gtrsim 750$ kpc, the  
CHVC halos would have to be quite different from those found
in the CDM collapse simulations (see \S 2.2). 
From a more observational perspective,
the CHVCs would then represent objects that are quite distinct from the 
dwarf galaxies.  For example,
for $r_{\rm gas}=2.5$ kpc
the required DM scale density in Burkert halos (see Table 6)
must be extremely small, with $n_{ds}\approx 5\times 10^{-2}$ cm$^{-3}$.
This is much smaller than the
characteristic DM densities of $\sim 1$ cm$^{-3}$ that  
we find for Leo A and Sag DIG,
or that have been inferred from the dwarf galaxy
rotation curve studies. 

As a specific example of a CHVC model for $d\approx 750$ kpc, we 
again consider a pressure-confined system.
We select a $-4\sigma$ Burkert halo with
$v_s=12$ km s$^{-1}$. The corresponding virial mass is 
$3\times 10^8 M_\odot$.
We set the confining pressure equal to the 
minimum possible HIM pressure for bound cloud solutions. For this model
the minimum pressure is $P_{\rm HIM}/k = 20$ cm$^{-3}$ K. We choose
a WM gas mass equal to $M_{\rm WM}=5.8\times 10^7M_\odot$, to give
a total HI column within the observed range.
The resulting
model structure is displayed in Figure~15, and the model parameters
are summarized in Table 4. We note that for this model
the concentration parameter $x_{\rm vir}=8.1$. 
This is extremely small for our assumed halo mass. 
For median halos, such concentrations are expected for
masses $\sim 10^{13}M_\odot$ (see eq.~[\ref{e:correl}]),
corresponding to the mass scale of an entire galaxy group. 

In our extragalactic model, the peak HI column density 
equals $1.0\times 10^{20}$ cm$^{-2}$. 
The intrinsic WM gas scale height is $r_{\rm gas}=3.0$ kpc, and
the projected $1/e$ HI gas scale height $r_{\rm {N_{\rm HI}}}=2.6$ kpc.
For a WNM core size of 690 arcsec, the implied distance is 777 kpc.
The pressure contrast $P_0/P_{\rm HIM}=4.4$, so that $P_0/k=88$ cm$^{-3}$ K.
The central pressure satisfies the condition for multi-phased behavior
$P_{\rm min}(0) < P_0 < P_{\rm max}(0)$ where $P_{\rm min}(0)=67$ and
$P_{\rm max}(0)=523$ cm$^{-3}$ K.  We note that due to the larger 
distance, the
central gas pressure is significantly lower than in our
circumgalactic model.  Large peak HI column densities are therefore required
to provide the opacity necessary to reduce
$P_{\rm min}$ and $P_{\rm max}$ to values that allow multi-phased gas
in the core. For $-4\sigma$ Burkert halos we find that multi-phased gas
is possible for columns between $9\times 10^{19}$ to 
$3\times 10^{20}$ cm$^{-2}$.
In our specific extragalactic model a small multiphase core is just 
possible out
to 1.2 kpc, consistent with the observed presence of CNM cores within
the WNM gas scale height. While the required WNM column
of $\sim 10^{20}$ cm$^{-2}$ is within the range of observed values,
most of the CHVCs are not as shielded as this (Burton et al. 2001).
At large distances, the multi-phased cores in most objects would vanish.

The WNM component extends to $r_{\rm WNM}=3.4$ kpc, and is surrounded
by a WIM envelope that extends to 7.2 kpc.
The total HI mass is $1.4\times 10^7M_\odot$.
The central gas density 
$n_{\rm H,0}=7.6\times 10^{-3}$ cm$^{-3}$ is sufficiently
large for the formation of a neutral core (since 
for $r_{\rm gas}\approx 2.5$ kpc
$n_{\rm H,0crit}=2\times 10^{-3}$ cm$^{-3}$). As expected, the 
ionization front is associated with a projected HI
column close to $10^{19}$ cm$^{-2}$.  
At 3.0 kpc from the cloud center,
$N_{\rm HI}=8\times 10^{18}$ cm$^{-2}$
and decreases below 10$^{18}$ cm$^{-2}$ at 4.0 kpc.
Extended low column density wings at a level of 
$\sim 10^{18}$ cm$^{-2}$ are not present.

If the CHVCs are at large distances, and they are
pressure-confined objects, their large extents
set limits on the intergalactic medium (IGM) pressure within the
Local Group. The bounding pressure $P_{\rm HIM}$ cannot
exceed $\sim 40$ cm$^{-3}$ K, since otherwise the clouds
would be compressed to radii less than 2.6 kpc.
Conversely, for $v_s=12$ km s$^{-1}$ the bounding pressure cannot
be smaller than 20 cm$^{-3}$ K, since otherwise the clouds would
become unbound. The lower limit on the pressure can
be reduced if more massive halos are considered.
If the IGM pressure is close to the required $\sim 20$ cm$^{-3}$ K
this gas must also be very hot.
For a uniform density HIM contained within a sphere of radius $d$ the
total HIM mass is
\begin{equation}
M_{\rm HIM}= 5.0\times 10^{11} \biggl({d\over 1 \ {\rm Mpc}}\biggr)
\biggl({P_{\rm HIM}/k\over 10 \ {\rm cm^{-3}} \ {\rm K}}\biggr)
\biggl({T_{\rm HIM} \over 10^6 \ {\rm K}}\biggr)^{-1} \ \ \ {\rm M}_\odot
\end{equation}
where $T_{\rm HIM}$ is the HIM temperature. Assuming
the HIM mass does not exceed a fraction $\sim 0.25$ 
of the total baryonic mass $\sim 3\times 10^{11}M_\odot$
of the LG (essentially the Milky Way plus M31) then for
$d=750$ kpc, and $P_{\rm HIM}/k=20$ cm$^{-3}$ K, it follows
that $T_{\rm HIM} \gtrsim 10^7$ K.
It is difficult to see how such hot gas could remain bound to
the Local Group.

Finally, another important feature of models for $d\gtrsim 750$ kpc is a
much larger WM to DM mass ratio.
In our extragalactic model, $M_{\rm WM}/M_{\rm vir}=0.22$
(more than twice the cosmic baryon fraction) as opposed to
0.011 in our circumgalactic model. For $v_s=12$ km s$^{-1}$,
$M_{\rm vir}$ grows by a factor of 3.2 as
$\sigma$ is reduced from 0 to -4.
In contrast, $M_{\rm WM} \simeq \pi m_{\rm H} N_{\rm HI}r_{\rm gas}^2$,
so that $M_{\rm WM}$ grows by a factor $\gtrsim 50$.
The gas-to-DM mass fraction
can be reduced if $v_s$ is increased. However, it is clear that
as the CHVC halos are placed at larger distances they must 
become increasingly gas rich. This poses an additional problem
for the extragalactic hypothesis, as it is difficult for low-mass
halos to retain their gas.

\section{Summary}

In this paper we examine the hypothesis that the
compact high velocity HI clouds (CHVCs) trace
sub-structure within the dark-matter halo of the Galaxy,
or within the entire Local Group system.

For this purpose
we carry out detailed computations
of the coupled hydrostatic and ionization structures
of pressure supported clouds that are confined by
gravitationally dominant dark-matter (DM) ``mini-halos''.
We focus on low-metallicity systems
that are ionized and heated by the metagalactic background field.
We provide a fit for this field (from the near-IR to the
hard X-ray regime) that is based on a combination of observations
and theoretical estimates.
We explicitly consider the effects of
external bounding pressures on the mini-halo cloud structure.
We consider bounding pressures provided by a hot ionized Galactic corona
or an intergalactic medium in the Local Group.

We consider dark-matter halos with either cuspy (NFW) or constant density
(Burkert) cores. We adopt a $\Lambda$CDM cosmological model
to parameterize the expected correlations
between the various halo scale parameters, as well as the
dispersion in the correlations.
We focus on low-mass
halos with virial masses $\sim 10^8$ to $10^{10}M_\odot$,
(or halo scale velocities between 
$\sim 10$ to 50 km s$^{-1}$) containing 
warm ($10^4$ K) clouds of neutral plus photoionized
gas with masses between $\sim 10^4$ and 10$^8M_\odot$.

We determine how the cloud sizes and hydrogen
gas distributions, as well as the gas 
phase-states -- ionized (WIM), neutral (WNM), or
multi-phased (WNM/CNM) --  depend on the halo parameters, gas masses,
and bounding pressures. 
We consider both gravitationally confined 
systems, where the gas is effectively restricted to within 
the halo scale radii (or halo cores), as well as
pressure-confined clouds in which the gas can extend to large halo radii. 
We find that gravitationally confined systems are those for which
$v_s/c_g \gtrsim 1.5$, where $v_s$ is the halo scale velocity and $c_g$ is the
gas sound speed. We determine the conditions 
required for the transition from
ionized to neutral clouds, and for the formation of thermally unstable
multi-phased (WNM/CNM) cores within neutral clouds.   
We also compute the emission measures
and H$\alpha$ recombination line surface brightnesses
that are produced in the WIM components of the halo clouds.
For optically thick clouds ionized by the metagalactic field,
we find that the H$\alpha$ surface brightness ranges from 4 mR at the cloud
centers to 7 mR at the limb-brightened edges.

As a step in our computations we construct pressure vs.~density phase
diagrams for low metallicity ($Z=0.1 Z_\odot$)
gas that is heated by the metagalactic field for a wide
range ($10^{18}-10^{21}$ cm$^{-2}$) of assumed HI shielding columns,
and we compute the critical phase-transition pressures
$P_{\rm min}$ and $P_{\rm max}$ as functions of the
hydrogen ionization rate. We use the results of these
computations to determine the thermal phase states of the
neutral gas in our halo cloud models.

We present mini-halo models for the pressure supported HI structures
observed in the Local Group dwarf irregular galaxies Leo A and Sag DIG.
We include the stellar contribution to the gravitational potentials. 
We identify the HI cloud boundaries observed in Leo A and Sag DIG
with the ionization fronts, and we derive an upper limit of 
$P_{\rm HIM}/k \lesssim 100$ cm$^{-3}$ K for the ambient 
pressure of the intergalactic medium in the Local Group.
The distances to Leo A and Sag DIG are well established, and 
the observed HI gas scale heights of $0.5$ kpc
in these objects imply 
characteristic DM densities of $1.2$ amu cm$^{-3}$
(or $0.03M_\odot$ pc$^{-3}$) for the DM halos. These densities are
consistent with those previously found via rotation curve studies of
rotationally supported
dwarfs and low-surface brightness galaxies.
Leo A and Sag DIG obey the halo correlations
that are expected for typical (``median'') 
DM halos in a $\Lambda$CDM cosmology, provided the
halos contain constant density cores.  NFW halos would have to
be extremely underconcentrated (by $-4\sigma$ relative
to median halos) given the observed gas scale heights. 

For the family of median Burkert (or $-4\sigma$ 
NFW) halos that obey the scalings and correlations
as fixed by the dwarf galaxies, we find that the transition from
ionized to neutral clouds occurs for central
WNM HI column densities equal to
$3\times 10^{18}$ cm$^{-2}$, independent of the halo virial
mass or total gas mass. The gas is fully neutral within projected radii where 
HI column exceeds $10^{19}$ cm$^{-2}$. A photoionization cut-off
is predicted near $10^{19}$ cm$^{-2}$ (similar to what
is found for flattened, rotationally supported,
disk systems) independent of the
halo mass. We find that for clouds
heated by the metagalactic field, multi-phased cores
occur for peak WNM columns between $2\times 10^{19}$ and
$2\times 10^{20}$ cm$^{-2}$. 
The peak WNM columns of $\sim 10^{21}$
cm$^{-2}$ observed in the dwarf galaxies require
additional (possibly internal) heat sources. 

We construct explicit ``mini-halo'' models for the multi-phased
(and low-metallicity) compact high-velocity HI clouds.
We consider the CHVCs as either ``circumgalactic'' objects
associated with the Milky Way halo, with implied characteristic
distances of 150 kpc, or as more distant $\gtrsim 750$ kpc
``extragalactic'' Local Group objects. 
If the CHVC halos are drawn from the same family 
of halos that successfully reproduce the dwarf galaxy observations,
then the CHVCs must be circumgalactic objects. The observed $1/e$
gas scale heights then correspond to sizes of 0.5 kpc
(as in the dwarfs). The typical HI mass for an individual CHVC
is then $\sim 3\times 10^5M_\odot$, and the total HI mass
in the entire population of CHVCs is $\sim 6\times 10^7M_\odot$.
The predicted multi-phased behavior that occurs for
peak WNM columns between $2\times 10^{19}$ and
$2\times 10^{20}$ cm$^{-2}$ is consistent with the observed
multi-phased behavior and range of peak WNM columns in the CHVCs.
If the large population of CHVCs represent ``missing
low-mass DM satellites'' of the Galaxy, then
these HI clouds must be pressure-confined to keep the gas neutral
within the weak DM potentials.
The observed number of CHVCs imply
typical CHVC mini-halo scale velocities of $v_s=12$ km s$^{-1}$.
For such objects the 
confining pressure must exceed $\sim 50$ cm$^{-3}$ K
to keep the gas bound.
We construct a simple model for a Galactic corona, and show that
a hot $2\times 10^6$ K Galactic corona (in pressure equilibrium 
with the Galactic disk)
could provide the required pressure at 150 kpc.

If the CHVCs are ``extragalactic'' objects
with distances $\gtrsim 750$ kpc, then their associated halos
must be very ``underconcentrated'', with characteristic
DM densities $\lesssim 0.08$ cm$^{-3}$, much
smaller than expected for their mass, and significantly
smaller than observed in the dwarf galaxies.
Multi-Phased cores are possible, but require
shielding columns that are generally higher than observed.
For $d=750$ kpc, the typical CHVC HI mass is $\sim 1\times 10^7M_\odot$,
and the total HI mass in the system of CHVCs is $\sim 2\times 10^9M_\odot$.

Our analysis favors the circumgalactic hypothesis for the location
of the CHVCs.  In this picture, the CHVCs represent pressure-confined 
clouds that are associated with
tidally truncated dark-matter sub-halos that have
survived the hierarchical formation process of the
Galactic halo.  The CHVC mini-halos are visible as neutral 21 cm
sources due the compression provided by an ambient Galactic corona.

Our analysis also appears to favor Burkert halos with constant density
dark-matter cores, as opposed to NFW halos with diverging central
DM densities. While Burkert halos are consistent with the
observed scale heights at circumgalactic distances of 150 kpc, 
NFW halos would have to be significantly underconcentrated 
to yield the observed cloud sizes for such distances. 

We compute the maximum masses of WNM gas that
can be maintained by the metagalactic field within the DM halos.
Our analysis suggests that in the mini-halo scenario for the CHVCs,
these objects are ``failed galaxies'' and did not form stars,
because most of the neutral gas was always
maintained as WNM by the metagalactic field.  In contrast the
halos associated the dwarf galaxies in the Local Group
(and elsewhere) contained sufficiently large masses of gas
such that most of the neutral component was converted to CNM,
a precondition for star-formation. 

Our hydrostatic mini-halo cloud models are able to account for many
properties of the CHVCs, including their observed
peak HI columns, core sizes, and multi-phased behavior. However,
important theoretical and observational 
difficulties remain. Theoretically, a question remains as to
the origin of the gas in the CHVC mini-halos. The gas is unlikely
to be purely primordial, since such gas is easily
lost from low-mass mini-halos as the bounding pressure becomes low.
The presence in some objects of
extended low column density HI wings, and
H$\alpha$ emission line fluxes in several CHVCs that are
significantly larger than expected, represent significant observational
challenges to the mini-halo models we have presented in this paper. 
Additional high resolution HI mapping observations
and sensitive H$\alpha$ line measurements are required
to determine the possible contributions of gas stripping
and perhaps collisional ionization to the CHVC cloud
structure.  Such observations will help establish,
or possibly refute, the hypothesis that the CHVCs
trace dark-matter substructure.

\vspace{0.8cm}

We thank Leo Blitz, Stu Bowyer, James Bullock, Butler Burton, Marc Davis, 
Orly Gnat, David Hollenbach, Tsafrir Kollat, and David Spergel
for many helpful discussions during the course of this work. 
We thank James Bullock for providing us with his CDM halo code.
We also thank the referee for comments and suggestions that improved
this paper.
C.F.M. was supported in part by NSF grant AST-0098365.
M.G.W. was supported in part by a NASA LTSA grant NAG5-9271,
and by NSF grant AST 95-29167. 
 
\appendix
\section{NFW AND BURKERT HALOS AND THE SMALL-X LIMIT}
In this appendix we derive some useful analytic formulae
for the cloud properties and gas distributions when the
clouds are confined to the ``small-$x$'' ($x<1$) regions of the
dark-matter halos.

We first summarize the basic properties of the
NFW (Navarro et al. 1997) and Burkert (1995)
dark-matter halos. 
In Table 5 we list the expressions for the (dimensionless)
density profile, $f_\rho$, the enclosed mass, $f_M$, and
the gravitational potential, $f_\varphi$, (as defined in
\S 2.1) for NFW and Burkert halos.

In NFW halos the central
densities diverge, with $f_\rho \approx 1/x$ for $x\ll 1$,
while the enclosed mass and 
potential remain finite, with $f_M \approx (3/2)x^2$
and $f_\varphi \approx (3/2)x$. 
The scale density $\rho_{ds}$ (see \S 2.1) equals the local
DM density at $x=0.46$. At $x=1$ the DM density
$\rho_d=\rho_{ds}/4$, and the enclosed mass $M_d=0.58M_{ds}$.
The enclosed mass $M_d=M_{ds}$ at $x=1.6$.
For $x \gg 1$, $f_\rho\approx 1/x^3$,  $f_M$ diverges 
logarithmically, and $f_\varphi$ approaches an
asymptotic value of 3.  The circular velocity $v=v_sf_v$ reaches
a maximum value $v_{\rm max}=0.8v_s$ at $x=2.2$.

Burkert halos are defined such that they
contain constant density dark-matter cores. 
Thus, for $x \ll 1$, $f_\rho \approx 1$, $f_M\approx x^3$,
and $f_\varphi \approx x^2/2$.
The scale density $\rho_{ds}$
equals the central (core) density $\rho_d(0)$. At $x=1$ the DM density 
$\rho_d=\rho_{ds}/4$, and the enclosed mass $M_d=0.38M_{ds}$.
The enclosed mass $M_d=M_{ds}$ at $x=1.7$.
For $x\gg 1$, $f_\rho \approx 1/x^3$, $f_M$
diverges logarithmically, and $f_\varphi$ approaches an
asymptotic value of $3\pi/4$.
The circular velocity $v=v_sf_v$ reaches
a maximum value $v_{\rm max}=0.8v_s$ at $x=3.3$.

We now consider simplified isothermal clouds with fixed ionization state
(e.g., fully ionized or completely neutral)
so that the gas is characterized by a constant
velocity dispersion $c_g$.
Writing equation~(\ref{e:hbasic}) as
\begin{equation}
{d\rho \over \rho} = -{1\over c_g^2}d\varphi
\end{equation}
it follows that the hydrostatic gas density profile
for a dark-matter dominated cloud is
\begin{equation}
f_{\rm gas}(x)\equiv {n_{\rm H}\over n_{\rm H,0}} = 
{\rm exp}\left(-{v_s^2 \over c_g^2}f_\varphi\right)
\label{e:fgas1}
\end{equation}
where $f_\varphi$ is the dimensionless halo potential.
For a given halo profile (NFW or Burkert) 
the gas density distribution depends on the
single parameter $v_s/c_g$. In Table A1 we include the
resulting analytic solutions for the gas distributions
(Makino, Sasaki \& Suto 1998; Wu 2000, Wu \& Xue 2000) 
as given by equation~(\ref{e:fgas1}).

For $x\ll 1$ the expressions for the gas distributions
simplify considerably, and simple analytic formulae for the
gas scale heights, column densities, masses, and cloud
emission measures may be derived. We list these
various formulae in Table 6.

In the small-$x$ limit, the gas density distribution
$f_{\rm gas}$ varies exponentially
in NFW halos. In Burkert halos the gas distribution is
a Gaussian. For both
halo types, we define the gas scale height, $x_{\rm gas}=r_{\rm gas}/r_s$,
as the radius at which the gas density decreases to a factor
$1/e$ times the central value. Expressions for
$r_{\rm gas}$ as functions of the DM halo parameters 
are listed in Table 6. 
In NFW halos $r_{\rm gas}$ depends on two (correlated) halo 
scale paramters, e.g.~$n_{ds}$ and $r_s$, or $r_s$ and $v_s$.
In Burkert halos, $r_{\rm gas}$ may be expressed as a function
of $n_{ds}$ only, that is, on the value of the uniform dark-matter
density in the $x \ll 1$ halo core.

In Figures 16 and 17 
we plot the small-$x$ values of $r_{\rm gas}$ as a function
of the halo scale velocity, $v_s$, for median, $+3\sigma$ over-concentrated,
and $-3\sigma$ under-concentrated NFW and Burkert halos. 
In Figures 16 and 17 we display
$r_{\rm gas}$ as a function of $M_{\rm vir}$.
In these plots we assume a (WNM) sound speed
of 8.2 km s$^{-1}$. Analytic expressions for $r_{\rm gas}$ as
functions of $\sigma$ and $v_s$, 
or as functions of $\sigma$ and $M_{\rm vir}$
may be obtained using the scaling 
relations~(\ref{e:ndsfit}), (\ref{e:rsfit}),
and (\ref{e:mvfit}) for $n_{ds}$, $r_s$ and $M_{\rm vir}$ (see \S 2.2).
We list these expressions for $r_{\rm gas}$ in Table 6.
In both NFW and Burkert halos the scale heights for median ($\sigma=0$) halos
are small.  For median NFW halos $r_{\rm gas}$
ranges from 0.1 to 0.04 kpc for $v_s$ between 10 to 50 km s$^{-1}$, or
virial masses between $10^8$ and $10^{10}M_\odot$.
For median Burkert halos $r_{\rm gas}$ ranges from 0.2 to 0.6 kpc
for $v_s$ from 10 to 50 km s$^{-1}$, or $M_{\rm vir}$ 
between $10^8$ and $10^{10}M_\odot$. We note that for Burkert
halos $r_{\rm gas}$ is a very weak function of $M_{\rm vir}$.
The scale heights may be increased by moving to under-concentrated halos.
We illustrate this in Figures 16c, 16d, 17c, and 17d where we
display level curves for $r_{\rm gas}$ in the range of
0.5 to 2.5 kpc, in $\sigma$ vs.~$v_s$, and 
$\sigma$ vs.~$M_{\rm vir}$ diagrams.  It is apparent
that large scale heights require extremely underconcentrated halos.

In the small-$x$ limit the total gas mass 
remains finite as $x_{\rm W/H}$ becomes arbitrarily large
(i.e. as the bounding pressure is reduced to zero).
Similarly, the gas column densities at all
impact parameters $b\equiv r_p/r_{\rm gas}$ 
(where $r_p$ is the projected radius) remain finite. Small-$x$
expressions for the cloud masses and columns
are listed in Table 6. In NFW halos the column densities
are proportional to $bK_1(b)$ where $K_1$ is the first-order
modified Bessel function. The product $bK_1(0)=1$
for $b=0$, and declines by a factor of $1/e$ at
$b=1.65$.  Thus, for NFW halos the projected $1/e$ scale height is
1.65 times larger than the intrinsic gas scale height.
In Burkert halos the column density distribution is a
Gaussian, and the projected $1/e$ 
scale height is equal to the intrinsic scale height.

Analytic conditions for the 
transition from ionized to neutral clouds may be derived
based on the behavior of the   
cloud emission measure
with varying optical depth.  The emission measure
depends on the fraction of the incident flux, $\pi J^*$, of
Lyman continuum photons that are absorbed by the cloud.
In optically thick clouds, the entire Lyc flux is absorbed
in a thin shell, and $EM (x=0)
=2\pi J^*/\alpha_B$. In the opposite limit, $EM$
must vanish in optically thin clouds as the cloud mass becomes 
sufficiently small.
Our radiative transfer computations show that $EM$
always reaches a maximum value of
\begin{equation}
\label{e:consb}
EM_{\rm max} (x=0)
 \approx 4\pi J^* / \alpha_B \ \ \ ,
\end{equation}
in {\it critical clouds}
where the fractional ionization $x_e=0.5$ at the cloud center.
Indeed, it is
straightforward to show analytically that for spherical clouds
(and with the neglect of secondary ionizations)
the emission measure must satisfy the upper and lower bounds,
$8\pi J^*/\alpha_B  \ge EM \ge 2\pi J^*f/\alpha_B$,
where $f$ is the fraction of the incident radiation that is absorbed.
Our radiative transfer computations are consistent with these limits.
We find that $EM$ vanishes when the cloud is optically
thin and $f$ is small. The emission measure then rises to a maximum value
of close to $4\pi J^*/\alpha_B$ in critical clouds, and then declines to
$EM=2\pi J^*/\alpha_B$ in optically thick ($f=1$) clouds.

The emission measures for fully ionized clouds are listed in Table 6.  
Crucially, these emission measures cannot exceed $EM_{\rm max}$. It follows
that to remain fully ionized the 
gas scale height must satisfy an upper bound, with
\begin{equation}
\label{e:ncorenfw}
r_{\rm gas} \le {4\pi J^* f\over \alpha_B n^2_{\rm H,0}} =
1.24 J_3^*f\biggl(\frac{n_{{\rm H},0}}{10^{-3}\ {\rm cm}^{-3}}\biggr)^{-2}
\end{equation}
where $f=1$ for NFW halos, $f=\sqrt{2/\pi}$ for Burkert halos,
and $J_3^*\equiv J^*/10^3$ photons cm$^{-2}$ s$^{-1}$ sr$^{-1}$
(for our assumed field $J_3^* = 1.02$).
Neutral cores must form when $r_{\rm gas}$ exceeds this limit.
We identify critical clouds as those for which equality holds
in equation~(\ref{e:ncorenfw}). 
Given our formulae for
$r_{\rm gas}$ equation~(\ref{e:ncorenfw})
implies a critical central gas density, $n_{\rm H,0crit}$
above which neutral cores must form. The resulting expressions
for $n_{\rm H,0crit}$, as functions of $v_s$ or $M_{\rm vir}$,
are listed in Table 6. 
We note that in evaluating these critical densities,
a WIM sound speed should be adopted since critical clouds are WIM dominated.
It is evident that for NFW halos the critical density is a weakly
increasing function of the halo scale velocity or mass, 
whereas in Burkert halos
the critical density is a very weakly decreasing function of
the scale velocity or mass.

For what values of $v_s/c_g$ is the small-$x$ approximation valid?
It follows from the analytic expressions for $f_{\rm gas}$ listed
in Table 5 that for $v_s/c_g \gtrsim 2$
the gas densities become vanishingly small for $x \gtrsim 1$,
while for $v_s/c_g \lesssim 2$ the gas densities remain
relatively large out to large halo radii.
Given the analytic expressions for $f_{\rm gas}$ listed in 
Table 5 we have numerically integrated
the total gas mass, $M_{\rm WM}/4\pi r_s^3n_{\rm H,0}$, 
peak gas column, $N_{\rm WMpeak}/2n_{H,0}r_s$, and emission measure
$EM/2n^2_{\rm H,0}r_s$ (assuming fully ionized clouds)
as functions of $v_s/c_g$. We then compared 
the results of the numerical integrations
to the values as given by the small-$x$ formulae.
We find that for $v_s/c_g > 1.5$ the small-$x$ expressions for
the gas column (and therefore also the
gas scale height $r_{\rm gas}$), and the emission measure, are
good to better than a factor of 2, for both NFW and Burkert profiles.
For $v_s/c_g > 2$ the accuracy
is better than a few percent. The total gas mass
diverges rapidly at small $v_s/c_g$, and the small-$x$
expressions for the total gas mass become accurate
only for $v_s/c_g > 3$. For such large values of $v_s/c_g$ the
gas mass is effectively independent of the bounding pressure
as $P_{\rm HIM}$ becomes arbitrarily small.

\section{METAGALACTIC RADIATION FIELD}

In this paper we construct models assuming the ($z=0$)
metagalactic field is the dominant source of radiation
impinging on the mini-halo clouds.

The external field, $J_\nu$
(ergs cm$^{-2}$ s$^{-1}$ Hz$^{-1}$ sr$^{-1}$, hereafter cgs units),
is a critical quantity controlling the overall cloud structure.
For a given total gas mass, the Lyc flux determines
the fractions contained in the WIM vs.~the shielded WNM components. 
The radiation field also determines the thermal phase 
state of the HI gas (see \S 3.2), since
hard ultraviolet (EUV) and X-ray photons can penetrate
into the shielded cores and heat the HI gas via photoionization. 
Non-ionizing radiation is also important. Far-ultraviolet
(FUV) grain photoelectric emission is a heat source, while
FUV photons produce the C$^+$ ions which dominate the
cooling of the CNM.  Molecular hydrogen (H$_2$)
is photodissociated by the FUV field, and the abundance of H$^-$,
a key intermediary in the gas phase
production of H$_2$, is limited by optical 
and infrared photodetachment.
 
In view of the above processes the metagalactic field 
must be specified from X-ray energies $\sim 1$ keV, 
above which clouds with the largest HI columns
we consider ($\sim 10^{22}$ cm$^{-2}$) become optically thin,
down to to the near-infrared H$^-$ 
photodetachment threshold at 1.64 $\mu$m.
We adopt a representation for the metagalactic 
field that is based on a combination of 
observational constraints and theoretical estimates.
We then estimate the distance from the Galactic disk at which the
metagalactic field dominates. 
We first consider the ionizing radiation, and then
the FUV through IR components.

The metagalactic Lyc flux is generally
believed to be produced by quasars with an
additional possible contribution from star-forming galaxies.
The background field is directly
observable at X-ray wavelengths (Gendreau et al. 1995;
Chen, Fabian \& Gendreau 1997;
Miyaji et al. 1998; Hasinger 2000),
but is unobservable near the Lyman limit
due to efficient absorption by Galactic HI.

The unobservable flux at $z=0$ can be constrained
via H$\alpha$ recombination
line measurements from (starless) intergalactic HI clouds 
(Vogel et al. 1995; Weiner et al. 2001), 
or via measurements of the HI ``truncation  column densities''
at the edges of galaxy disks
(Sunyaev 1969; Silk 1971; Maloney 1993). 
Vogel et al. (1995) determined a (2$\sigma$) upper limit of 
20 mR for the H$\alpha$ surface brightness in the intergalactic 
Giovanelli \& Haynes cloud.  This measurement sets an upper
limit on $4\pi J^*$ (eq.[\ref{e:lint}]). 
For opaque spherical clouds, $4\pi J^* = \alpha_B\langle EM\rangle$ where
$\langle EM\rangle$ is the emission measure averaged over the projected
surface area.  We conclude that 
$4\pi J^* < 4.5\times 10^4$ photons cm$^{-2}$ s$^{-1}$
in the local IGM.

Theoretical calculations of the metagalactic ionizing
radiation field at low redshifts have been presented by
Haardt \& Madau (1996), Okoshi \& Ikeuchi (1996), and Shull et al.~(1999)
with similar results.
At $z=0$ Haardt \& Madau estimate that $J_{\nu 0}=2.0\times 10^{-23}$ cgs
at the Lyman limit, while
Shull et al.~find $J_{\nu 0}=1.3\times 10^{-23}$ cgs.
The Lyc intensity drops sharply (as an approximate power-law)
to a value of $\sim 3\times 10^{-25}$ cgs
at the HeII edge (4 ryd), and flattens at higher energies where the IGM
becomes less opaque. 

In estimating the background field below 4 ryd to the Lyman limit we rely on
on the Haardt \& Madau calculations. At energies greater than 4 ryd
including EUV and X-ray wavelengths we rely on observations.
Chen et al. (1997) (see also Gendreau et al. 1995)
find that between 1 and 7 keV the X-ray background (as observed
by ROSAT and ASCA) 
is well fit by a simple
power-law, $AE_{\rm kev}^{-\Gamma}$, where $E_{\rm kev}$ is the photon
energy in kev, $A=10.5\pm 0.3$ photons s$^{-1}$ cm$^{-2}$ kev$^{-1}$ sr$^{-1}$,
and the photon index $\Gamma=1.46\pm 0.06$.  A very similar result was
found by Miyaji et al. (1998).
At soft X-ray energies ($\lesssim 1$keV) it becomes difficult to
disentangle the background from thermal emission
produced in the (distant) Galactic halo and (nearby) Local Bubble of hot gas. 
However, in an analysis of soft X-ray ``shadows''
(i.e. anticorrelations on the sky between soft X-ray flux
and absorbing HI column density) Snowden et al. (2000) argue that  
the Chen et al.~power-law plausibly extends down to 1/4 keV, where
additional flux (of distant origin) at 1/4 keV is due to a
thermal 10$^{6.4}$ K Galactic component.

We note that the Chen et al.~power-law gives 
$J_\nu=2.7\times 10^{-25}$ cgs
at the HeII edge (0.054 keV). This is only slightly smaller than
the theoretical Haardt \& Madau value of
$3.2\times 10^{-25}$ cgs at 4 ryd. We therefore assume that
the Chen et al. power-law extends down to 4 ryd. For lower energies 
we adopt a second power-law that
matches Chen et al.~at 4 ryd, and Haardt \& Madau at the Lyman limit.  
This gives,
\begin{equation}
\label{e:haardt}
J_\nu = J_{{\nu_0}}\Biggl({\nu \over \nu_0}\Biggr)^{-3.13}
\end{equation}
for $1 < \nu/\nu_0 < 4$, and
\begin{equation}
\label{e:chen}
J_\nu = 2.512\times 10^{-2} J_{{\nu_0}}
\Biggl({\nu \over \nu_0}\Biggr)^{-0.46}
\end{equation}
 for $\nu/\nu_0 > 4$, where $J_{{\nu_0}}=2.0\times 10^{-23}$ cgs
and $\nu_0$ is the Lyman limit frequency.
Most of the ionizing
energy is emitted near the Lyman limit, although it is evident 
that the total energy in the
Chen et al.~field (with $\nu J_\nu \sim \nu^{0.54}$)
increases with photon energy (see Fig.~5). The Chen et al.~field is
truncated at a turn-over energy near 100 keV (see Hasinger 2000).

Assuming the Vogel et al.~(1995) H$\alpha$ limit, and
assuming that the ionizing field varies 
with frequency as given by equations~(\ref{e:haardt}) and (\ref{e:chen}),
we derive an upper limit
of $J_{{\nu_0}} < 7.0\times 10^{-23}$ cgs for the metagalactic
field intensity at the Lyman limit.
The upper limit (shown
in Fig.~8) is 3.5 times larger than the value computed
by Haardt and Madau. 

We now consider the FUV and optical extragalactic background.
This background is generally believed to be produced by the
integrated light of faint and distant galaxies. Martin and Bowyer (1989)
and Martin, Hurwitz, \& Bowyer (1991) carried out sounding
rocket and space-shuttle observations, and detected
a 1600 \AA \ background flux
$F_{1600}^*
=40 \pm 10$ photons cm$^{-2}$ s$^{-1}$ \AA$^{-1}$ sr$^{-1}$,
corresponding to 
$J_\nu=4.2\pm 1.1\times 10^{-22}$ cgs at 1600 \AA.
Armand, Milliard \& Deharveng (1994) found that 
at 2000 \AA \ $J_\nu$
lies in the range of $5.3\times 10^{-22}$ to $1.7\times 10^{-21}$ cgs,
consistent with the Martin \& Bowyer observations.

Bernstein (1999) (see also Bernstein, Freedman \& Madore 2002)
has recently reported the first (HST) detections of the
extragalactic background light at optical wavelengths. She finds that
$J_\nu$ equals 
$1.2\pm 0.57\times 10^{-20}$, 
$2.8\pm 0.8\times 10^{-20}$, 
and $4.9\pm 1.3\times 10^{-20}$ cgs, 
at 3000, 5500, and 8000 \AA \ respectively.
The measured fluxes are consistent with
the upper limit of $1.2\times 10^{-19}$ cgs
at 4000 \AA \ previously set by Matilla (1989).

We fit the FUV and optical measurements with a two-piece power-law
expression
\begin{equation}
\label{e:bowyer}
J_\nu = J_{{\nu_0}}\Biggl({\nu \over \nu_0}\Biggr)^{-5.41}
\end{equation}
for $1 > \nu/\nu_0 > 0.3$ (i.e. from 912 to 3000 \AA),
and 
\begin{equation}
\label{e:bernstein}
J_\nu = 1.051\times 10^2 J_{{\nu_0}}
\Biggl({\nu \over \nu_0}\Biggr)^{-1.5}
\end{equation}
for $\nu/\nu_0 < 0.3$, where again $J_{\nu_0}=2\times 10^{-23}$ cgs.
This fit is consistent with the FUV and optical observations
within the errors, (see Fig.~5).
At $1.25$ $\mu$m equation~(\ref{e:bernstein}) 
gives $1.1\times 10^{-19}$ cgs, consistent with the (COBE)
upper limit of $2.8\times 10^{-19}$ cgs found by Dwek \& Arendt (1998)
at 1.25 $\mu$m.
We assume that equation~(\ref{e:bernstein}) 
remains valid down to the $H^-$ threshold at 1.65 $\mu$m.

For our fit to the metagalactic field
the mean intensity in the 6-13.6 eV FUV range 
is $6.7\times 10^{-6}$ ergs cm$^{-2}$ s$^{-1}$. 
This is $0.42\%$ of
the mean FUV intensity in the solar neighborhood 
as estimated by Habing (1968). 

At what distance from the Galaxy does the metagalactic 
background dominate the radiation field? 
The Galaxy produces Lyc photons (mainly in OB associations)
at a rate of $Q_{\rm Gal}\sim 2.6\times 10^{53}$
photons s$^{-1}$. A small but uncertain fraction, 
$f_{\rm esc} \sim 5\%$,
of this radiation may escape the optically thick disk
(Dove \& Shull 1994; Bland-Hawthorn \& Maloney 1999;
Dove, Shull \& Ferrara 2000). For $f_{\rm esc}=5\%$,
the Galactic Lyc flux  at the surface
of an opaque spherical cloud (on the side facing the Galaxy),
$Q_{\rm Gal}f_{\rm esc}
/4\pi d^2$,
equals the metagalactic Lyc flux of
$\pi J^*=3.2\times 10^3$ photons cm$^{-2}$ s$^{-1}$
at a distance $d=180$ kpc.
The Galaxy may also be
a source of soft X-rays emitted by cooling hot gas
in supernova remnants (Slavin, McKee \& Hollenbach 2000).
These authors estimate that for photon energies greater than 100 eV
a soft X-ray luminosity of
$2.2\times 10^{39}$ erg s$^{-1}$ escapes the Galaxy. This implies that the
Galactic and metagalactic X-ray fluxes (between 0.1 and 1 keV)
are equal at $d=16$ kpc. It is most straightforward to estimate
the non-ionizing 
(IR through UV) luminosity of the Galaxy via direct
observations of an analogous galaxy. For example, Kennicutt (2001)
has noted that the SBbc galaxy NGC 3992 is very similar
to the Galaxy in size, shape, luminosity, and star-formation properties.  
The distance to NGC 3992 is 15 Mpc, and its
IR (1.65$\mu$m), optical (5500 \AA), and FUV (1600 \AA)
luminosities are equal to
$2.2\times 10^{29}$, $1.5\times 10^{29}$ 
and $2.0\times 10^{27}$ erg s$^{-1}$ Hz$^{-1}$
respectively (data from the NED database\footnote
{nedwww.ipac.caltech.edu}).
Assuming these values for the Milky Way, the 
non-ionizing Galactic and metagalactic
fluxes are equal at $\sim 100$ kpc.  
Given the above estimates 
we conclude that the metagalactic field begins to
dominate the ionization and heating of the clouds at a distance of
100 to 200 kpc from the Galaxy.

\vfill\eject
\centerline{\bf References}
\vskip 0.1 true in 
{\hoffset 20pt
\parindent = -20pt

Aaronson, M.\ 1983, ApJ, 266, L11



Armand, C., Milliard, B., \& Deharveng, J.M. 1994, AA, 284, 12


Bahcall, N., Ostriker, J.P., Perlmutter, S., \& Steinhardt, P.J. 
1999, Science, 284, 1481


Bajaja, E., Morras, R,. \& Poppel, W.G.L. 1987, 
    Pub. Astr. Inst. Czech. Ac. Sci., 69, 237


Bakes, E.L.O., \& Tielens, A.G.G.M. 1994, ApJ, 427, 822

Ballantyne, D.R., Ferland, G.J., \& Martin, P.G. 2000, ApJ, 536, 773

Balucinska-Church, M., \& McCammon, D. 1992, ApJ, 400, 699

Barkana, R., \& Loeb, A. 1999, ApJ, 523, 54

Bernstein, R.A. 1999, in The Hy Redshift Universe, ASP Conf.\ vol.\ 193,
eds.\ A. J. Bunker and W. J. M. van Breugel (ASP: San Francisco), 48

Bernstein, R.A., Freedman, W.L., \& Madore, B.F. 2002, ApJ in press

Blais-Ouellette, S., Amram, P., \& Carignan, C. 2001, AJ, 121, 1952

Bland-Hawthorn, J., Veilleux, S., Cecil, G.N., Putman, M.E.,
Gibson, B.K., \& Maloney, P.R. 1998, MNRAS, 299, 611

Bland-Hawthorn, J., \& Maloney, P.R. 1999, ApJ, 510, L33

Blitz, L., Spergel, D., Teuben, P., Hartmann, D., \& Burton, W.B. 
   1999, ApJ, 514, 818

Blitz, L., \& Robishaw, T. 2000, ApJ, 541, 675

Blumenthal, G.R., Faber, S.M., Primack, J.R., \& Rees, M.J. 
    1984, Nature, 311, 517


Bowen, D.V., \& Blades, J.C. 1993, ApJ, 403, L55

Braun, R., \& Burton, W.B., 2000, 354, 853

Bregman, J. 1996, in The Interplay between Massive Star Formation, 
Galaxy Evolution, and the ISM, D. Kunth et al. eds.,
11th IAP Astrophysics meeting, Institut d'Astrophysique, Paris,
Gif-Sur-Yvette, Editions Frontieres, 211.

Br\"uns, C., Kerp, J., Kalberla, P.M.W., \& Mebold, U. 2000, AA, 357, 120

Br\"uns, C., Kerp, J., \& Pagels, A. 2001, AA 370, L26

Bruzual, G., \& Charlot, S. 1993, ApJ, 466, 254

Bullock, J.S., Kollat, T.S., Sigad, Y., Somerville, R.S., 
    Kravtsov, A.V., Klypin, A.A., Primack, J.R., \& Dekel, A.
    2001, MNRAS, 321, 559

Bullock, J.S., Kravtsov, A.V., \& Weinberg, D.H. 2000, ApJ, 539


Burkert, A. 1995, ApJ, 447, L25

Burton, W.B., Braun, R., \& Chengalur, J.N. 2001, AA 369, 316

Cardelli, J.A., Meyer, D.M., Jura, M., Savage, B.D. 1996, ApJ, 467, 334


Chen, L.-W., Fabian, A.C., \& Gendreau, K.C. 1997, MNRAS, 285, 449

Cook, K.H. 1988, BAAS, 20, 742

Corbelli, E. \& Salpeter, E.E. 1993, ApJ, 419, 104



Dalal, N., \& Kochanek, C. 2002, ApJ in press

Danly, L., Albert, C.E., \& Kuntz, K.D. 1993, ApJ, 416, L29


de Blok, W.J.G., \& McGaugh, S.S. 1997, MNRAS, 290, 533

de Blok, W.J.G., McGaugh, S.S., \& Rubin, V. 2001, AJ, 122, 2396

Dehnen, W., \& Binney, J.J, 1998, MNRAS, 298, 387

de Jong, T. 1977, AA, 55, 137


de Heij, V., Braun, R., \& Burton, W.B. 2002a, AA in press, astro-ph/0206306

de Heij, V., Braun, R., \& Burton, W.B. 2002b, AA in press, astro-ph/0201249

de Heij, V., Braun, R., \& Burton, W.B. 2002c, AA in press, astro-ph/0206333

Dekel, A., \& Silk, J. 1986, ApJ 303, 39

Dove, J.B., \& Shull, J.M. 1994, 430, 222

Dove, J.B., Shull, J.M., \& Ferrara, A. 2000, ApJ 531, 846

Draine, B.T. 1978, ApJS, 36, 595

Dwek, E., \& Arendt, R. G. 1998, ApJ, 508, L9


Eichler, D. 1976, ApJ, 208, 694

Einasto, J., Saar, E., Kaasik, A., \& Chernin, A.D. 1974, Nature, 252, 111

Efstathiou, G. 1992, MNRAS, 256, 43P

Eke, V. R., Navarro, J.F., \& Steinmetz, M. 2001, ApJ, 554, 114

Ferland, G.J., Korista, K.T., Verner, D.A., Ferguson, J.W.,
Kingdon, J.B., \& Verner E.M. 1998, PASP, 110, 761

Ferrara, A., \& Field, G. 1994, ApJ, 423, 665

Ferriere, K.M., Zweibel, E.G., \& Shull, J.M. 1988, ApJ, 332, 984


Field, G.B., 1975, in ``Atomic and Molecular Physics and the 
Interstellar Medium'', Les Houches lectures, 
eds. R. Balian, P. Encrenaz, J. Lequeux,
(North Holland Publishing), 467

Field, G.B., Goldsmith, D.W., \& Habing, H.J. 1969, ApJL, 155, L149

Firmani, C., D'Onghia, E., Avila-Reese, V., Chincarini, G., \& Hernandez, X.
2000, MNRAS, 321, 713

Gallagher III, J.S.\ 1994, PASP, 106, 1225

Gardiner, L.T., \& Noguchi, M. 1996, MNRAS, 278, 191

Gendreau, K. et al. 1995, PASJ, 47, L5


Gibson, B.K., Giroux, M.L., Stocke, J.T., \& Shull, J.M. 2000,
in Gas \& Galaxy Evolution, ASP conference series, Hibbard et al. eds.

Gibson, B.K., Fenner, Y., Maddison, S.T., Kawata, D. 2002
ASP Conference proceedings, eds. J.S. Mulchaey and J. Stocke, 254, p.~225

Giovanelli, R. 1981, AJ, 86, 1468

Gunn, J., \& Gott, R. 1972, ApJ, 176, 1

Haardt, F. \& Madau, P. 1996, ApJ 461, 20

Hartmann, D., \& Burton,  W.B. 1997, 
 Atlas of Galactic Neutral Hydrogen, (Cambridge: Cambridge Univ. Press)

Hasinger, G. 2000, in ISO Surveys of a Dusty Universe,
eds. D. Lemke, M. Stickel, and K. Wilke (Springer)
astro-ph/0001360

Hoessel, J.G., Saha, A., Krist, J., \& Danielson, G.E. 1994, AJ, 108, 654


Hulsbosch, A.N.M., \& Wakker, B.P. 1988, AAS, 75, 191


Jenkins, E.B., Jura, M., \& Loewenstein, M. 1983, 270. 88

Garmire, G.P., Nousek, J.A., Apparao, K.M.V., Burrows, D.N., 
Fink, R.L., \& Kraft, R.P. 1992, ApJ, 399, 694

Kaufman, M.J., Wolfire, M.G., Hollenbach, D.J., \& Luhman, M.L.
1999, ApJ, 527, 795

Kennicutt, R.C. 2001, in Galactic Structure, Stars, \& the
Interstellar Medium C.E. Woodward, M.D. Bicay, and J.M. Shull eds.,
ASP Conference Series, Vol. 231, 2

Kepner, J.V., Babul, A., \& Spergel, D. 1997, ApJ, 487, 61

Kitayama, T., \& Ikeuchi, S. 2000, ApJ, 529, 615

Klypin, A.A., Kravtsov, A.V., Valenzuela, O. \& Prada, F. 1999, ApJ, 522, 82

Kulkarni, S.R., \& Heiles, C. 1987, in Interstellar Processes, 
ed. D. Hollenbach, \& H.A. Thronson, Jr. (Dordrecht, Reidel),  87

Kulkarni, S.R., \& Heiles, C. 1988, in Galactic and Extragalactic Radio
Astronomy, ed. G.L. Verschuur, \& K.I. Kellermann (New York: Springer) 95


Kuntz, K.D., \& Snowden, S.L. 2000, ApJ, 543, 195




Lockman, F.J., Murphy, E.M., Petty-Powell, S., \& Urick, V.J.
2002, ApJS, 140, 331

Lu, L., Savage, B.D., Sembach, K.R., Wakker, B.P., Sargent, W.L.W., \&
Oosterloo, T.A. 1998, AJ, 115, 162



Madsen, G.J., Reynolds, R.J., Haffner, L.M., Tufte, S.L., \& Maloney, P.R.
2001, ApJ, 560, L135

Makino, N. Sasaki, S., \& Suto, Y. 1998, ApJ, 497, 555

Maloney, P.R.\ 1993, ApJ, 414, 41


Martin, C., \& Bowyer, S. 1989, ApJ, 338, 667

Martin, C., Hurwitz, M., \& Bowyer, S. 1991, ApJ, 379, 549


Mateo, M. 1998, ARAA, 36, 435


Mattila, K. 1990, in Galactic and Extragalactic Background Radiation, 
    IAU Symp 
    No.\ 139, eds.\ S. Bowyer and Ch. Leinert (Kluwer: Dordrecht), 257\

Mao, S., \& Schneider, P. 1998, MNRAS, 295, 587

McKee, C.F., \& Ostriker, J.P. 1977, ApJ, 218, 148


Metcalf, R.B., \& Madau, P. 2001, ApJ, 563, L9

Meyer, D.M., Jura, M., \& Cardelli, J.A. 1998, ApJ 493, 222

Miralda-Escude, J., \& Rees, M.J.\ 1993, MNRAS, 260, 617

Miyaji, T., Ishisaki, Y., Ogasaka, Y., Freyberg, M.J., Hasinger, G., 
\& Tanaka, Y. 1998, AA, 334, L13

Moore, B. 1994, Nature, 370, 629

Moore, B., \& Davis, M. 1994, MNRAS, 270, 209

Moore, B., Ghigna, S,. Governato, G., Lake, G., Quinn, T, Stadel, J.,
    \& Tozzi, P. 1999, ApJ, 524, 19

Moore, B., \& Putman, M.E. 2001, astro-ph/0110417

Mori, M., \& Burkert, A. 2000, 538, 559

Morrison, R., \& McCammon, D. 1983, ApJ 270, 119

Muller, C.A., Oort, J.H., \& Raimond, E. 
1963, C.R. Acad. Sci. Paris, 257, 1661

Murakami, I., \& Ikeuchi, S.\ 1990, PASJ, 42, L11

Murphy, E.M., et al. 2000, ApJ, 538, 35


Navarro, J.F., Frenk, C.S., \& White, S.D.M. 1996, ApJ, 462, 563

Navarro, J.F., Frenk, C.S., \& White, S.D.M. 1997, ApJ, 490, 493

Navarro, J.F., \& Steinmetz 2000, ApJ, 528, 607

Okoshi, K., \& Ikeuchi, S. 1996, PASJ, 48, 441

Olive, K.A., \& Steigman, G. 1995, ApJS, 97, 49

Oort, J. 1966, Bull. Astr. Inst. Netherlands, 18, 421

Oort, J. 1970, AA, 7, 381


Peebles, P.J.E. 1982, ApJ, 263, L1

Press, W.H., \& Schechter, P.\ 1974, ApJ, 187, 425

Putman, M.E., et al. 2002, AJ, 123, 873


Quilis, V., \& Moore, B. 2001, ApJ 555, L95


Rees, M.J.\ 1986, MNRAS, 218, 25P

Roberge, W.G., Dalgarno, A., \& Flannery, B.P. 1981, ApJ, 245, 817



Rosenberg, J.L., \& Schneider, S.E. 2002, ApJ 567 247


Sembach, K.R. 2002, in Extragalactic Gas at Low Redshift,
ASP Conference proceedings, eds. J.S. Mulchaey and J. Stocke, 254, p.~283

Sembach, K.R., Savage, B.D., Lu, L., \& Murphy, E.M. 1999, ApJ, 515, 108

Sembach, K.R., et al. 2000, ApJ, 538, L31

Shapiro, P.R., \& Field, G.B. 1976, ApJ 205, 762

Shull, J.M., Roberts, D., Giroux, M.L., Penton, S.V., \& Fardal, M.A.
1999, AJ, 118, 1450

Shull, J.M., \& van Steenberg, M.E. 1985, ApJ 298, 268

Silk, J., \& Sunyaev, R.A. 1976, Nature, 260, 508

Skillman, E.D., Kennicutt, R.C., \& Hodge, P.W. 1989, ApJ, 347, 875

Skillman, E.D., Terlevich, R., \& Melnick, J. 1989, MNRAS, 240, 563 

Slavin, J.D., McKee, C.F., \& Hollenbach, D.J. 2000, ApJ, 541, 218

Snowden, S.L., Freyberg, M.J., Kuntz, K.D., \& Sanders, W.T. 2000, ApJS, 128, 171


Spitzer, L. 1987, in Dynamical Evolution of Globular Clusters
(Princeton University Press, Princeton) 13.

Sternberg, A. 1998, ApJ, 506, 721

Sunyaev, R.A., 1969, ApL, 3, 33


Tegmark, M., Silk, J., \& Evrard, A. 1993, AJ, 417, 54 

Thoul, A.A., \& Weinberg, D.\ 1996, ApJ, 465, 608

Tielens, A.G.G.M., Hony, S., van Kerckhoven, C. \& Peeters, E. 1999,
in The Universe as Seen by ISO, eds. P. Cox and M.F. Kessler, ESA-SP 427,
(Noordwijk: ESA), 579

Tolstoy, E. et al. 1998, AJ, 116, 1244

Tufte, S.L., Reynolds, R.J., \& Haffner, L.M. 1998, ApJ, 504, 773

Tufte, S.L., Wilson, J.D., Madsen, G.J., Haffner, L.M., \& Reynolds, R.J.
2002, ApJL, in press (astro-ph/0206198)


van den Bergh, S. 1999, AAR, 9, 273

van den Bosch, F.C., Robertson, B.E., Dalcanton, J.J., \& de Blok, W.J.G.
2000, ApJ, 119, 1579

van den Bosch, F.C., \& Swaters, R.A. 2001 MNRAS, 325, 101

van Woerden, H., Schwarz, U.J., Peletier, R., Wakker, B.P. \&
   Kalbera, P.M.W. 1999, Nature 400, 138




Verschuur, G.L. 1969, ApJ, 156, 771

Vogel, S.N., Weymann, R. Rauch, M., \& Hamilton, T. 1995, ApJ, 441, 162

Wakker, B.P., \& Boulanger, F. 1986, AA, 170, 84

Wakker, B.P., \& Schwarz, U.J. 1991, AA, 250, 484

Wakker, B.P., \& van Woerden, H. 1991, AA, 250, 509


Wakker, B.P., \& van Woerden, H. 1997, ARA\&A, 35, 217
in Stromlo Workshop on High-Velocity Clouds, 
eds. Gibson, B.K. and Putman, M.E.,
ASP conference series, 166, 311

Wakker, B.P., van Woerden, H., \& Gibson, B.K. 1999,
in Stromlo Workshop on High Velocity Clouds, eds.
B.K. Gibson, and M.E. Putman, ASP Conference series, 166, p. 311 

Wakker, B.P., et al. 1999, Nature, 402, 388

Weiner, B.J., \& Williams, T.B. 1996, AJ, 111, 1156

Weiner, B.J., Vogel, S.N., \& Williams, T.B. 2001,
in Gas \& Galaxy Evolution, eds. J.E. Hibbard,
M.P. Rupen and J.H. van Gorkom, ASP Conference series, 
San Francisco, Astronomical Society of the Pacific, 515

Weymann, R.J., Vogel, S.N., Veilleux, S., \& Epps, H.W. 2001, ApJ, 561, 559

White, S.D.M., \& Rees, M.J., 1978, MNRAS, 183, 341

Wishart, A.W. 1979, MNRAS, 187, 59

Wolfire, M.G., Hollenbach, D., McKee, C.F., Tielens, A.G.G.M., 
\& Bakes, E.L.O. 1995a, ApJ, 443, 152

Wolfire, M.G., McKee, C.F., Hollenbach, D., \& Tielens, A.G.G.M. 
1995b, ApJ, 453, 673

Wolfire, M.G., McKee, C.F., Hollenbach, D., \& Tielens, A.G.G.M. 
2002, ApJ submitted

Wu, X.-P., \& Xue, Y.-J. 2000, ApJ, 542, 578

Wu, X.-P. 2000, MNRAS, 316, 299

Young, L.M., \& Lo, K.Y. 1996, ApJ, 462, 203

Young, L.M., \& Lo, K.Y. 1997, ApJ, 490, 710

Zeldovich, Y.B., \& Pilkelner, S.B. 1969, Soviet Physics JETP, 29, 170

Zwaan, M., \& Briggs, F.H. 2000, ApJ, 530, L61
}

\clearpage

\listoffigures

\clearpage
\begin{figure}[htpb]
\plotone{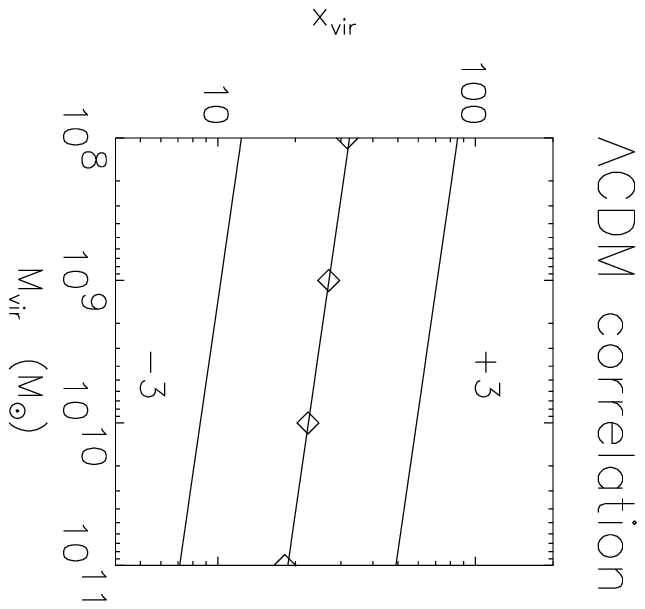}
\vspace{25cm}
\caption{The $\Lambda$CDM correlation relation, $x_{\rm vir}(M_{\rm vir})$,
between the halo concentration and virial mass
for low-mass halos. The square symbols indicate the values
computed using the Bullock et al.
(2001) model. The solid lines are our fits (as given by eq.~[\ref{e:fit}])
 for median, $+3\sigma$ over-concentrated, and
$-3\sigma$ under-concentrated, halos.}
\end{figure}

\clearpage
\begin{figure}[htpb]
\plotone{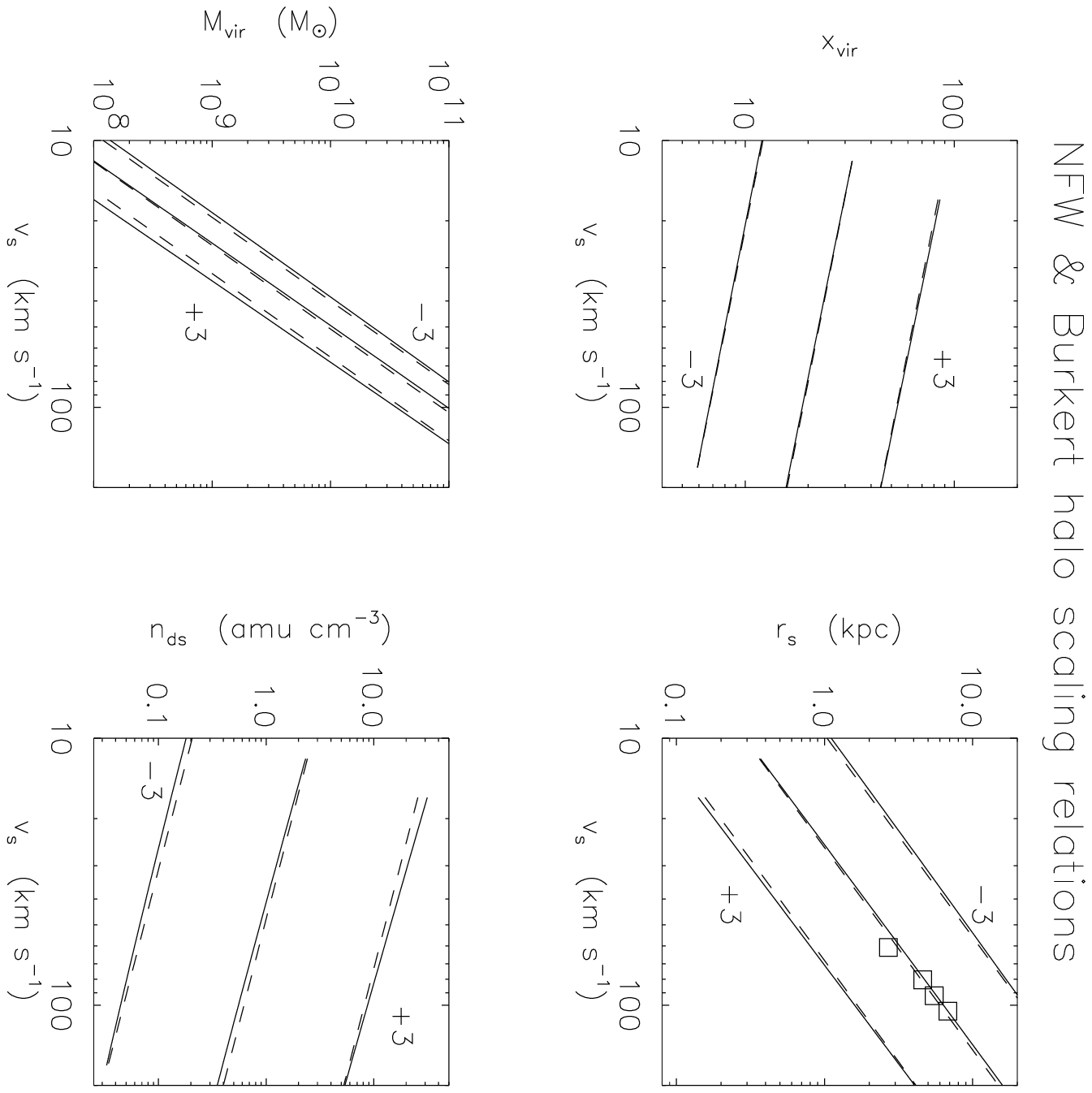}
\vspace{25cm}
\caption{$\Lambda$CDM scalings for NFW and Burkert halos. The halo
parameters, for median $+3\sigma$ and $-3\sigma$ halos, 
are displayed as function of the scale velocity $v_s$:
(a) concentration parameter $x_{\rm vir}$, 
(b) virial mass $M_{\rm vir}$, (c) scale radius $r_s$, and (d)
characteristic density $n_{ds}$. The dashed lines show the
analytic scalings given by equations~(\ref{e:xvfit}), (\ref{e:mvfit}),
(\ref{e:ndsfit}) and (\ref{e:rsfit}).
The square symbols in panel (c) show the 
observed correlation bewteen $r_s$ and $v_s$ for the dwarf galaxies
DDO 154, DDO 179, NGC 3109, and DDO 105 (Burkert 1995).}
\end{figure}

\clearpage
\begin{figure}[htpb]
\plotone{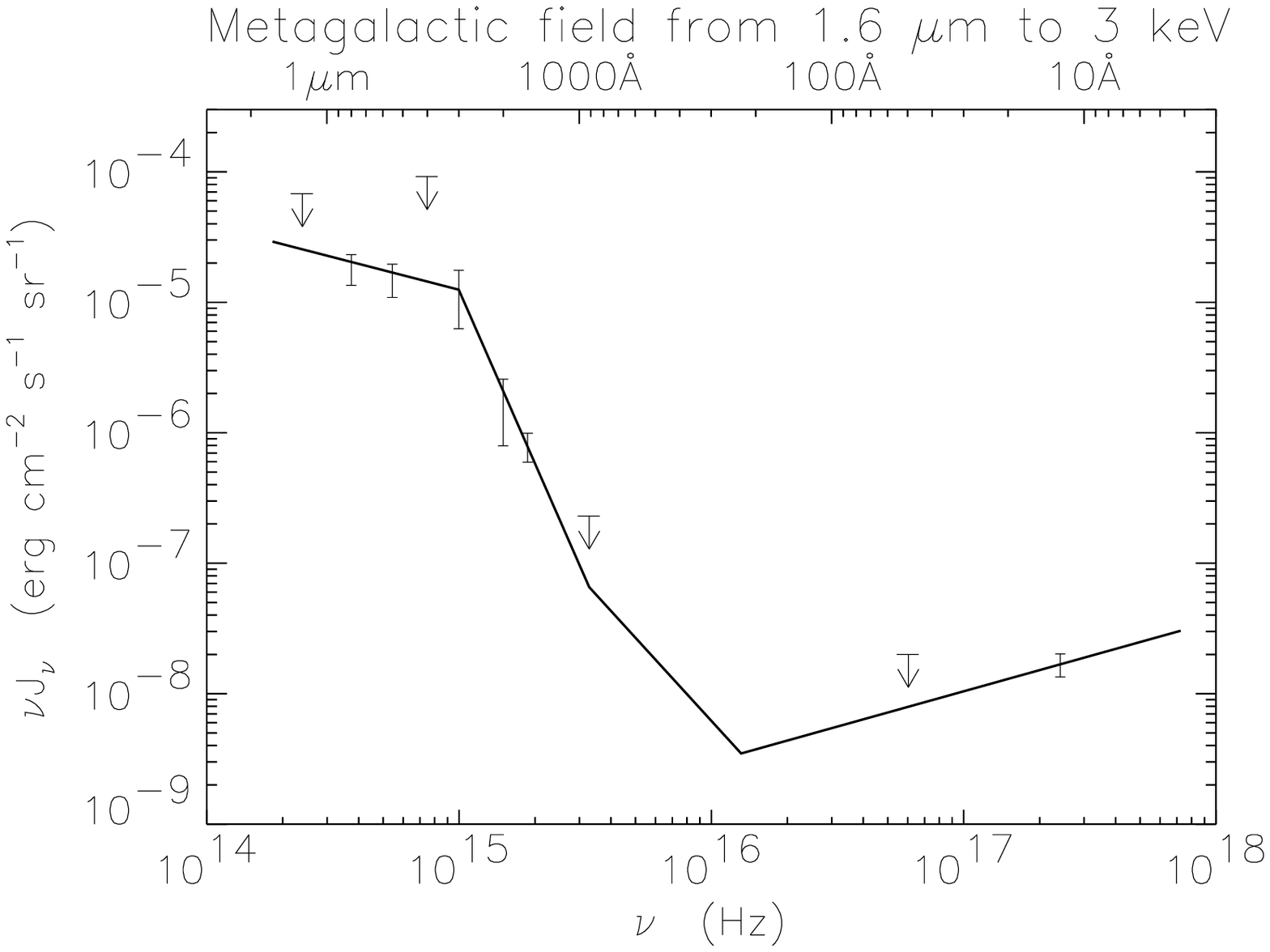}
\vspace{25cm}
\caption{Metagalactic radiation field from 1.6$\mu$m to 3 keV.
See Appendix B for a discussion of this fit.}
\end{figure}

\clearpage
\begin{figure}[htpb]
\plotone{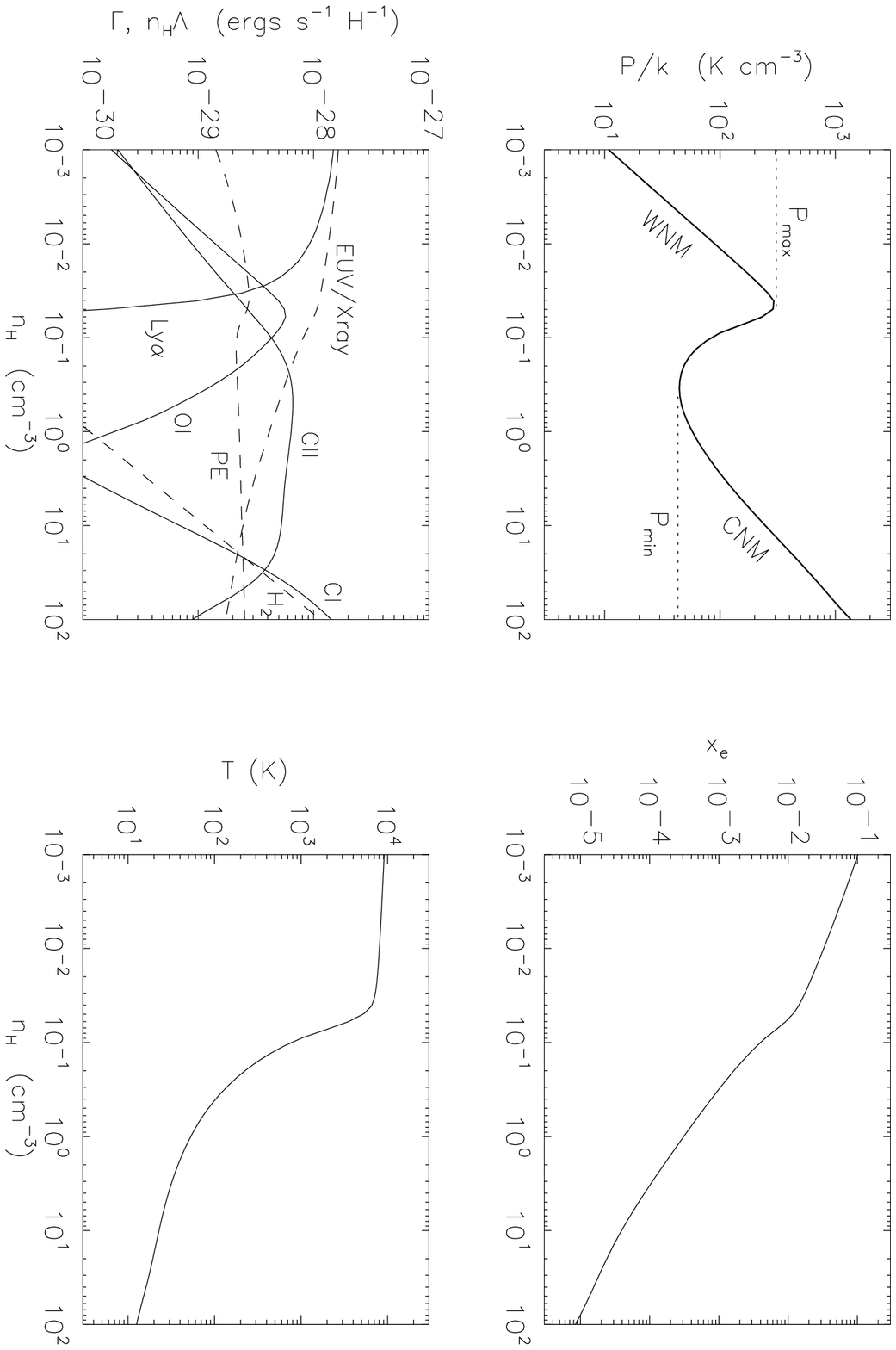}
\vspace{25cm}
\caption{Thermal properties of low metallicity ($Z=0.1Z_\odot$) HI gas
heated by a radiative flux equal to $2\pi J_\nu$, where
$J_\nu$ is the metagalactic radiation intensity (see Appendix B). 
The assumed HI shielding column is 
$N_{\rm HI}=5\times 10^{19}$ cm$^{-2}$, and the thermal properties
are displayed as functions of the hydrogen gas density $n_{\rm H}$:
(a) pressure vs. density phase diagram indicating the WNM and CNM
branches, and the critical pressures, $P_{\rm min}$ and $P_{\rm max}$, within
which thermal instability and
multi-phased behavior occurs, (b) gas heating rates (dashed) and
cooling rates (solid), (c) fractional ionization, and 
(d) HI gas temperature.}
\end{figure}
\clearpage
\begin{figure}[htpb]
\plotone{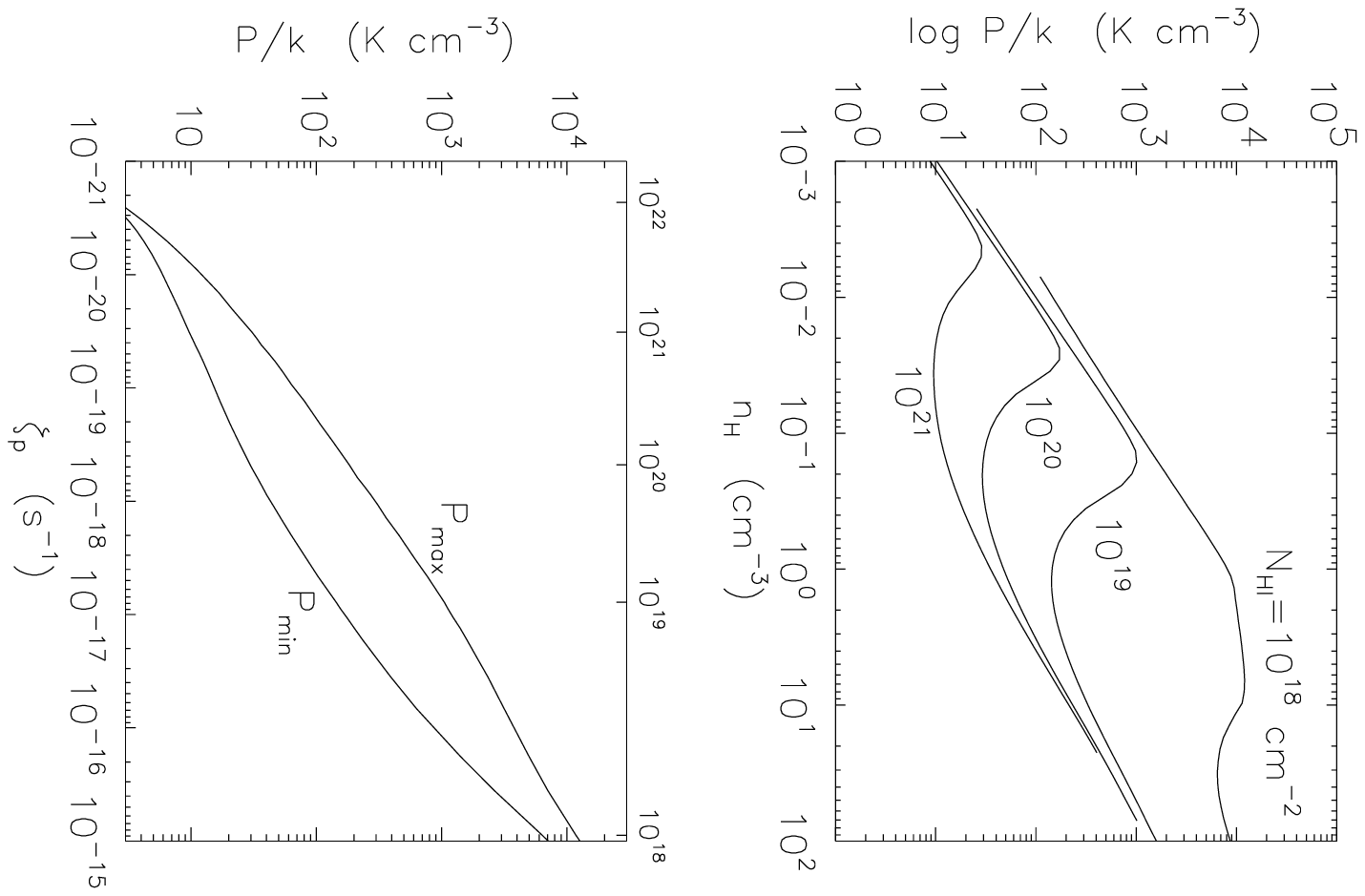}
\vspace{25cm}
\caption{The HI thermal phase properties as functions of
the shielding column $N_{\rm HI}$:
(a) phase diagrams for $N_{\rm HI}$ between 10$^{18}$ and
10$^{21}$ cm$^{-2}$, (b) the critical pressures
$P_{\rm min}$ and $P_{\rm max}$ as functions
of the primary ionization rate $\zeta_p$ (s$^{-1}$). 
The upper axis in panel (b) indicates the associated
HI shielding columns (cm$^{-2}$) for a plane-parallel geometry.}
\end{figure}

\clearpage
\begin{figure}[htpb]
\plotone{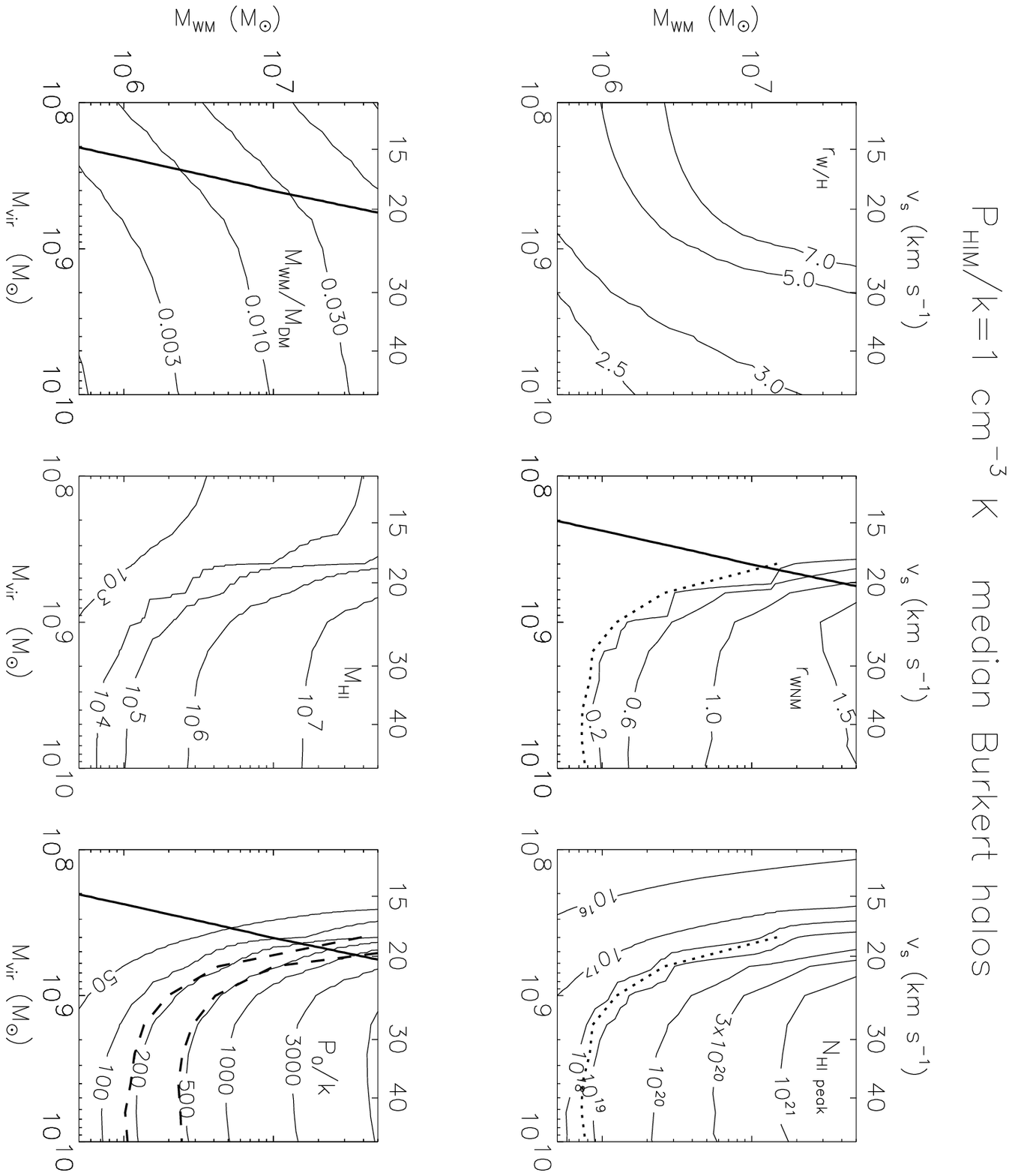}
\vspace{25cm}
\caption{$M_{\rm WM}$ vs. $M_{\rm vir}$ diagrams for $10^4$ K WM clouds in
median ($\sigma=0$) Burkert halos,
and a bounding pressure $P_{\rm HIM}=1$ cm$^{-3}$ K: (a) total cloud radius
$r_{\rm W/H}$, (b) the ratio,  $M_{\rm WM}/M_{\rm DM}$,
of the WM cloud mass and enclosed DM mass, (c) WNM radius
$r_{\rm WNM}$, (d) total HI gas mass,
(e) peak HI column density, (f) central pressure.
The dotted curves in panels (c) and (e) indicate the locus of critical clouds.
The thick solid curves in panels (b), (c) and (f) indicate
marginally bound clouds. The dashed curves in panel (f) 
indicate the ``multi-phased strip'' within which multi-phased behavior is
possible in the cloud cores.}
\end{figure}

\clearpage
\begin{figure}[htpb]
\plotone{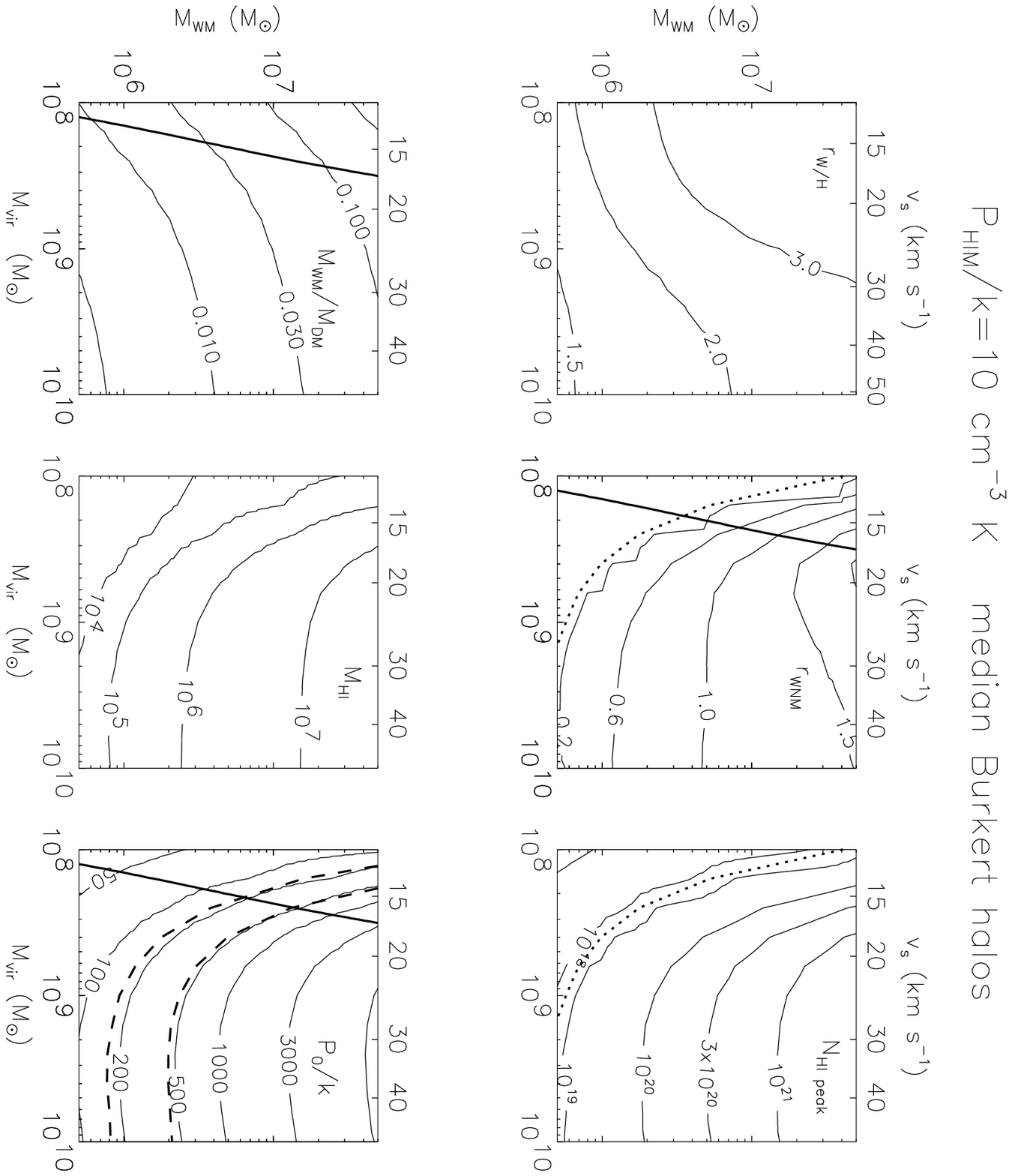}
\vspace{25cm}
\caption{$M_{\rm WM}$ vs. $M_{\rm vir}$ diagrams for median
($\sigma=0$) Burkert halos,
and $P_{\rm HIM}=10$ cm$^{-3}$ K.}
\end{figure}

\clearpage
\begin{figure}[htpb]
\plotone{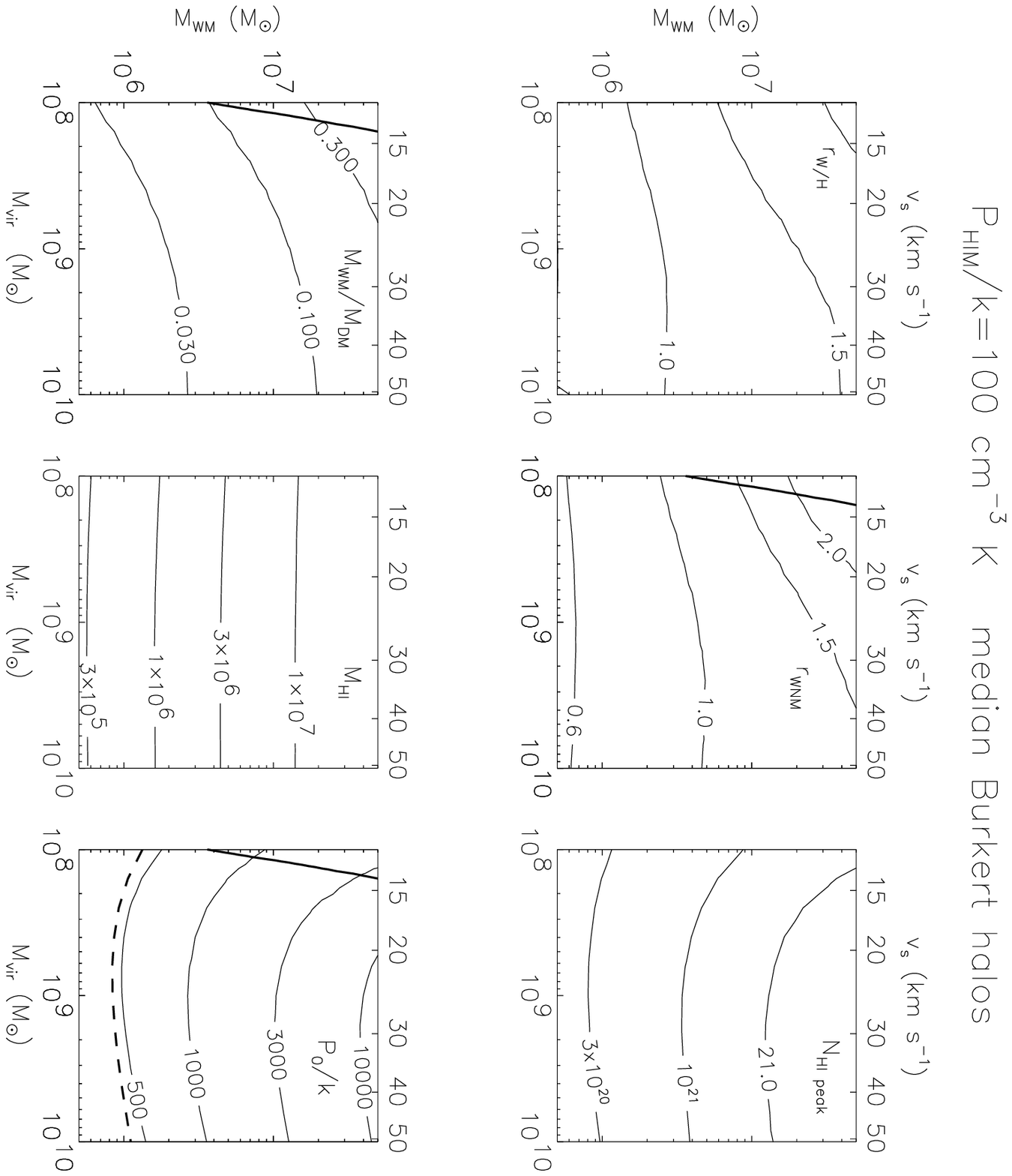}
\vspace{25cm}
\caption{$M_{\rm WM}$ vs. $M_{\rm vir}$ diagrams for median 
($\sigma=0$) Burkert halos,
and $P_{\rm HIM}=100$ cm$^{-3}$ K.}
\end{figure}

\clearpage
\begin{figure}[htpb]
\plotone{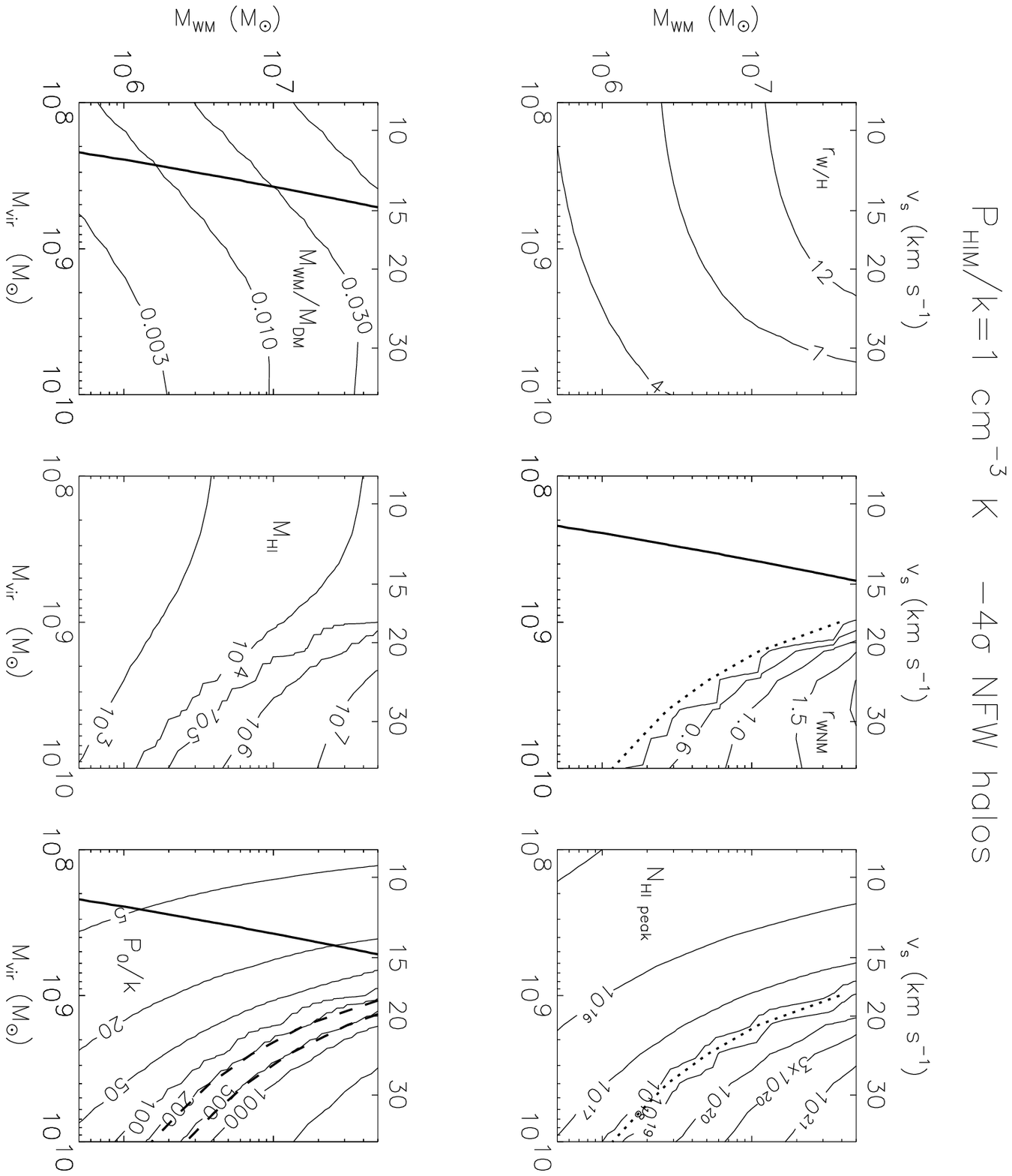}
\vspace{25cm}
\caption{$M_{\rm WM}$ vs. $M_{\rm vir}$ diagrams for $-4\sigma$ NFW halos,
and $P_{\rm HIM}=1$ cm$^{-3}$ K.}
\end{figure}

\clearpage
\begin{figure}[htpb]
\plotone{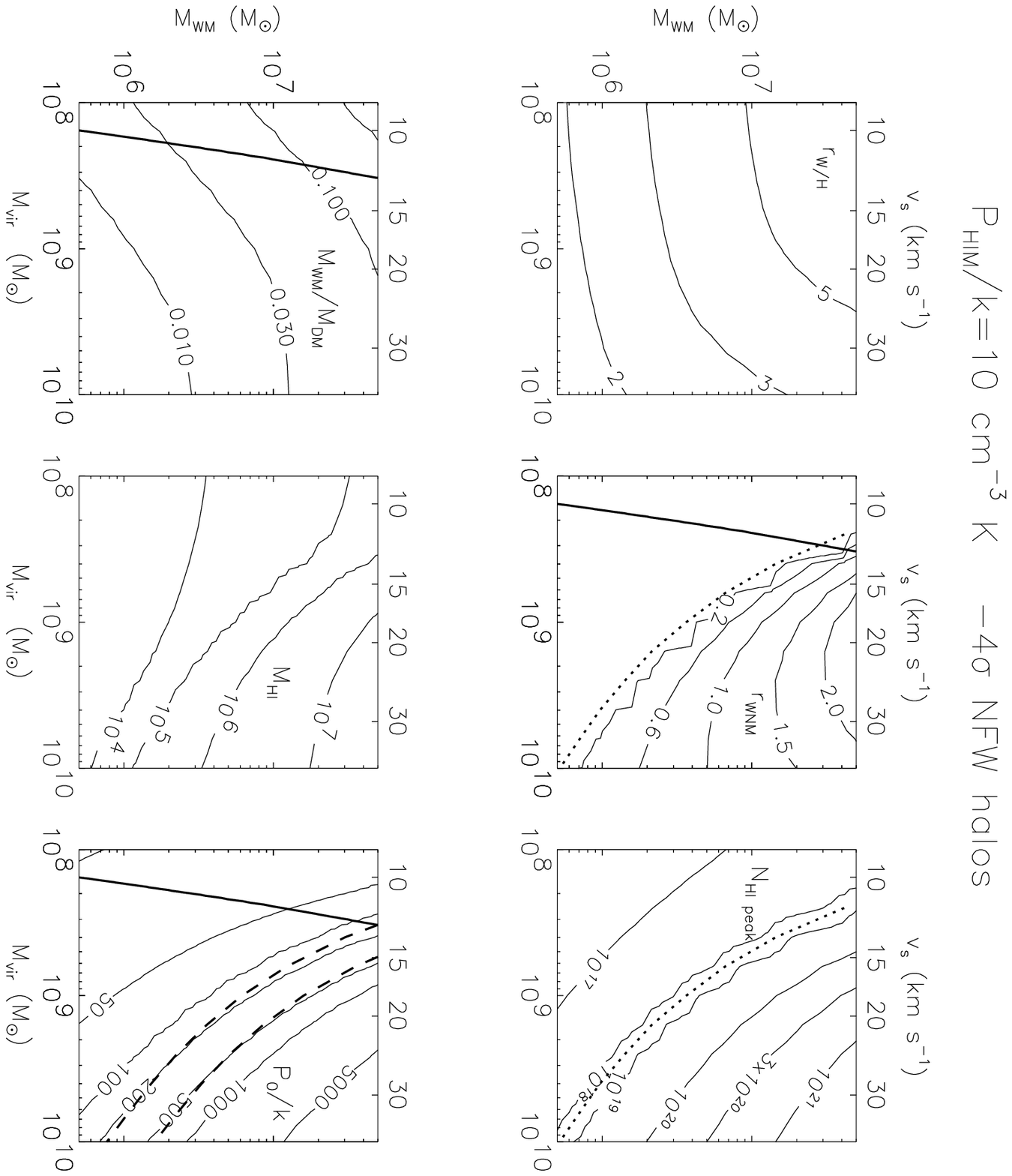}
\vspace{25cm}
\caption{$M_{\rm WM}$ vs. $M_{\rm vir}$ diagrams for $-4\sigma$ NFW halos,
and a bounding pressure $P_{\rm HIM}=10$ cm$^{-3}$ K.}
\end{figure}

\clearpage
\begin{figure}[htpb]
\plotone{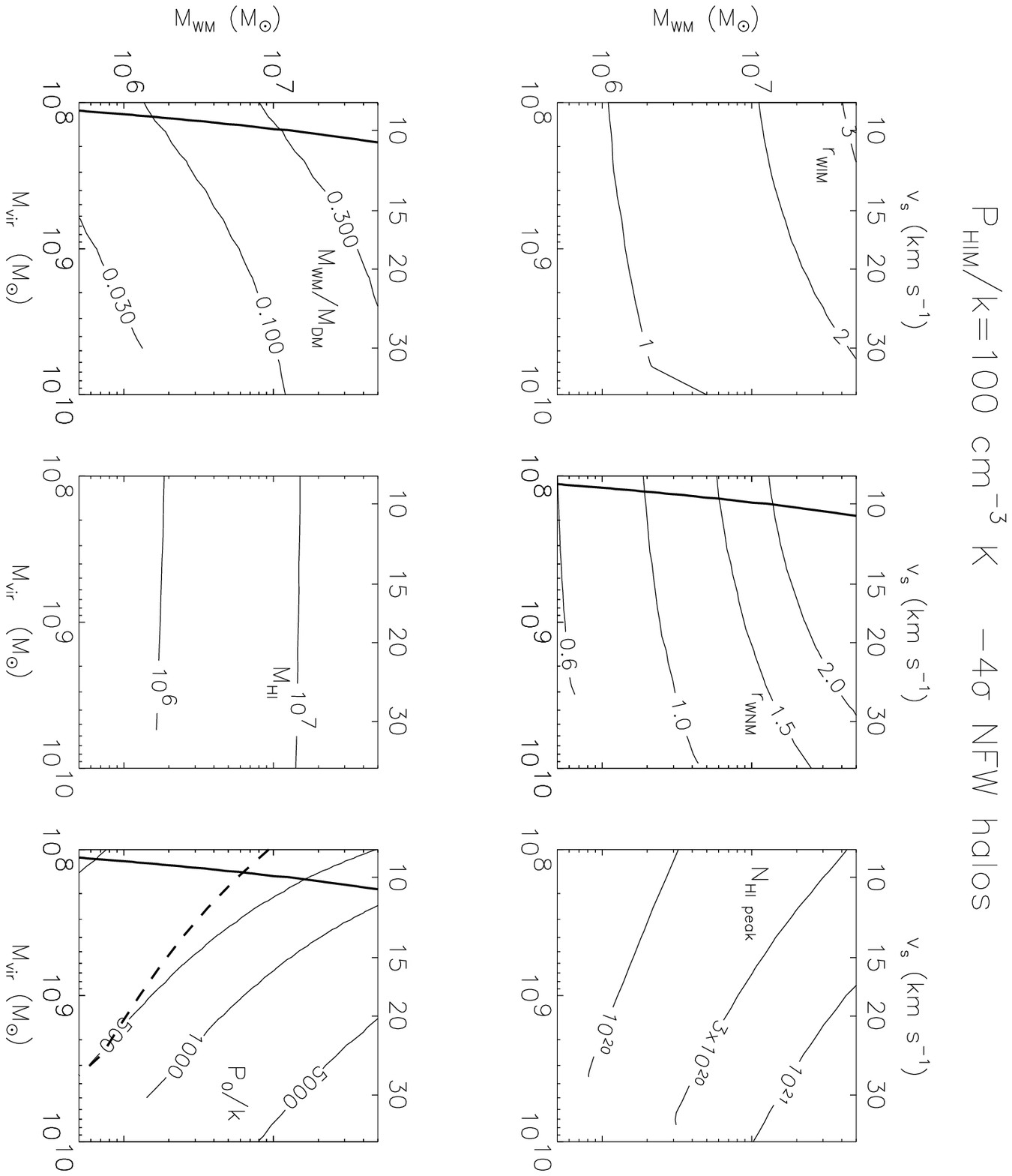}
\vspace{25cm}
\caption{$M_{\rm WM}$ vs. $M_{\rm vir}$ diagrams for $-4\sigma$ NFW halos,
and $P_{\rm HIM}=100$ cm$^{-3}$ K.}
\end{figure}

\clearpage
\begin{figure}[htpb]
\plotone{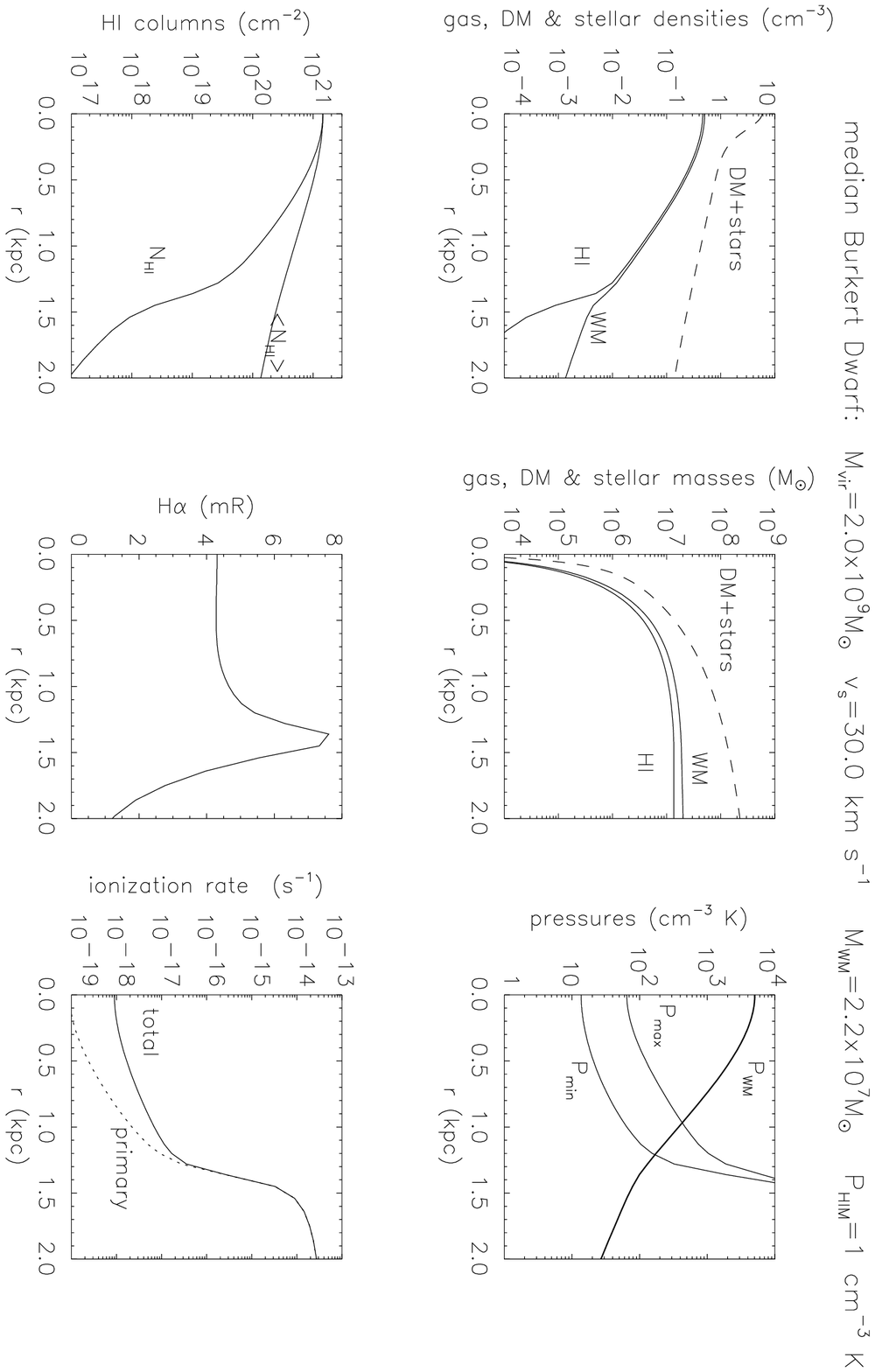}
\vspace{25cm}
\caption{Median $M_{\rm vir}=2\times 10^9$ $M_{\odot}$ 
(or $v_s=30$ km s$^{-1}$) Burkert halo model for the 
dwarf galaxies Leo A and Sag DIG:
(a) WM and HI gas densities (solid curves), 
and the sum of the DM and
stellar densities (dashed curve) as functions of the cloud radius $r$, 
(b) projected HI column density $N_{\rm HI}$,
and the average HI column, $<N_{\rm HI}>$,
within $r$, (c) enclosed WM and HI gas
masses (solid), and enclosed DM plus stellar masses (dashed),
(d) H$\alpha$ recombination line surface brightness, and (e) WM gas pressure,
$P_{\rm WM}$, and the critical pressures $P_{\rm min}$ and $P_{\rm max}$, and
(f) total (solid) and primary (dashed) hydrogen ionization rates.}

\end{figure}

\clearpage
\begin{figure}[htpb]
\plotone{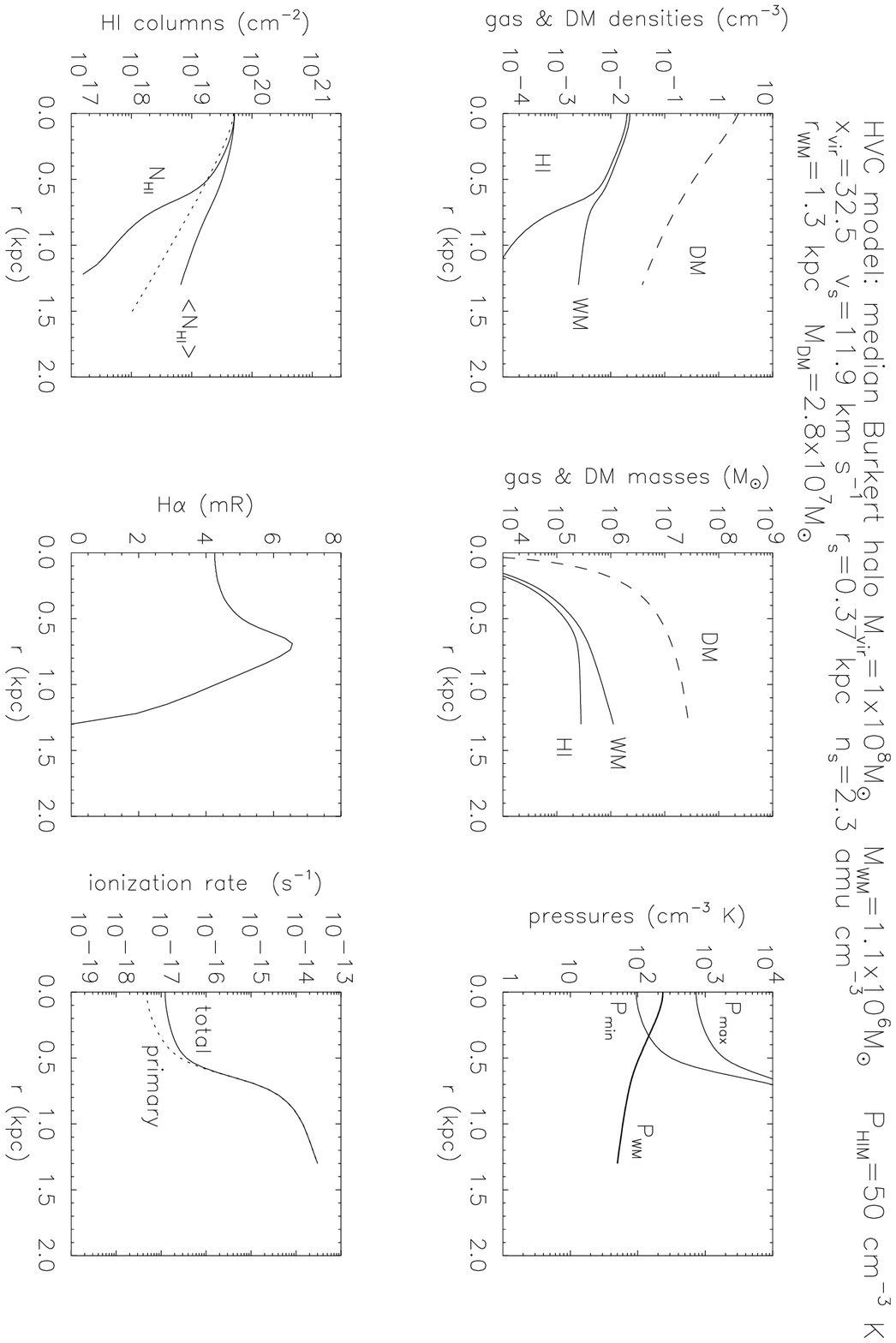}
\vspace{25cm}
\caption{Circumgalactic CHVC model. Median Burkert halo,
with $M_{\rm vir}=1\times 10^8M_\odot$ (or $v_s=12$ km s$^{-1}$),
and $P_{\rm HIM}/k=50$ cm$^{-3}$ K.
(a) WM and HI gas densities (solid curves), and the DM
density (dashed curve) as functions of the cloud radius $r$, 
(b) projected HI column density $N_{\rm HI}$,
and the average HI column within $r$,
the dotted curve is the HI column density for an exponential
HI gas density distribution (see text), (c) enclosed WM and HI gas
masses (solid), and DM mass (dashed),
(d) H$\alpha$ recombination line surface brightness, (e) WM gas pressure,
$P_{\rm WM}$, and the critical pressures $P_{\rm min}$ and $P_{\rm max}$,
and (f) total (solid) and primary (dashed) 
hydrogen ionization rates.}

\end{figure} 

\clearpage
\begin{figure}[htpb]
\plotone{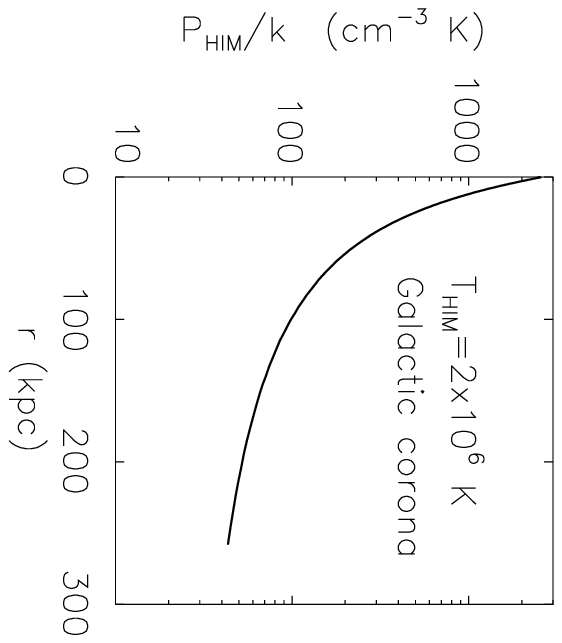}
\vspace{25cm}
\caption{Galactic HIM pressure for a $T_{\rm HIM}=2\times 10^6$ K corona.}
\end{figure}

\clearpage
\begin{figure}[htpb]
\plotone{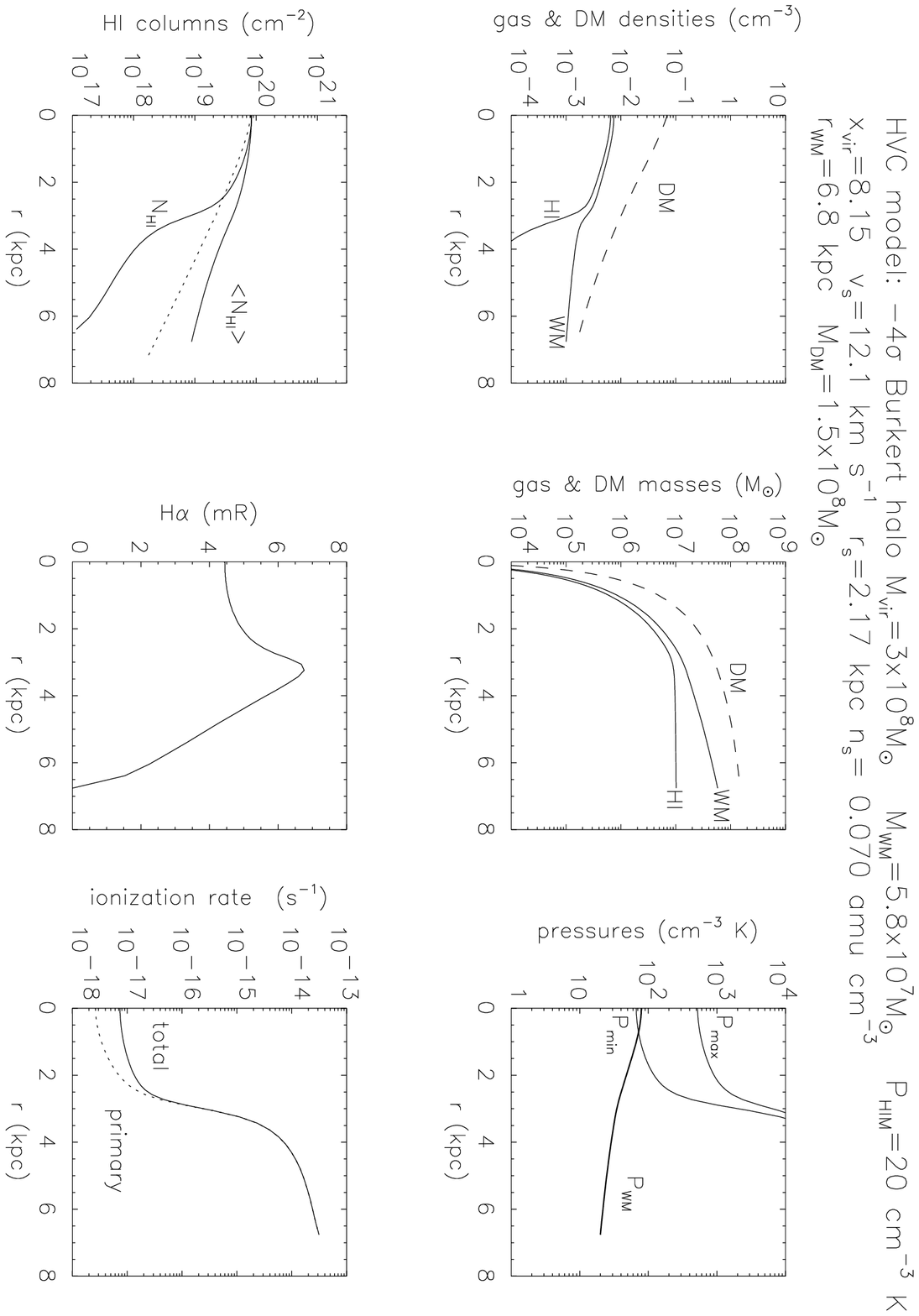}
\vspace{25cm}
\caption{Extragalactic CHVC model. $-4\sigma$ Burkert halo,
with $M_{\rm vir}=1\times 10^8M_\odot$ (or $v_s=12$ km s$^{-1}$),
and $P_{\rm HIM}/k=20$ cm$^{-3}$ K.
(a) WM and HI gas density distributions (solid curves), and the DM
densities (dashed curve), (b) projected HI column density $N_{\rm HI}$,
and the average HI column within $r$,
the dotted curve is the HI column density for an exponential
HI gas density distribution (see text), (c) enclosed WM and HI gas
masses (solid), and enclosed DM plus stellar masses (dashed),
(d) H$\alpha$ recombination line surface brightness, (e) WM gas pressure,
$P_{\rm WM}$, and the critical pressures $P_{\rm min}$ and $P_{\rm max}$,
and (f) total (solid) and primary (dashed) 
hydrogen ionization rates.}
\end{figure}

\clearpage
\begin{figure}[htpb]
\plotone{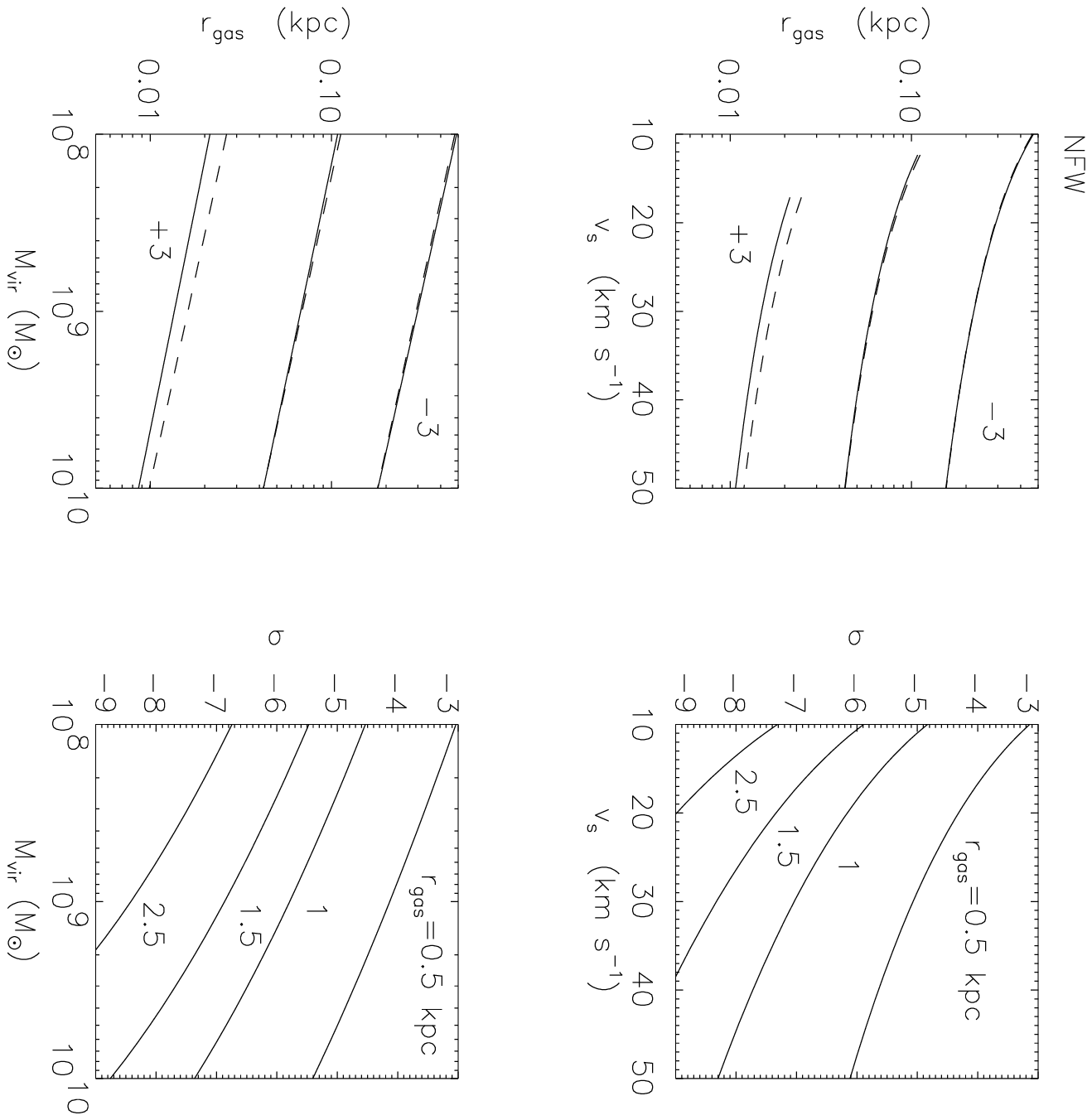}
\vspace{25cm}
\caption{The small-$x$ 1/e gas scale height $r_{\rm gas}$ for
a 10$^4$ K WNM in NFW halos:
(a) $r_{\rm gas}$ for median, $+3\sigma$, and $-3\sigma$ halos
as a function of $v_s$, (b) $r_{\rm gas}$ as a function of $M_{\rm vir}$,
(c) level curves for $r_{\rm gas}$ as
a function of $\sigma$ and $v_s$, and (d) level curves for $r_{\rm gas}$
as a function of $\sigma$ and $M_{\rm vir}$.
The dashed curves in panels (a) and (b)
display the scale height as given by the
analytic expressions in Table 6. 
}
\end{figure}

\clearpage
\begin{figure}[htpb]
\plotone{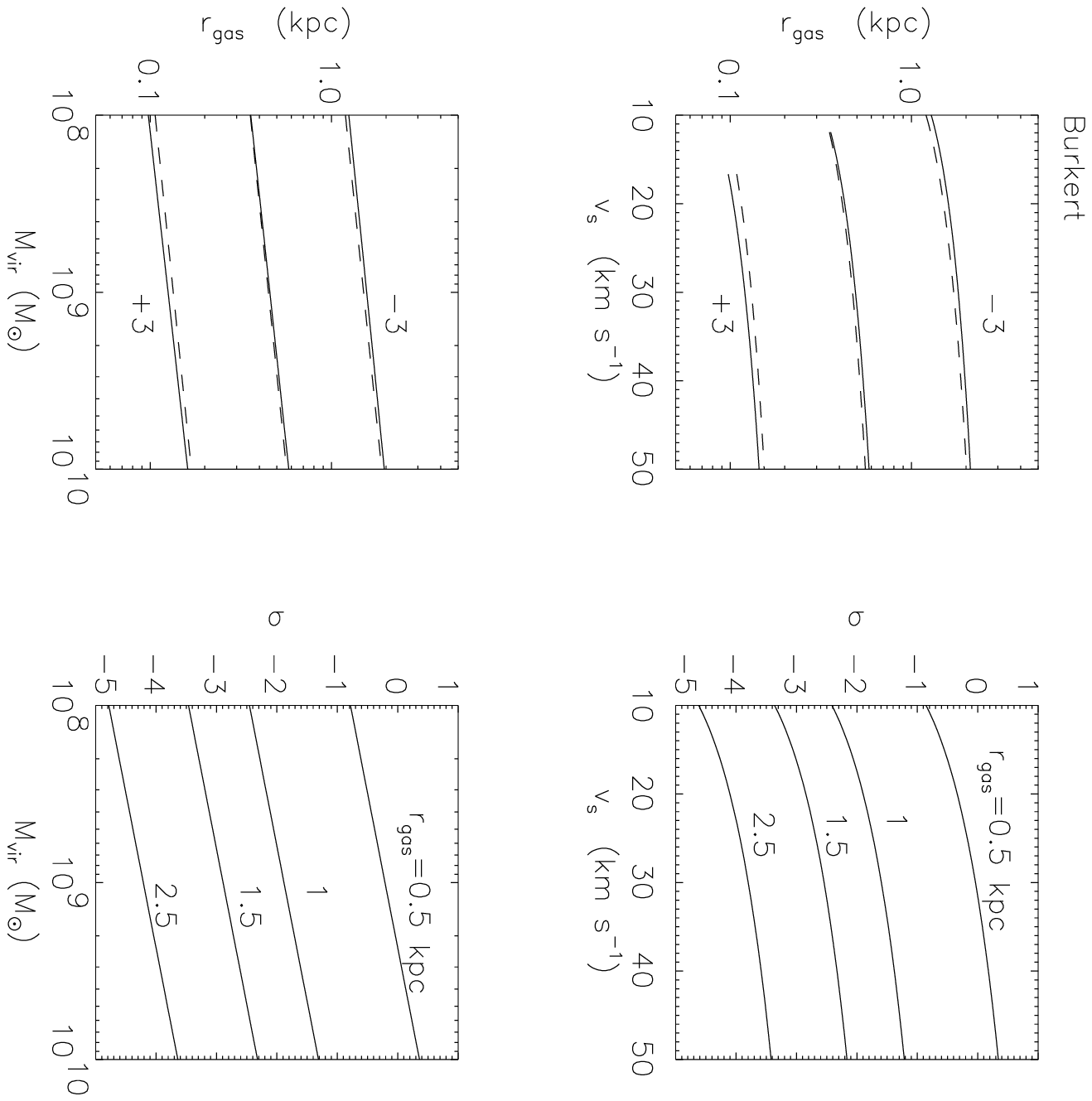}
\vspace{25cm}
\caption{The small-$x$ 1/e gas scale height $r_{\rm gas}$ for
a 10$^4$ K WNM in Burkert halos:
(a) $r_{\rm gas}$ for median, $+3\sigma$, and $-3\sigma$ halos
as a function of $v_s$, (b) $r_{\rm gas}$ as a function of $M_{\rm vir}$,
(c) level curves for $r_{\rm gas}$ as
a function of $\sigma$ and $v_s$, and (d) level curves for $r_{\rm gas}$
as a function of $\sigma$ and $M_{\rm vir}$.
The dashed curves in panels (a) and (b)
display the scale height as given by the
analytic expressions in Table 6.
}
\end{figure}

\clearpage
 
\begin{deluxetable}{lllll}
\footnotesize
\tablecaption{Dwarf Galaxies: Observations\label{tbl-dwarfs}}
\tablehead{
\colhead {} & unit & & \colhead{Leo A}  & \colhead{Sag DIG}
}
\startdata
distance             & kpc &        \ \ \ \         & 690   & 1061  \\
optical core radius  & kpc &        \ \ \ \         & 0.185  & 0.125 \\
$L_V$                &$10^7 \ L_\odot$ &   \ \ \ \         & 3.0    & 6.9 \\
single dish 21 cm flux & Jy km s$^{-1}$ &  \ \ \ \ & 69     & 33   \\
VLA 21 cm flux       &Jy km s$^{-1}$ &   \ \ \ \    & 65     & 30   \\
WNM velocity dispersion & km s$^{-1}$ &  \ \ \ \    & 9.0    & 9.5  \\
WNM mass fraction       &                &  \ \ \ \    & 0.80    & 0.83  \\
$N_{\rm HIpeak}$ (WNM) & cm$^{-2}$ & & $2.2\times 10^{21}$ & $1.0\times 10^{21}$\\
$\theta_{\rm gas}$ (projected)  & arcsec &         \ \ \ \      & 160    & 70  \\
$r_{\rm {N_{\rm HI}}}$ & kpc &                         & 0.5   & 0.4 \\
$N_{\rm HImin}$ & cm$^{-2}$ &   \ \ \ \    & $2\times 10^{19}$  & $5\times 10^{18}$ \\
$r_{\rm HImin}$ & kpc &         \ \ \ \           & 1.1    & 1.6  \\ 

\enddata


\end{deluxetable}

\clearpage
 
\begin{deluxetable}{lllll}
\footnotesize
\tablecaption{Dwarf Model: median Burkert halo\label{tbl-dwarfmodel}}
\tablehead{
\colhead {} & {} & {} & {} & {}
}
\startdata
\cutinhead{Halo Parameters}\
$x_{\rm vir}$    &    & &       & 25.6 \\
$M_{\rm vir}$  & M$_\odot$ & & & $2.0\times 10^9$ \\
$r_{\rm vir}$  & kpc  & &    & 32.5 \\
$n_{ds}$    &  amu cm$^{-3}$    & &    & 1.24 \\
$r_s$       &  kpc    & &    & 1.27 \\
$v_s$       &  km s$^{-1}$ & &     & 30.0 \\
$M_{ds}$    &M$_\odot$  & &   & $2.67\times 10^8$ \\
$r_{\rm gas}$ (small-$x$) & kpc & & & 0.49 \\
\cutinhead{Gas Input Parameters}
$P_{\rm HIM}/k$  & cm$^{-3}$ K & &  & 1.0 \\
$M_{\rm WM}$     & M$_\odot$  & &    & $2.2\times 10^{7}$ \\
$M_{\rm WM}/M_{\rm vir}$ &    & &    & $1.1\times 10^{-2}$ \\
\cutinhead{Gas Output Parameters}
$r_{\rm W/H}$  & kpc   & &    & 4.44      \\
$r_{\rm gas}$ (numerical) & kpc   & &   &  0.56         \\
$r_{\rm {N_{HI}}}$ & kpc  & &   &  0.59       \\
$r_{\rm WNM}$ & kpc   & &   & 1.39       \\
$N_{\rm HIpeak}$ & cm$^{-2}$ & & & $1.42\times 10^{21}$ \\
$M_{\rm HI}$  & M$_\odot$  & &  & $1.34\times 10^7$ \\
${\cal W}/{\cal T}$        & &   &         & 4.4 \\
$P_0/P_{\rm HIM}$  &    & &    & $4.97\times 10^3$ \\
$n_{\rm H,0}$        &  cm$^{-3}$   & &     & 0.46 \\

\enddata


\end{deluxetable}

\clearpage
 
\begin{deluxetable}{lll}
\footnotesize
\tablecaption{CHVC Circumgalactic median Burkert halo
\label{tbl-chvclg}}
\tablehead{
\colhead {} & {} & {}
}
\startdata
\cutinhead{Halo Parameters}
$x_{\rm vir}$    &          & 32.5 \\
$M_{\rm vir}$  & M$_\odot$ & $1.0\times 10^8$ \\
$r_{\rm vir}$  & kpc      & 12.0 \\
$n_{ds}$    &  amu cm$^{-3}$        & $2.33$ \\
$r_s$       &  kpc              & 0.37 \\
$v_s$       &  km s$^{-1}$      & 11.9 \\
$M_{ds}$    &M$_\odot$     & $1.22\times 10^7$ \\
$r_{\rm gas}$ (small-$x$) & kpc & 0.36 \\
\cutinhead{Gas Input Parameters}
$P_{\rm HIM}/k$  & cm$^{-3}$ K & 50.0 \\
$M_{\rm WM}$     & M$_\odot$     & $1.1\times 10^{6}$ \\
$M_{\rm WM}/M_{\rm vir}$ &       & $1.1\times 10^{-2}$ \\

\cutinhead{Gas Output Parameters}
$r_{\rm W/H}$  & kpc       & 1.30      \\
$r_{\rm gas}$ (numerical) & kpc      & 0.55          \\
$r_{\rm {N_{HI}}}$ & kpc     & 0.50        \\
$r_{\rm WNM}$ & kpc      & 0.67      \\
$r_{\rm CNM}$ & kpc      & 0.33      \\
$N_{\rm HIpeak}$ & cm$^{-2}$ & $5.10\times 10^{19}$ \\
$M_{\rm HI}$  & M$_\odot$  & $2.81\times 10^5$ \\
${\cal W}/{\cal T}$      &              & 1.0 \\
$P_0/P_{\rm HIM}$  &        & 4.7 \\
$n_{\rm H,0}$        &  cm$^{-3}$        & $2.08\times 10^{-2}$ \\

\enddata


\end{deluxetable}

\clearpage
 
\begin{deluxetable}{lll}
\footnotesize
\tablecaption{CHVC extragalactic Model: $-4\sigma$ 
              Burkert halo\label{tbl-chvccircum}}
\tablehead{
\colhead {} & {} & {}
}
\startdata
\cutinhead{Halo Parameters}
$x_{\rm vir}$    &          & 8.1 \\
$M_{\rm vir}$  & M$_\odot$ & $3.2\times 10^{8}$ \\
$r_{\rm vir}$  & kpc      & 17.7 \\
$n_{ds}$    &  amu cm$^{-3}$    & $7.0\times 10^{-2}$ \\
$r_s$       &  kpc              & 2.17 \\
$v_s$       &  km s$^{-1}$      & 12.1 \\
$M_{ds}$    &M$_\odot$     & $7.43\times 10^7$ \\
$r_{\rm gas}$ (small-$x$) & kpc & 2.1 \\
\cutinhead{Gas Input Parameters}
$P_{\rm HIM}/k$  & cm$^{-3}$ K & 20.0 \\
$M_{\rm WM}$     & M$_\odot$     & $7.1\times 10^{7}$ \\
$M_{\rm WM}/M_{\rm vir}$ &    & 0.22 \\
\cutinhead{Gas Output Parameters}
$r_{\rm W/H}$  & kpc       & 7.15      \\
$r_{\rm gas}$ (numerical) & kpc      & 3.07         \\
$r_{\rm {N_{\rm HI}}}$ & kpc     & 2.63         \\
$r_{\rm WNM}$ & kpc      & 3.40      \\
$r_{\rm WNM/CNM}$ & kpc      & 1.16      \\
$N_{\rm HIpeak}$ & cm$^{-2}$ & $9.99\times 10^{19}$ \\
$M_{\rm HI}$  & M$_\odot$  & $1.44\times 10^7$ \\
${\cal W}/{\cal T}$      &              & 1.0 \\
$P_0/P_{\rm HIM}$  &        & 4.4  \\
$n_{\rm H,0}$        &  cm$^{-3}$        & $7.65\times 10^{-3}$ \\

\enddata


\end{deluxetable}

\clearpage

\begin{table}
\caption{Halo Profiles}
\begin{tabular}{lll}
\tableline
\tableline
\multicolumn{1}{c}{NFW} \\
\tableline

$f_\rho$  $=$  $\frac{1}{x(1+x)^2}$ & & \\

$f_M$ $=$ $3\biggl[{\rm ln}(1+x)-{x\over 1+x}\biggr]$ & & \\

$f_\varphi$  $=$  $3\biggl[1-{{\rm ln}(1+x) \over x}\biggr]$ & & \\

$f_{\rm gas}$ $=$ $e^{{-3(v_s/c_g)^2}}(1+x)^{{(v_s/c_g)^2}/x}$ & & \\

\tableline
\multicolumn{1}{c}{Burkert}\\
\tableline

$f_\rho$  $=$  $\frac{1}{(1+x)(1+x^2)}$ & & \\

$f_M$ $=$ ${3\over 2}\biggl[{{\rm ln}(1+x^2)\over 2} + {\rm ln}(1+x)
- {\rm tan}^{-1}x\biggr]$ & & \\

$f_\varphi$ $=$ ${3\over 2}\biggl\lbrace{{\rm tan}^{-1}x)(1+{1\over x}) 
- [{\rm ln}(1+x)](1+{1\over x}) 
+{1\over 2}[{\rm ln}(1+x^2)](1-{1\over x})\biggr\rbrace}$ & & \\

$f_{\rm gas}$ $=$ $\biggl[e^{{-(1+1/x){\rm tan}^{-1}x}} (1+x)^{(1+1/x)}
(1+x^2)^{(1/2)(1/x-1)}\biggr]^{(3/2)(v_s/c_g)^2}$ & & \\

\tableline

\end{tabular}

\end{table}

\clearpage
\begin{figure}[htpb]
\plotone{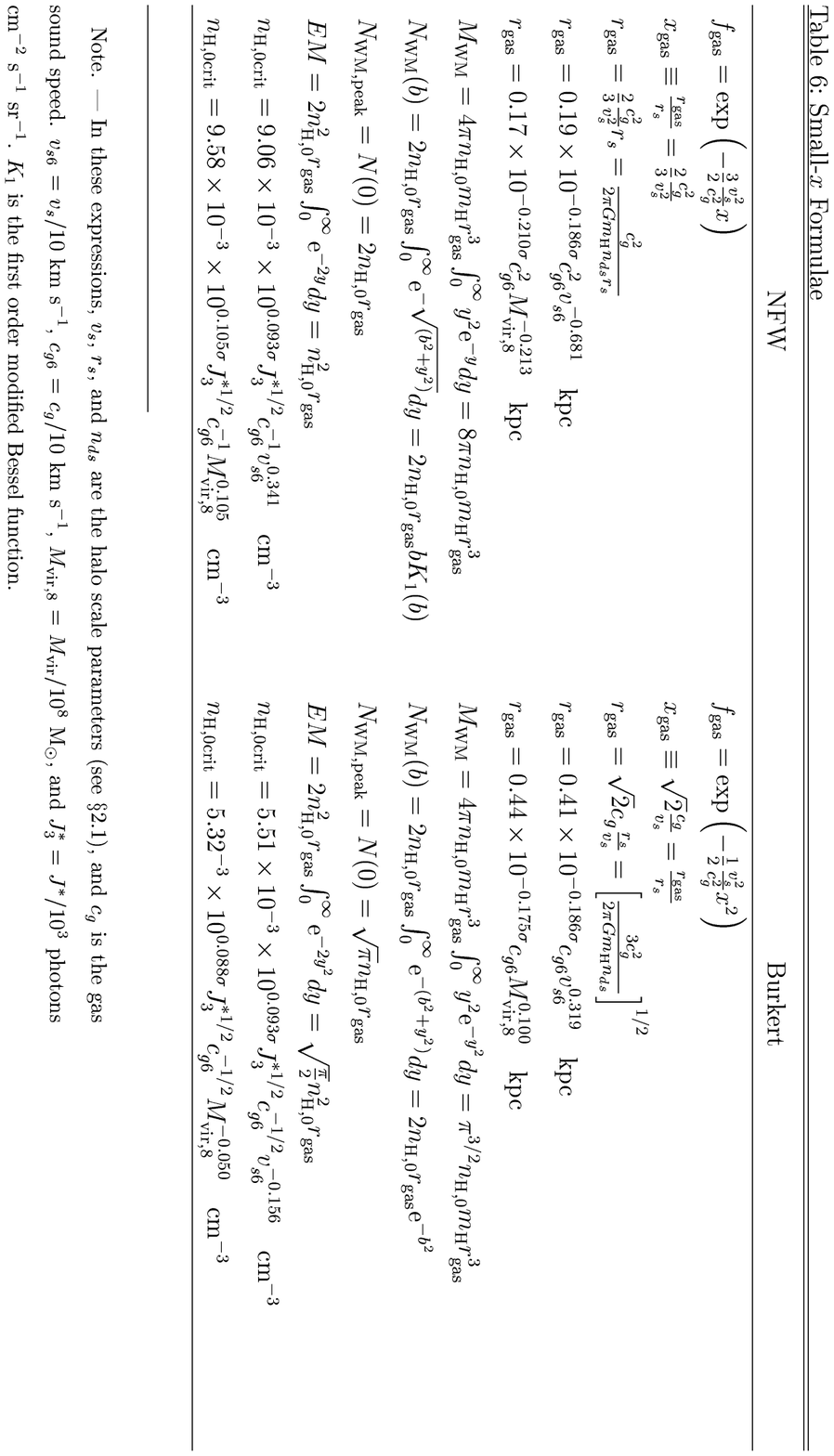}
\end{figure}

\end{document}